\documentclass[colorlinks=true,urlcolor=blue,linkcolor=blue,citecolor=blue,12pt,a4paper,reqno]{amsart}
\usepackage{amssymb}
\usepackage[usenames]{xcolor}
\usepackage[normalem]{ulem}
\usepackage{hyperref}
\title{On the Reconstruction of Static and Dynamic Discrete Structures}
\author{Andreas Alpers and Peter Gritzmann}
\address{Zentrum Mathematik, Technische Universit\"at M\"unchen, D-85747 Garching bei M\"unchen, Germany}
\email{alpers@ma.tum.de, gritzmann@tum.de}

\usepackage[utf8]{inputenc}%(only for the pdftex engine)
\usepackage{microtype}
\usepackage{enumitem}
\usepackage{subfigure}
\usepackage{tikz}

\newtheorem{thm}{Theorem}

\newtheorem{question}{Problem}

\newcommand{\inte}{{\mathrm{int}}}
\newcommand{\lin}{{\mathrm{lin}}}
\newcommand{\conv}{{\mathrm{conv}}}

\newcommand\NP{\mathbb{N}\mathbb{P}}
\newcommand\R{\mathbb{R}}
\newcommand\Z{\mathbb{Z}}
\newcommand\N{\mathbb{N}}

\renewcommand\P{\mathbb{P}}
\newcommand\Q{\mathbb{Q}}
\newcommand\CP{\mathcal{P}}
\newcommand\CO{\mathcal{O}}
\newcommand\CF{\mathcal{F}}
\newcommand\CS{\mathcal{S}}
\newcommand\CI{\mathcal{I}}
\newcommand\CL{\mathcal{L}}
\newcommand\CC{\mathcal{C}}
\newcommand\DS{\displaystyle}
\newcommand{\1}{1 \kern -.41em 1 }

\newenvironment{problem}[1]{
\vspace{.4cm}
\begin{minipage}[c]{0.9\textwidth}
       \noindent
       {#1.}
        \nopagebreak
        \it
        \begin{list}{}
          {\setlength{\labelwidth}{16mm}% {2cm}fuer 12pt
           \setlength{\leftmargin}{18mm}% {25mm}
           \topsep2pt
           \itemsep2pt
           
          }
        \nopagebreak
  }
{\end{list}\bigskip%\vspace{3mm plus 1mm minus 1mm}
\end{minipage}
}

\begin{document}
\date{}

\maketitle

\begin{abstract}
We study inverse problems of reconstructing static and dynamic discrete structures from tomographic data
  (with a special focus on the `classical' task of reconstructing finite point sets in $\R^d$).
  The main emphasis is on recent mathematical developments and new applications, which emerge in scientific areas such as physics 
  and materials science, but also in inner mathematical fields such as number theory, optimization, and imaging. 
  Along with a concise introduction to the field of discrete tomography, we give pointers 
  to related aspects of computerized tomography in order to  contrast the worlds of continuous and discrete inverse problems.
\end{abstract}

\tableofcontents

\section{Introduction}

We begin with an informal definition of the general field of discrete tomography. As a comprehensive treatise of the general field would,
however, go far beyond the scope of the present paper, and as we want to limit the overlap with surveys in the literature as much as 
possible, we will focus on  the most fundamental case of reconstructing finite point sets in $\R^d$ from some of their 
discrete X-rays.
Our special emphasis will be on (a subjective selection of) recent developments and applications. We conclude the introduction
with a few comments on the structure of the paper and its bibliography.

\subsection{What is discrete tomography?}

Discrete tomography deals with the problem of retrieving knowledge about an otherwise unknown 
discrete object from information about its interactions with certain query sets. Of course, this is not a formal
definition and, as a matter of fact, the occuring terms leave room for different interpretations. 
For instance, the {\em discrete object} can be any set in some $\R^d$ that allows a finite encoding, e.g.,  
a finite point set, a polytope or even a semialgebraic set. Also functions with finite support are included.
{\em Knowledge} may mean full reconstruction, the detection of certain properties and measures of the object or 
just a `yes/no' decision whether the object equals (or is close to) a given blueprint. 
The {\em query sets} may be windows, affine spaces or certain families of more general manifolds, and {\em interaction} may
simply mean intersection but could also refer to a very different probing procedure.

While results which, in retrospective, belong to this area go back a long time, the name
{\em discrete tomography} and the establishment of the so-named field is of more recent origin. 
In the past decades the focus has been on the issues of {\em uniqueness}, {\em computational complexity}, and {\em algorithms}, first under the theoretical
assumption that exact X-ray data were available. Later, {\em stability} and {\em instability} questions were pursued, 
and the effect of noise was studied. 

Discrete tomography has important applications in physics, materials science, and many other fields.
It has, however, also been applied in various other contexts, including scheduling~\cite{ggp-2000}, 
data security~\cite{irving-jerrum-94}, image processing~\cite{slump-gerbrands-82}, data compression~\cite{datacompression}, 
combinatorics~\cite{ryserbook, permutationmatrices08, planepartitions, flrs-91}, and graph theory~\cite{graphsinDT, ferrara}. 
Closely connected are also several recreational games such as nonograms~\cite{nonograms}, path puzzles~\cite{pathpuzzles}, 
sudokus~\cite{sudoku}, and color pic-a-pix~\cite{colorpicapix}. 

\subsection{Scope of the present paper}

In the following we will concentrate mainly on the `classical' task of reconstructing a finite point set 
in $\R^d$ from the cardinalities of its intersections with the lines parallel to a (small) finite number 
of directions. Already in this restricted form, discrete tomography displays important features also known 
from the continuous world.
In particular, discrete tomography is ill-posed in the sense of Hadamard~\cite{hadamard-23}: the data may be inconsistent,
the solution need not be unique~(Thm.~\ref{thm:nonunique}), and 
small changes in the data may result in dramatic changes of the solutions~(Thm.~\ref{thm:instabDT3}).

As will become clear, discrete tomography is not simply the discretization of continuous tomography.
It derives its special characteristics from the facts that, on the one hand, there are only data in very few 
directions available, but on the other hand, the classes of objects that have to be reconstructed are rather restricted. 
Therefore discrete tomography is based on methods from combinatorics, 
discrete optimization, algebra, number theory, and other more discrete subfields of mathematics and 
computer science.

There exist already books and articles, which give detailed accounts of various aspects of 
discrete tomography and its applications. We single out \cite{gardnerbook,grimmgritzmannhuck, gritzmann-97,gritzmann-devries-2001, kubaherman1,kubaherman2,ryser}.
The present article differs from these surveys in various ways: the mathematical focus will be
on recent developments (which have not been covered in previous surveys). Further, new applications will play a
significant role, i.e., applications to other scientific areas like physics and materials science, but also
to inner mathematical fields such as number theory, optimization, and imaging.
We will, moreover, include pointers to certain related aspects of computerized tomography in order to 
contrast the worlds of continuous and discrete inverse problems. As general sources for the continuous case and inverse problems, see~\cite{hermanbook, markoebook, nattererbook, nattererbook2} and \cite{discreteinversepbook2, hansenbook, kirschbook, siltanenbook, tarantolabook}, respectively.

As a service to the reader and with a view towards a more complete picture we will restate some 
aspects which are basic for the present article but have been covered before. In order to
limit the overlap to other surveys we will, however, neither elude on the tomographic reconstruction of 
quasicrystals (see \cite{grimmgritzmannhuck}) or polyominoes (see~\cite{polyominoes-99}), nor on
the polyatomic case (see~\cite{gardnergritzmann99, duerr-Guinez-matamala-12}) or point X-rays (see~\cite{duliogardner}). Also, we will
not study general $k$-dimensional X-rays (see~\cite{grimmgritzmannhuck, gritzmann-devries-2002}) but
concentrate on the case $k=1$. This means that our exposition will be based on the \emph{X-ray transform}
rather than on the \emph{Radon transform} (which is the case~$k=d-1)$.

\subsection{Organization of the present paper and its bibliography}

After introducing the basic notation in Sect.  \ref{h-DT-basics} we will briefly survey well-known structural results related to the 
ill-posedness of the problems (Sect. \ref{sec-structural}) and their computational complexity 
(Sect. \ref{sec:compaspects}). Turning to recent results and applications, Sect. \ref{sect:superresolution} will illustrate some quite unexpected 
complexity jumps in (a related basic model of) superresolution. 
Particular emphasis will then be placed on new developments in {\em dynamic discrete tomography} 
involving the movement of points over time which are only accessible by very few of their X-ray images. 
As Sect. \ref{sect:dynamics} will show aspects of discrete tomography and particle tracking interact deeply. 
Another more recent issue, which comes up in materials science, is that of multi-scale tomographic imaging.
Sect. \ref{sec:tom-grain-mapping} will indicate how  different aspects of the reconstruction of polycrystalline materials based on 
tomographic data lead to very different techniques involving methods from the geometry of numbers, combinatorial 
optimization and computational geometry.
Sect. \ref {sect:numbertheory} deals with some inner mathematical connections between discrete tomography and 
the Prouhet-Tarry-Escott problem from number theory, and Sect. \ref{sec:concluding-remarks} concludes with some final remarks.

Let us point out that (with the exception of some new interpretations and simple observations) 
the results stated here have all been published in original research papers (which are, of course, cited appropriately).
Even more, since we want to use the standard notation and, in particular, a standard framework for expressing the
results, some overlap with the above mentioned surveys is unavoidable. 

Finally, let us close the introcuction with a comment on the bibliography. Due to the character of the present paper
we included references of different kinds. Of course, we listed all original work quoted in the main body of the
paper. However, we felt that for the generally interested reader it would be worthwhile to add sources for 
general reading. On the other hand, in terms of the included applications we focussed mainly on outlining 
those aspects to which discrete tomography can potentially contribute. While this is in line with the scope of the
present paper, readers interested in these fields of applications may appreciate pointers to sources for additional
reading. Hence we organized the bibliography in six different parts, namely general reading,
papers in tomography, and further reading on particle tracking, tomographic grain mapping, 
macroscopic grain mapping, and the Prouhet-Tarry-Escott problem, respectively.

\section{Basic notation}\label{h-DT-basics}

As pointed out before we will focus on the `classical' inverse problem 
of reconstructing a finite point set $F$ in $\R^d$ or $\Z^d$ from the cardinalities of its intersections 
with the lines parallel to a finite number of directions.  There are, however, certain aspects 
which involve weights on the points of $F$. Hence we will introduce the basic notions
for appropriate generalizations of characteristic functions of finite point sets, partly following
 \cite{grimmgritzmannhuck}. 

As usual, let $\mathbb{N}_0$, $\mathbb{N}$, $\mathbb{Z},$ $\mathbb{Q},$ and $\mathbb{R}$ 
denote the sets of non-negative integers, natural numbers, integers, rationals, and reals, respectively. 
Further, for $n\in\mathbb{N},$ we will often use the notation $[n] = \{1,\dots,n\}$ and
$[n]_0=[n]\cup\{0\}$.

In the following, let $d,m \in \mathbb{N}$; $d$ denotes the dimension of the space $\R^d$, and~$m$ is the number of directions in which images are taken. To exclude trivial cases, we will
usually assume that $d,m\ge 2$.

In oder to describe the objects of interest, we fix nonempty sets~$D \subset \mathbb{R}^d$ and~$C\subset \mathbb{R}$ and consider functions $\psi\!:\, D\rightarrow C$ 
with finite support $\textnormal{supp}(\psi)=\{x\in D : \psi(x)\neq 0\}$.  In  our context,
the most relevant pairs $(D,C)$ of a domain and a codomain are those where  $D=\mathbb{R}^d$ 
or $D=\mathbb{Z}^d$ and $C=\{0,1\}$.
Other standard codomains are $C=\mathbb{N}_0$, $C=\mathbb{Z},$ and also their relaxations
$[0,1]$,  $[0,\infty[$, and $\R$.

For any  pair $(D,C)$, let $\mathcal{F}(D,C)$ denote the class of all functions $\psi\!:\, D \rightarrow C$ with finite
support.  Of course, for $C=\{0,1\}$, such a function $\psi$ can be viewed as the indicator or characteristic function of a 
finite set $F$ and can therefore be identified with $\textnormal{supp}(\psi)$.  
We will write $\mathcal{F}(D)$ for $\mathcal{F}(D,\{0,1\})$ and identify it with the set of all finite subsets of~$D$.
In particular, the case $\mathcal{F}(\mathbb{Z}^d)$ encodes the classical \emph{finite lattice sets}.
Since this case is particularly important we will often abbreviate $\mathcal{F}(\mathbb{Z}^d)$ by $\mathcal{F}^d$.

Further, let $\mathcal{S}^{d}$ denote the set of all $1$-dimensional subspaces of $\mathbb{R}^d$, 
while $\mathcal{L}^{d}$ is the set of $1$-dimensional \emph{lattice lines}, i.e., lines through the origin spanned by an integer vector.
For $S\in\mathcal{S}^{d}$, we use the notation ${\mathcal{A}}(S)$ for the set of all
affine lines in $\mathbb{R}^d$ that are parallel to $S$.
The situation of $(\mathcal{F}^d,\mathcal{L}^{d})$ will be referred to as the
\emph{lattice case}.

\begin{figure}[htb]
\centering
\includegraphics[width=0.33\textwidth]{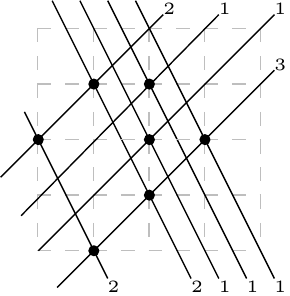}
\caption{A finite lattice set (black dots) and its 1-dimensional X-rays in the two directions~$(1,1)^T$ and~$(1,-2)^T.$ } \label{fig:Xrays}
\end{figure}

Now, let $\psi\in \mathcal{F}(D,C)$ and $S\in\mathcal{S}^{d}$. The \emph{discrete X-ray} of $\psi$ parallel to
$S$ (or, in a slight abuse of language, in the direction $S$) is the function 
$X_{S} \psi\!:\, \mathcal{A} (S)\rightarrow \mathbb{R}$ defined by
\[
  T \, \longmapsto \, \bigl(X^{}_{S}\psi\bigr) (T) \, = 
  \sum_{x\in T}\psi(x).
\] 
Since $\psi$ has finite support all sums are finite.
In the case of $C=\{0,1\}$ where $\psi$ can be identified with $F=\textnormal{supp}(\psi)$, we will 
often write $X_{S} F$. 
See Fig.~\ref{fig:Xrays} for an illustration. 

The mapping $\mathcal{X}_\psi$ on $\mathcal{S}^{d}$ defined by $S\mapsto X^{}_{S} \psi$
is called the \emph{discrete X-ray transform} of~$\psi$. (In typical applications only very
few values of $\mathcal{X}_\psi$ are available.)

Note that it is straightforward to extend this notation to $k$-dimensional X-rays.
Accordingly, for $k=d-1$, we obtain the \emph{discrete Radon transform} of~$\psi$ whose
measurements come from \emph{hyperplane X-rays}. 
We will, however, focus on the X-rays defined above,
which provide $1$-dimensional measurements.
The basic task of discrete tomography is then to reconstruct an otherwise unknown function~$\psi\in \mathcal{F}(D,C)$ from its X-rays with respect to a finite number~$m$
of given lines~$S\in\mathcal{S}^{d}$. 

The X-ray information is encoded by means of \emph{data functions}.
In fact, the lines $T\in {\mathcal{A}}(S)$ can be parametrized by vectors $t \in S^{\perp}$ 
such that $T=t+S$. Hence, one may regard $X^{}_{S}\psi$ as a function on $S^{\perp}$.
For algorithmic purposes, it is often preferable to use other representations and encode
$X^{}_{S}\psi$ as a finite set of pairs $(x,\beta)$ with $x\in D$, $\beta \in C$ and
$X^{}_{S}\psi(x+S)=\beta$.

\section{Ill-posedness}\label{sec-structural}
We begin with some results that deal with the basic issues of uniqueness and stability.

\subsection{Uniqueness and non-uniqueness}\label{subsec-uniqueness}

Given a subset $\CF$ of $\CF(D)$, and $\CS \subset \CS^d$. We say that two different sets $F_1,F_2\in \CF$ 
are {\em tomographically equivalent} with respect to $\CS$ if $X_{S} F_1 = X_{S} F_2$ for all $S\in \CS$.
The pair $(F_1,F_2)$ will then also be referred to as a {\em switching component}.
Further, a set $F\in \CF$ is {\em uniquely determined} within $\CF$ by its X-rays parallel to the lines in~$\CS$ 
if there does not exist any other set $F'$ in $\CF$ that is {\em tomographically equivalent} to $F$ with respect to $\CS$.
If the context is clear we will simply say that $F\in \CF$ is {\em uniquely determined}.

The following classical non-uniqueness result, usually attributed to \cite{lorentz49}, has been rediscovered several times.

\begin{thm} \label{thm:nonunique}
For any finite subset  $\mathcal{L}$ of $\mathcal{L}^d$ there exist sets in $\mathcal{F}^d$ that cannot be 
determined by $X$-rays parallel to the lines in $\mathcal{L}.$
\end{thm}

\begin{figure}[htb]
\centering
\subfigure[] {\includegraphics[width=0.28\textwidth]{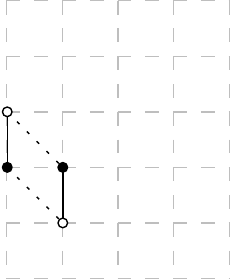}}\hspace*{4ex}
\subfigure[] {\includegraphics[width=0.28\textwidth]{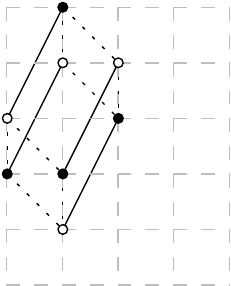}}\hspace*{4ex}
\subfigure[] {\includegraphics[width=0.28\textwidth]{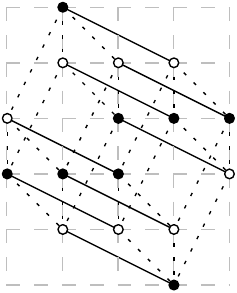}}
\caption{Construction of different lattice sets (black and white points) with equal X-rays in (a) two, (b) three, and (c) four directions. } \label{fig:zonotopeconstruction}
\end{figure}

Figure~\ref{fig:zonotopeconstruction} gives an illustration of the typical construction process to obtain different lattice sets with equal X-rays.

Note that Thm.~\ref{thm:nonunique} is in accordance with similar results in continuous or geometric tomography.
In fact, let $\CS \subset \CS^d$ be finite, and let $C,K\in \R^d$ be compact and $C\subset \inte(K)$.
Further let $f:\R^d\rightarrow \R$ be infinitely often differentiable with support $K$. 
Then there is a function $g$ with support in $K$, infinitely often differentiable, but otherwise arbitrary on 
$C$ such that the continuous X-rays of $f$ and $g$ with respect to all lines in $\CS$ coincide; 
for a proof, see~\cite{solomonwagner}.
Also, characteristic functions of compact sets, i.e., functions $f:\mathbb{R}^d\to\{0,1\}$ with compact support 
are not determined by their (continuous) X-rays in any finite number of directions; see~\cite[Thm.~2.3.3]{gardnerbook}.

While non-uniqueness is an undesirable feature for many applications it will play a positive role for 
applications in number theory later; see Sect.~\ref{sect:numbertheory}.

In contrast to Thm.~\ref{thm:nonunique}, uniqueness of the reconstruction can sometimes be guaranteed when 
certain prior knowledge is available. 

\begin{thm} \label{thm:irrationalunique}
There exists a line $S\in \CS^d$ such that every set $F\in \CF^d$ is uniquely determined by its one
X-ray $X_SF$.
\end{thm}

At first glance, this result may seem surprising. However, the used a priori information is that the set is
contained in $\Z^d$. Then, indeed, the X-ray $X_SF$ for any one line in~$\CS^d\setminus \CL^d$ 
determines $F$ uniquely, simply because no translate of $S$ can contain more than one lattice point.
As this argument shows, Thm.~\ref{thm:irrationalunique} can easily be extended to $\CF(\Z^d,C).$

While Thm.~\ref{thm:irrationalunique} does not seem to be of great practical use, it shows nonetheless
that no matter how fine the lattice discretization might be, one X-ray suffices. In the limit, however,
finitely many X-rays do in general not suffice. In this sense, discrete tomography does not behave like
a discretization of continuous tomography. Note that Thm.~\ref{thm:irrationalunique} is
quite different in nature from a result of \cite{solomonwagner} (see also \cite[Thm.~3.148]{markoebook}) that for almost any finite dimensional space of objects its elements can
be distinguished by a single X-ray in almost any direction. In fact, $\CF^d$ is not finite-dimensional,
and moreover, any line $S\in \CS^d\setminus \CL^d$ works for any set $F\in \CF^d,$ i.e., $S$ does not 
depend on the set but is given beforehand.

The next result is due to  R{\'e}nyi~\cite{renyi52} for $d=2$, who attributes an algorithmic proof to 
Haj{\'o}s. The generalization to $d\geq 2$ was given by Heppes~\cite{heppes}. 

\begin{thm}[\cite{renyi52} R{\'e}nyi] \label{thm:renyi}
Let $\mathcal{S}$ be a finite subset of $\mathcal{S}^d$. 
Then every set $F\in\mathcal{F}(\R^d)$ with $|F|\le | \CS| -1$ is uniquely determined by its X-rays 
parallel to the lines in $\mathcal{S}$.
\end{thm}

Since the two color classes of the two-coloring of the vertices of the regular $2m$-gon in the plane 
are tomographically equivalent with repect to the lines parallel to its edges, Thm.~\ref{thm:renyi}
is best possible. A strengthening for (mildly) restricted sets of directions is given in
\cite{bianchilonginetti}. But even more: generic directions are much better.

\begin{thm}[\cite{matousek08}] \label{thm:matousek}
There exist constants $c>0$ and $m_0\in\mathbb{N}$ such that for all $m\ge m_0$ the following holds:
For almost all sets  $\mathcal{S}\subset\mathcal{S}^2$ of~$m$ directions any $F\in\mathcal{F}(\mathbb{R}^2)$ 
with $|F|\leq2^{cm/\log(m)}$ is uniquely determined by its X-rays parallel to the lines of $\mathcal{S}$.
\end{thm}

Let us point out that in continuous tomography it is well-known that a compactly supported, infinitely 
differentiable function $f:\mathbb{R}^d\to\mathbb{R},$ which does not contain `details of size $2\pi/b$' 
or smaller can be recovered reliably from its X-rays on special sets of~$m$ directions, provided that 
$m>b.$ For a precise statement and proof, see~\cite[Thm.~2.4]{nattererbook}. See also~\cite{louisbookchapter}.
Such a result can be viewed, to some extent, as an analogue to R{\'e}nyi's theorem. 
(Note, however, that the difference to~$f$ is measured in an integral norm and hence the 
difference may get arbitrarily small without ever reaching $0$. 
For a characterization of the null-space, see~\cite{louisbookchapter, pmaass}.)

\begin{figure}[htb]
\centering
\includegraphics[width=0.25\textwidth]{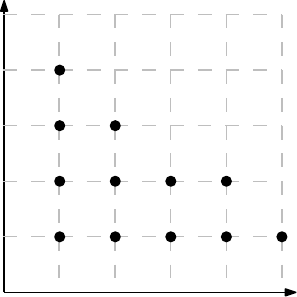}
\caption{A lattice set (black points) uniquely determined by its X-rays in horizontal and vertical directions.} \label{fig:renyiexample}
\end{figure}

For the special case $m=|\mathcal{S}|=2,$ uniqueness within $\mathcal{F}^2$ is characterized by the work of 
Ryser~\cite{ryser}. See Fig.~\ref{fig:renyiexample} for an example of a set of~$12$ points in $\Z^2$ that is uniquely determined by its two X-rays in the coordinate directions. 
For uniqueness results from two directions in geometric tomography, see~\cite{kubavolcic88, lorentz49}.

In the lattice case, the specialization of R{\'e}nyi's theorem that any set $F\in \CF^d$ is determined by any set of
$|F|+1$ lattice lines is only best possible for the cardinalities $|F|=m \in \{1,2,3,4,6\}$.
For other~$m$ the result can be improved at least by $1$. 

\begin{thm}[\cite{alpers-larman15}] \label{thm:renyiimprovement}
Let $m \in\mathbb{N}\setminus \{1,2,3,4,6\}$ and let $\mathcal{F}^d(m)$ be the class of sets in $\mathcal{F}^d$ 
of cardinality less than or equal to $m.$ Let $\mathcal{L}\subset \mathcal{L}^d$ with $|\mathcal{L}|\geq m.$ 
Then the sets in $\mathcal{F}^d(m)$ are determined by their X-rays parallel to the lines in $\mathcal{L}.$
\end{thm}

The question of smallest switching components is widely open in the lattice case.

\begin{question}
What is the smallest number $n=n(d,m)$ such that there exist $\mathcal{L}\subset \mathcal{L}^d$ 
with~$|\mathcal{L}|= m$ and two different lattice sets $F_1,F_2 \in \CF^d$ of cardinality $n$
that are tomographically equivalent with respect to the lines in $\mathcal{L}$?
\end{question}

Probabilistic arguments of \cite{alpers-larman15} show that, in the lattice case, switching components 
of a size that is polynomial in $m$ exist for each $d$. All deterministic constructions so far lead to exponential 
size switching components. Several small switching components are depicted in Fig.~\ref{fig:smallswitch}. 

We remark that switching components seem to have appeared first in the work of Ryser~\cite{ryser}. Later work on switching components includes~\cite{katzbook,chang, hajdutijdeman, kongherman98, matousek08, shilferstein}. Computational investigations related to the explicit construction of switching components can be found in \cite{brunettighosts}. Switching components for other projection models are considered in~\cite{svalbe3, svalbe1, svalbe0, svalbe2, zopf}.  
%The first computational study on generating minimal switching components seems to have been performed by Kiermaier~\cite{kiermaier-04}. 
Special types of switching components in the context of superresolution imaging, $hv$-convex polynomioes, and, in a more algebraic setting, are studied in~\cite{agsuperresolution}, \cite{bdfr-17}, and~\cite{kongherman98}, respectively. 

Quite strong uniqueness results exist for a geometrically motivated more restricted class of lattice sets.
A lattice set $F\in \CF^d$ is called {\em convex}, if 
$$
F=\conv(F)\cap \Z^d.
$$ 

\begin{thm}[\cite{gardnergritzmann97}]\label{thm:uniquenessconv}
There are $S_1,S_2,S_3,S_4\in \mathcal{L}^d$ such that every finite convex lattice set $F$ is uniquely determined 
by $X_{S_1}F,\ldots,X_{S_4}F$. Further, every set of at least seven coplanar lattice lines always suffices.
\end{thm}

Let us point out that the `good' sets of directions with respect to Thm.~\ref{thm:uniquenessconv} 
need not have coordinates of large absolute value. Examples for $d=2$ are 
\[\{(1,0)^T, (1,1)^T, (1,2)^T, (1,5)^T\} \quad \textnormal{and}\quad \{(1,0)^T, (2,1)^T, (0,1)^T, (-1,2)^T\}.\]
In fact, the sets of four good lattice lines in Thm. \ref{thm:uniquenessconv} are those whose cross-ratio of
their slopes does not lie in $\{4/3, 3/2, 2,3,4\}$. A converse result for the more general class of \emph{hv-convex lattice sets} 
is given in~\cite{barcucciconvex}. Generalizations of Thm.~\ref{thm:uniquenessconv} to so-called \emph{Q-convex lattice sets} and \emph{convex algebraic Delone sets} (in the context of quasicrystals) can be found in~\cite{daurat05} and \cite{huckspiess13}, respectively.

As a matter of fact, the directions in Thm.~\ref{thm:uniquenessconv} are all coplanar. It is not known how exactly
the situation changes if we insist that the lines are in {\em general position,} i.e., each~$d$ of them span~$\mathbb{R}^d.$

\begin{question}\label{quest-general}
Let $d\geq 3$. Is there a finite subset $\CL$ of lines in $\mathcal{L}^d$ in general position such that each convex 
set in $\CF^d$ is uniquely determined by its X-rays parallel to the lines in~$\CL?$
If so, what is the smallest cardinality? 

Is there a smallest number~$m$ such that any set $\CL\subset \mathcal{L}^d$ 
of $m$ lines has the property, that each convex set in $\CF^d$ is uniquely determined by its X-rays parallel to the lines in $\CL$?
\end{question}

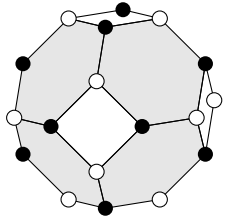
\begin{figure}[htb]
\centering
\begin{tikzpicture}[line join=bevel,z=-5.5, scale=0.6]

\coordinate (A0) at (1,0,2);
\coordinate (A1) at (1,0,-2);
\coordinate (A2) at (-1,0,2);
\coordinate (A3) at (-1,0,-2);
\coordinate (A4) at (2,1,0);
\coordinate (A5) at (2,-1,0);
\coordinate (A6) at (-2,1,0);
\coordinate (A7) at (-2,-1,0);
\coordinate (A8) at (0,2,1);
\coordinate (A9) at (0,2,-1);
\coordinate (A10) at (0,-2,1);
\coordinate (A11) at (0,-2,-1);
\coordinate (A12) at (0,1,2);
\coordinate (A13) at (0,1,-2);
\coordinate (A14) at (0,-1,2);
\coordinate (A15) at (0,-1,-2);
\coordinate (A16) at (2,0,1);
\coordinate (A17) at (2,0,-1);
\coordinate (A18) at (-2,0,1);
\coordinate (A19) at (-2,0,-1);
\coordinate (A20) at (1,2,0);
\coordinate (A21) at (1,-2,0);
\coordinate (A22) at (-1,2,0);
\coordinate (A23) at (-1,-2,0);

\draw  [fill opacity=0.1,fill=black!10!black] (A0) -- (A14)  -- (A10) -- (A21) -- (A5) -- (A16) -- cycle;
%\draw  [dashed] (A1) -- (A13) -- (A9) -- (A20) -- (A4) -- (A17) -- cycle;
\draw  [fill opacity=0.1,fill=black!10!black] (A2) -- (A12) -- (A8) -- (A22) -- (A6) -- (A18) -- cycle;
%\draw  [dashed] (A3) -- (A15) -- (A11) -- (A23) -- (A7) -- (A19) -- cycle;
\draw  [fill opacity=0.1,fill=black!10!black] (A4) -- (A20) -- (A8) -- (A12) -- (A0) -- (A16) -- cycle;
%\draw  [dashed] (A5) -- (A21) -- (A11) -- (A15) -- (A1) -- (A17) -- cycle;
\draw  [fill opacity=0.1,fill=black!10!black] (A7) -- (A23) -- (A10) -- (A14) -- (A2) -- (A18) -- cycle;
%\draw  [dashed] (A6) -- (A22) -- (A9) -- (A13) -- (A3) -- (A19) -- cycle;
\draw   (A0) -- (A12) -- (A2) -- (A14) -- cycle;
%\draw  [dashed] (A1) -- (A15) -- (A3) -- (A13) -- cycle;
\draw   (A4) -- (A16) -- (A5) -- (A17) -- cycle;
%\draw  [dashed] (A6) -- (A19) -- (A7) -- (A18) -- cycle;
\draw   (A8) -- (A20) -- (A9) -- (A22) -- cycle;
%\draw  [dashed] (A10) -- (A23) -- (A11) -- (A21) -- cycle;

\node[draw=none,shape=circle,fill, inner sep=2pt] (d1) at (A0){}; 
\node[draw=none,shape=circle,fill, inner sep=2pt] (d1) at (A2){}; 
\node[draw=none,shape=circle,fill, inner sep=2pt] (d1) at (A4){}; 
\node[draw=none,shape=circle,fill, inner sep=2pt] (d1) at (A5){}; 
\node[draw=none,shape=circle,fill, inner sep=2pt] (d1) at (A6){}; 
\node[draw=none,shape=circle,fill, inner sep=2pt] (d1) at (A7){}; 
\node[draw=none,shape=circle,fill, inner sep=2pt] (d1) at (A8){}; 
\node[draw=none,shape=circle,fill, inner sep=2pt] (d1) at (A9){}; 
\node[draw=none,shape=circle,fill, inner sep=2pt] (d1) at (A10){}; 

\node[draw,shape=circle,black, fill=white, inner sep=2pt] (d1) at (A12){};  
\node[draw,shape=circle,black, fill=white, inner sep=2pt] (d1) at (A14){};  
\node[draw,shape=circle,black, fill=white, inner sep=2pt] (d1) at (A16){};  
\node[draw,shape=circle,black, fill=white, inner sep=2pt] (d1) at (A17){};  
\node[draw,shape=circle,black, fill=white, inner sep=2pt] (d1) at (A18){};  
\node[draw,shape=circle,black, fill=white, inner sep=2pt] (d1) at (A20){};  
\node[draw,shape=circle,black, fill=white, inner sep=2pt] (d1) at (A21){};  
\node[draw,shape=circle,black, fill=white, inner sep=2pt] (d1) at (A22){};  
\node[draw,shape=circle,black, fill=white, inner sep=2pt] (d1) at (A23){};  
\end{tikzpicture}
\caption{Two-coloring of the vertices of the truncated octahedron.}\label{fig:truncoct}
\end{figure}

The color classes of a $2$-coloring of the vertices of the permutahedron in $\mathbb{R}^d$  provide a lower bound on 
such a universal number, which grows at least quadratically in~$m$~\cite{ghiglione}. Fig.~\ref{fig:truncoct} depicts 
the $3$-dimensional permutahedron, which is a truncated octahedron.

An analogue to Thm.~\ref{thm:uniquenessconv} also holds in the realm of geometric tomography: 
convex subsets of~$\mathbb{R}^d$ are determined by their continuous X-rays from sets of 
four `good' directions, see~\cite{GarM80}. 

Let us point out that it is the codomain $\{0,1\}$ which makes the problem difficult.
In fact, the case of functions in $\mathcal{F}(\mathbb{Z}^d,\Z)$ or, equivalently, lattice sets with integer weights is 
much simpler, as linear diophantine equations can be solved via the {\em Hermite normal form}; see e.g. \cite[Sect.~4\&5]{schrijverbook}).
But this also implies that the study of uniqueness for functions $\psi\in\mathcal{F}(\mathbb{Z}^d,\N_0)$ is
much easier. In fact, suppose we are given a finite set~$\CL\subset \CL^d$ and a bounded subset $B$ of $\Z^d$ which 
will act as a superset of all supports we are allowing. Then the corresponding X-ray problem with data functions all 
identical to~$0$ can be formulated as a homogenous system of linear diophantine equations and solved efficiently. 
Let $\psi$ be a non trivial solution, define $\psi^+: \mathbb{Z}^d\to \R$ by
$$
\psi^+(x)= \begin{cases} \psi(x) & \mbox{if $\psi(x) >0$;}\\
0 & \mbox{if $\psi(x)\le 0$;}
\end{cases}
$$
for $x\in \Z^d$, and set 
$$
\psi^- = \psi^+ - \psi.
$$
Then, of course, $\psi^+, \psi^-\in \mathcal{F}(\mathbb{Z}^d,\N_0)$, $\psi = \psi^+- \psi^-$, and
$X_S\psi\equiv 0$ for all $S\in \CL$. 
Hence, $\psi^+, \psi^-$ are tomographically equivalent with respect to $\CL$.

The uniqueness problem for functions $\psi\in\mathcal{F}(\mathbb{Z}^d,\mathbb{N}_0)$ 
also permits an algebraic characterization. The subsequently stated result of Hajdu and Tijdeman~\cite{hajdutijdeman-07,hajdutijdeman} uses the following notation.
A vector $v=(\nu_1,\dots,\nu_d)^T\in\mathbb{Z}^d$ is \emph{reduced} if $\textnormal{gcd}(\nu_1,\dots,\nu_d)=1.$ 
Let $v^+$ and~$v^-$ denote the vectors whose $j$th component is $\nu_j^+=\max\{0,\nu_j\}$ and $\nu_j^-=\max\{0,-\nu_j\},$ respectively. 
With~$\boldsymbol{X}$ we abbreviate the vector $(X_1,\dots,X_d)$ of indeterminants. 
Accordingly, for $a=(\alpha_1,\dots,\alpha_d)^T\in\mathbb{N}_0^d$, the monomial 
$X_1^{\alpha_1}\cdot\ldots\cdot X_d^{\alpha_d}\in \mathbb{Z}[\boldsymbol{X}]$ is 
denoted by~$\boldsymbol{X}^a$. 

\begin{thm}[\cite{hajdutijdeman-07, hajdutijdeman}] \label{thm:hajdutijdeman}
Let $\psi,\varphi\in\mathcal{F}(\mathbb{N}_0^d,\mathbb{N}_0),$ let $v\in\mathbb{Z}^d$ be reduced, and set~$S=\lin\{v\}$. 
Then $X_S\psi=X_S\varphi$ if, and only if, the polynomial
\[
\sum_{a\in \textnormal{supp}(\psi)}\boldsymbol{X}^a-\sum_{b\in \textnormal{supp}(\varphi)}\boldsymbol{X}^b
\] 
is divisible by $\boldsymbol{X}^{v+}-\boldsymbol{X}^{v^-}.$ 
\end{thm}

Note that the assumption that the functions are defined on $\N_0^d$ rather than on~$\Z^d$ is no restriction of generality. 

Let $v^{(1)},\dots,v^{(m)}\in \Z^d\setminus \{0\}$ be reduced,  $\mathcal{S}=\{\lin\{v^{(1)}\},\dots,\lin\{v^{(m)}\}\}$ 
and
\[f_\mathcal{S}=\prod_{v\in\{v^{(1)},\dots,v^{(m)}\}}(\boldsymbol{X}^{v^+}-\boldsymbol{X}^{v^-}).
\]
A consequence of Thm.~\ref{thm:hajdutijdeman} is that $\psi,\varphi\in\mathcal{F}(\mathbb{N}_0^d,\mathbb{N}_0)$ are tomographically equivalent with respect to~$\mathcal{S}$ if, and only if, there is a polynomial $p$ in $\mathbb{Z}[\boldsymbol{X}]$ such that
\[
\sum_{a\in \textnormal{supp}(\psi)}\boldsymbol{X}^a-\sum_{b\in \textnormal{supp}(\varphi)}\boldsymbol{X}^b
=p\cdot f_\mathcal{S}.
\]

The algebraic representation by polynomials can be utilized in various ways; examples will be given 
in Sect.~\ref{subsec-stability} (stability) and \ref{sect:numbertheory} (number theory). Additional aspects of uniqueness, in particular, concepts of \emph{additivity}, are discussed in~\cite{aharoniherman, flrs-91, glw11}. For uniqueness results for functions in~$\mathcal{F}(\mathbb{N}_0^d,\{0,1\})$ with several different types of bounded support, see~\cite{brunettiboundedsets} and the references cited therein.

\subsection{Stability and instability}\label{subsec-stability}

The results of the previous section were based on the assumption 
that the data functions are given exactly. We will now consider the case that the X-rays may contain errors. 

In the following we will measure the size of a function $\psi\in \CF(D,C)$ in terms of its~$\ell_1$-norm, i.e.,
$$
\|\psi\|_1=\sum_{x\in D} |\psi(x)|.
$$
In particular, given a finite set $\CS\subset \CS^d$ of lines and two sets $F_1,F_2 \in \CF(\R^d)$ their
{\em X-ray difference} will be
$$
\Delta_\CS(F_1,F_2)=\sum_{S\in \CS}||X_{S}F_1-X_{S}F_2||_1.
$$

The first result in this section shows that at least some (however marginal) stability is present. In fact, the 
X-ray difference, if not $0$, must jump to at least~$2(m-1)$. 
This means that either two sets are tomographically equivalent or their X-ray difference grows at least linearly
in $m$. 

\begin{thm}[\cite{ag06}]\label{thm:stability1} 
Let $\CS\subset \mathcal{S}^d$, $|\CS|=m$, and $F_1, F_2 \in \mathcal{F}(\mathbb{R}^d)$ with $|F_1|=|F_2|$. If 
$\Delta_\CS(F_1,F_2)<2(m-1)$, then $F_1$ and $F_2$ are tomographically equivalent with respect to $\CS$.
The same statement holds in the lattice case. 
\end{thm} 

As we will see in Thm.~\ref{thm:instabDT3}, this result is, in fact, best possible. First, we use it to give `noisy' 
variants of some of the uniqueness results of the previous section.
We begin with a stable version of Thm. \ref{thm:renyi}.

\begin{thm}[\cite{ag06}] \label{thm:stability:renyi}
Let  $\CS\subset \CS^d$, $F_1,F_2 \in \mathcal{F}(\mathbb{R}^d)$, and $\Delta_{\CS}(F_1,F_2)<2|F_1|$.
Further, let $|F_1|=|F_2|$ and $|F_1|+1 \le |\CS|$, or let $|F_1|\le |F_2|$ and $2|F_1| \le |\CS|$. Then
$F_1=F_2$. 
The statement persists in the lattice case.
\end{thm}

The following result is a stable version of Thm.~\ref{thm:uniquenessconv}.

\begin{thm}[\cite{ag06}]\label{thm:stabilityconv}
There are sets $\CS \subset \mathcal{L}^d$ of cardinality $4$ for which the following is true:
If $F_1, F_2 \in \mathcal{F}^d$ are convex, and $|F_1|=|F_2|$, but $F_1\ne F_2$, then
$\Delta_{\CS}(F_1,F_2) \ge 6$.
Further, for any set $\CS \subset \mathcal{L}^d$ of at least $7$ coplanar lattice lines, and sets 
$F_1, F_2$ as before, $\Delta_{\CS}(F_1,F_2) \ge 2(|\CS|-1)$.
\end{thm}

The following theorem uses the known characterization of the (rather rare) cases of uniqueness 
in the special case $d=m=2$ to quantify the deviation of solutions for noisy data; it
generalizes a previous result from~\cite{ab-05}.

\begin{thm}[\cite{dalen}] \label{thm:stabDTDalen}
Let $\CS\subset \mathcal{L}^2$ with $|\CS|=2$, let $F_1,F_2\in\mathcal{F}^2$ with $|F_1|=|F_2|$.
Further, suppose that $F_1$ is uniquely determined by $X_{S}F_1$ for $S\in \CS$,
and set $\beta=\Delta_{\CS}(F_1,F_2)$. Then
\[ 
4|F_1\cap F_2|+(\beta+2)\Bigl(\beta-1 +\sqrt{8|F_1\cap F_2|+(\beta-1)^2}\Bigr)\geq 4|F_1|.
\] 
\end{thm}

Stability results in the continuous case with finitely many X-rays typically rely on bounds of the variation of the functions, measured in some weighted Sobolev norms; see~\cite[Sect.~4]{nattererbook} and~\cite[Sect.~5.9]{markoebook} (and references therein). In the realm of geometric tomography, Vol\v{c}i\v{c} \cite{volcicwellposed} showed that the problem of reconstructing a convex body from its X-rays in four `good' directions (which guarantee uniqueness) is well-posed. 
Some further stability estimates are given in~\cite{stabilitylonginetti}.

In contrast to Thm. \ref{thm:stabDTDalen}, the task of reconstructing finite lattice sets from X-rays 
taken along $m\geq 3$ directions is highly instable. In particular the following result shows that
Thm. \ref{thm:stability1} is sharp.

\begin{thm}[\cite{alpers-d03, agt-01}] \label{thm:instabDT3}
Let $\CS \subset \mathcal{S}^d$ with $|\CS|\geq 3$, and let $\alpha \in \mathbb{N}$.
Then there exist $F_1,F_2\in\mathcal{F}(\R^d)$ with the following properties:
\begin{enumerate}[label=(\roman*), itemindent=4ex]
%\begin{enumerate}[label=(\roman*), itemindent=1ex]
\item $F_1$ is uniquely determined by $X_{S}F_1$ for $S\in \CS$;
\item $F_2$ is uniquely determined by $X_{S}F_2$ for $S\in \CS$;
\item $\Delta_{\CS}(F_1,F_2)=2(m-1)$;
\item $|F_1|=|F_2| \ge \alpha$;
\item $F_1\cap F_2=\emptyset$.
\end{enumerate}
The statement also holds in the lattice case.
\end{thm} 

The proof for~$d=2$ is due to \cite{agt-01}, while~\cite{alpers-d03} extends the construction to general~$d$.
It is actually possible to show that not even affine transformations help much to increase the overlap 
of the two sets.

\section{Computational aspects} \label{sec:compaspects}

Next we deal with algorithmic aspects of actually reconstructing the, one or all sets that are 
consistent with the given X-ray data. We will restrict the exposition to functions in $\CF(\Z^d,C)$ with
$C\subset \Q$ and lines in $\CL^d$ since all computational issues can then be studied in the 
well-known \emph{binary Turing machine model}; see~\cite{gareyjohnsonbook, compcomplbook} 
for background information. Again, emphasis will be placed on the lattice case.

\subsection{Algorithmic problems}\label{subsec-basic}

Let $\CS\subset \mathcal{L}^{d}$ be finite.  From an algorithmic point of view the following questions 
are basic: Are the data consistent? If so, reconstruct a solution! Is this solution unique?  
We will now introduce the correponding problems more precisely.

\begin{problem}{\textsc{Consistency$_{\mathcal{F}(\mathbb{Z}^d,C)}(\CS)$}}
\item[Instance] Data functions $f_{S}$ for $S\in \CS$.
\item[Question] Does there exist $\psi\in \mathcal{F}(\Z^d,C)$ such
that\/ $X_{S}\psi=f_{S}$ for all $S\in \CS$?
\end{problem}

\vspace{-.5cm}

\begin{problem}{\textsc{Reconstruction$_{\mathcal{F}(\mathbb{Z}^d,C)}(\CS)$}}
\item[Instance] Data functions $f_{S}$ for $S\in \CS$.
\item[Task] Determine a function $\psi\in \mathcal{F}(\Z^d,C)$ such that\/
$X_{S}\psi=f_{S}$ for all $S\in \CS$, or decide that no such function exists.
\end{problem}

\vspace{-.5cm}

\begin{problem}{\textsc{Uniqueness$_{\mathcal{F}(\Z^d,C)}(\CS)$}}
\item[Instance] A function\/ $\psi \in \mathcal{F}(\Z^d,C)$.
\item[Question] Does there exist $\varphi\in\mathcal{F}(\Z^d,C)\setminus \{\psi\}$ such that
$X_{S}\psi=X_{S}\varphi$ for all $S\in \CS$?
\end{problem}

Of course, \textsc{Reconstruction}$_{\mathcal{F}(\Z^d,C)}(\CS)$ cannot be
easier than \textsc{Consistency}$_{\mathcal{F}(\Z^d,C)}(\CS)$. 
Further, note that \textsc{Uniqueness$_{\mathcal{F}(\Z^d,C)}(\CS)$} actually asks for nonuniqueness
in order to place the problem into the class $\NP$; see Thm. \ref{thm:complexity}.

For certain codomains such as $C=\{0,1\}$ it is reasonable to actually ask for the number of solutions even in the case
of non-uniqueness. We will introduce the following problem for geneneral $C$ with the understanding
that the (not really interesting) answer~`$\infty$' is permitted.

\begin{problem}{$\#$\textsc{Consistency}$_{\mathcal{F}(\Z^d,C)}(\CS)$}
\item[Instance] Data functions $f_{S}$ for $S\in \CS$.
\item[Task] Determine the cardinality of the set of functions $\psi\in\mathcal{F}(\Z^d,C)$ 
such that $X_{S}\psi=f_{S}$ for all $S\in \CS$.
\end{problem}

Observe that a given instance $\mathcal{I}=(f_S: S\in \CS)$ can be consistent only if 
$\|f_{S}\|_{1}$ does not depend on $S$. Since this condition can be
checked efficiently we will in the following often tacitly assume that this is the case and set
$$
n=n({\mathcal{I}})=\|f_{S}\|_{1}.
$$
Further, for any given instance $\mathcal{I}=(f_S: S\in \CS)$, the support of all
solutions is contained in the \emph{grid}
\[
   G\, =\, G({\mathcal{I}})\, =\, \bigcap_{S\in \CS}\bigl(\textnormal{supp}(f_S)+S\bigr)
\] 
associated with $\mathcal{I}$. Of course, $G({\mathcal{I}})$ can be computed from
$\CI$ by solving polynomially many systems
of linear equations. Hence we can associate a variable $x_g$ with every grid point and formulate
\textsc{Consistency$_{\mathcal{F}(\Z^d,C)}(\CS)$} as a linear (feasibility) problem with 
the additional constraints that $x_g\in C$ for all $g\in G$. This simple observation shows
already that \textsc{Consistency$_{\mathcal{F}(\Z^d,C)}(\CS)$} is algorithmically easy for
$C\in \{[0,1]\cap \Q,\Q\}$ simply because linear programming can be solved in polynomial time,
and also for $C=\Z$ since systems of linear diophantine equations can be solved in polynomial time;
see e.g. \cite[Sect.~4,5,13--15]{schrijverbook}.

Next we are turning to the other relevant codomains, with a special emphasis on the lattice case. 

\begin{thm}[\cite{ryser, gale57, ggp-99}] \label{thm:complexity}\hfill\\
\textsc{Consistency}$_{\mathcal{F}(\mathbb{Z}^d,C)}(\mathcal{S})$ and \textsc{Uniqueness}$_{\mathcal{F}(\mathbb{Z}^d,C)}(\mathcal{S}),$ $C\in\{\{0,1\},\mathbb{N}_0\},$  are both in $\mathbb{P}$ if~$|\mathcal{S}|\le 2$ 
whereas they are $\mathbb{N}\mathbb{P}$-complete if $|\mathcal{S}|\ge 3.$
Also, the problem $\#$\textsc{Consistency}$_{\CF^d}(\mathcal{S})$ is $\#\NP$-complete for $|\mathcal{S}|\ge 3$.
\end{thm}

The complexity status of the counting problem for $|\mathcal{S}|=2$ is still open.

\begin{question}
Is $\#$\textsc{Consistency}$_{\CF^d}(\mathcal{S}),$ $|\mathcal{S}|=2,$ a  $\#\mathbb{P}$-complete problem?
\end{question}

Let us now return to the  R{\'e}nyi setting. 

\begin{thm}[\cite{heppes, renyi52}] \textsc{Reconstruction}$_{\CF^d}(\mathcal{S})$ is in~$\mathbb{P}$ if the input is restricted to those instances $\mathcal{I}=(f_S: S\in \CS)$ with $n(\mathcal{I})<|\mathcal{S}|.$
\end{thm}

A similar result holds for convex lattice sets when the lattice lines are chose according to
Thm.~\ref{thm:uniquenessconv}. So, let $\CC^d$ denote the subset of $\CF^d$ of convex lattice set, and let
\textsc{Reconstruction}$_{\CC^d}(\mathcal{S})$ signify the correponding reconstruction task.

\begin{thm}[\cite{qconvex,brunettidaurat08}]
For any set $\mathcal{S}\subset \mathcal{L}^d$ of at least seven coplanar directions and for suitable such sets of cardinality 
four \textsc{Reconstruction}$_{\CC^d}(\mathcal{S})$ can be solved in polynomial-time. 
\end{thm} 

Let us now turn to the following `noisy' versions of \textsc{Consistency}$_{\CF^d}(\mathcal{S})$ 
and \textsc{Uniqueness}$_{\CF^d}(\mathcal{S}).$ 

\begin{problem}{\textsc{X-Ray-Correction$_{\mathcal{F}^d}(\mathcal{S})$}}
\item[Instance]  Data functions $f_{S}$ for $S\in \CS.$
\item[Question]  Does there exist $F\in \mathcal{F}^d$ such that \\
$\DS \sum_{S \in \mathcal{S}}||X_SF-f_S||_1\le m-1?$
\end{problem}

\vspace{-.5cm}

\begin{problem}{\textsc{Similar-Solution$_{\mathcal{F}^d}(\mathcal{S})$}}
\item[Instance]  A set $F_1\in \mathcal{F}^d.$
\item[Question]  Does there exist $F_2\in \mathcal{F}^d$ 
with $|F_1|=|F_2|$ and $F_1\ne F_2$ such that $\Delta_\CS(F_1,F_2)\le 2m-3?$
\end{problem}

\vspace{-.5cm}

\begin{problem}{\textsc{Nearest-Solution$_{\mathcal{F}^d}(\mathcal{S})$}}
\item[Instance] Data functions $f_{S}$ for $S\in \CS.$
\item[Task]  Determine a set $F^* \in \mathcal{F}^d$ such that \\ 
$\DS \sum_{S\in\mathcal{S}}||X_SF^*-f_S||_1=\min_{F \in \mathcal{F}^d}\sum_{S\in\mathcal{S}}||X_SF-f_S||_1.$
\end{problem}

Note that \textsc{X-Ray-Correction$_{\mathcal{F}^d}(\mathcal{S})$}
can also be viewed as the task of deciding, for given data functions $f_S,$ $S\in\mathcal{S},$ whether there exist
`corrected' data functions $f'_S,$ $S\in\mathcal{S},$ that are consistent and do not differ from the given functions 
by more than a total of~$m-1$. \textsc{Nearest-Solution$_{\mathcal{F}^d}(\mathcal{S})$} asks for a set 
$F^*\in\mathcal{F}^d$ that fits the potentially noisy measurements best. 

The computational complexity of these tasks is as follows.

\begin{thm}[\cite{ag06}] \label{thm:algstability}\hfill\\
The problems \textsc{X-Ray-Correction$_{\mathcal{F}^d}(\mathcal{S})$},
\textsc{Similar-Solution$_{\mathcal{F}^d}(\mathcal{S})$}, and
\textsc{Nearest-Solution$_{\mathcal{F}^d}(\mathcal{S})$}
are in $\mathbb{P}$ for $|\mathcal{S}|\le 2$ but are $\mathbb{N}\mathbb{P}$-complete for~$|\mathcal{S}|\ge 3.$
\end{thm}

\subsection{Algorithms} \label{sect:algorithms}

Several polynomial-time algorithms for \textsc{Reconstruction}$_{\mathcal{F}(\mathbb{Z}^d,C)}(\mathcal{S}),$ 
$C\in\{\{0,1\},\mathbb{N}_0\},$ $|\mathcal{S}|=2,$ can be found in the literature. In addition to Ryser's 
algorithm~\cite{ryser} for $C=\{0,1\},$ there are approaches based on network-flows~\cite{slump-gerbrands-82} 
or matroid intersections~\cite{h-GGP96}. Moreover, the problem can be modeled as an integer linear program, 
which involves a totally unimodular coefficient matrix, and which can therefore be solved as a linear program 
(see, for instance,~\cite[Sect.~16\&19]{schrijverbook}). For further comments, see~\cite[Sect.~1]{kubaherman1}. 

In the presence of $\NP$-hardness, one cannot expect to find generally efficient algorithms. There are, however,
various techniques from combinatorial optimization that can and have been applied to solve instances to
optimality up to certain sizes; see~\cite{gpvw-98}. Similarly as for $|\mathcal{S}|=2,$ the reconstruction problem 
\textsc{Reconstruction}$_{\mathcal{F}(\mathbb{Z}^d,C)}(\mathcal{S}),$ $C\in\{\{0,1\},\mathbb{N}_0\},$ 
can be formulated als integer linear program for arbitrary $|\mathcal{S}|.$ However, for $|\mathcal{S}|\geq 3$ the
coefficient matrix is in general no longer totally unimodular. Of course, we can still solve the 
corresponding linear programming relaxation (where $\{0,1\}$ is replaced by $[0,1]$ or $\N_0$ by $[0,\infty[$)
efficiently. Unless $\mathbb{P}=\mathbb{N}\mathbb{P},$ Thm.~\ref{thm:complexity} implies that it will, however, in general 
not be efficiently possible to convert the obtained fractional solution into a required integer one.

Since, in general, measured data are noisy anyway, research focused on approximate solutions. 
It is quite natural to try to solve \textsc{Reconstruction}$_{\CF^d}(\mathcal{S})$ even if $|\mathcal{S}|\ge 3$ 
by using the available polynomial-time algorithms for $|\mathcal{S}|=2$ in an alternating
approach. First, two of the given $|\mathcal{S}|$ data functions are selected and a solution~$F_0$ is computed which is
consistent with these. In the $j$th step, at least one of the two directions is replaced by a different one from $\CS,$ and a
solution is constructed which satisfies the corresponding two constraints and is closest to~$F_j$.
While each step of such an {\em alternating direction approach} can be performed in polynomial time, there are
severe limitations on the guaranteed quality of the produced solution. For an analysis of this and
other approaches, see~\cite{gpvw-98}.

Despite their theoretical limitations there are several approaches that are reported to work very well in practice. 
Among these are BART~\cite{bart2} and DART~\cite{DART2, TVRDART}. The former, which is a variant of ART as 
described in%
~\cite{hermanbookchapter}, 
%~\cite{gordonbender}, 
is implemented in the open-source software 
SNARK14~\cite{snark09} (example code can be found in~\cite{alpersgardner13}), the latter is implemented in  the open-source ASTRA toolbox~\cite{astratoolbox}. Further algorithms are discussed in~\cite[Sect.~8-14]{kubaherman1} 
and~\cite[Sect.~8-11]{kubaherman2}. 
For applets illustrating several algorithmic tasks in discrete and geometric tomography, see~\cite{discretetomographyapp} and~\cite{geometrictomographyapp}, respectively.

\begin{figure}[htb]
\centering
\includegraphics[width=0.7\textwidth]{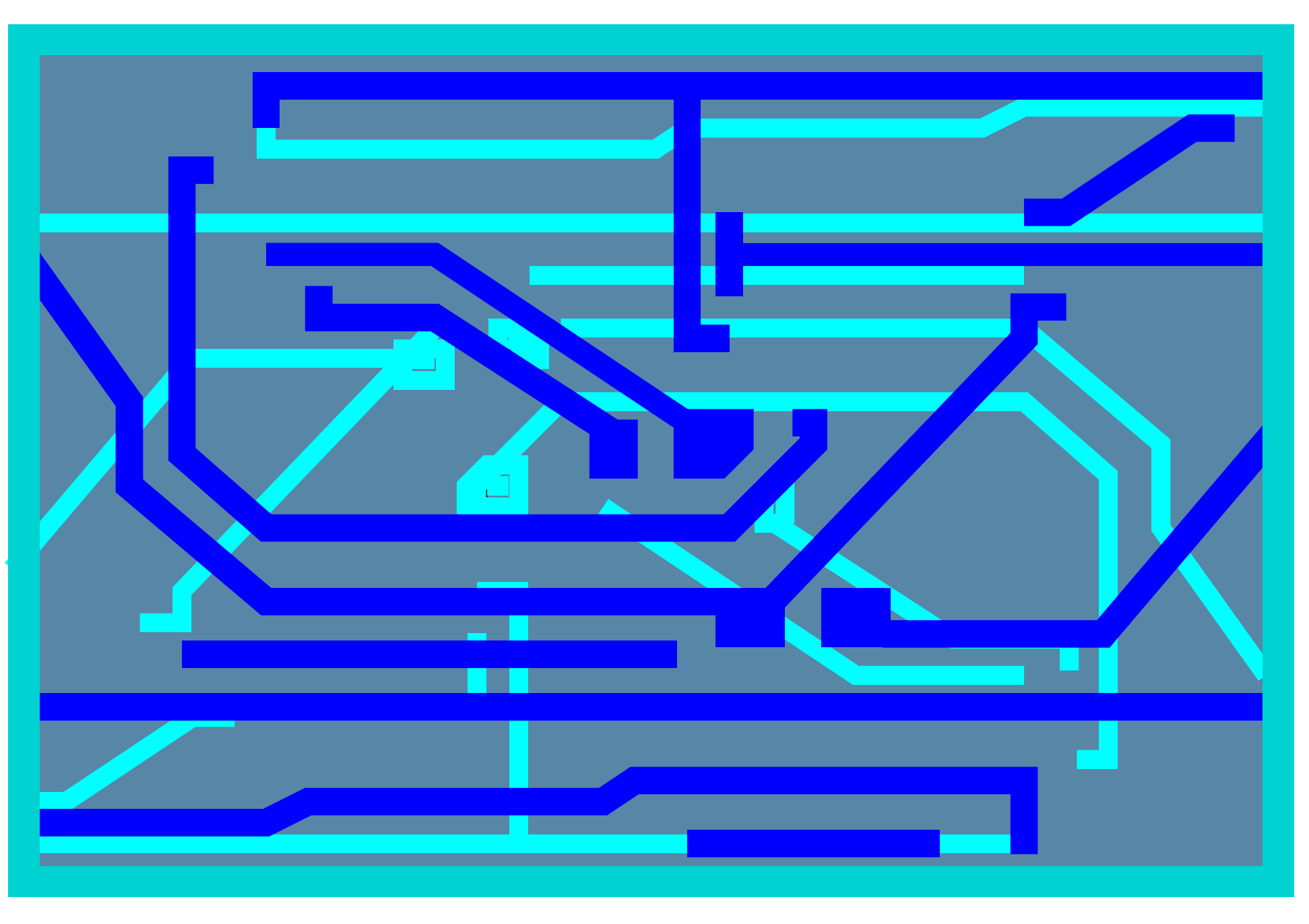}
\caption{(From \cite{gr-10}) An idealized circuit board.} \label{fig:circuit}
\end{figure}

Let us, finally, point out that for certain applications full reconstructions are not needed. For instance, in quality control for circuit board productions (see Fig.~\ref{fig:circuit}) one may want to certify that the production process actually produced 
a desired blueprint structure (`verification'). Then one can, of course, compute data functions from the blueprint and compare them with 
the measured data from the produced board. If the difference is large one would report an error. If, however, the
difference is small, the produced board can still be quite different from the blueprint (particularly if the data do not
determine the image uniquely). This ambiguity can be reduced by applying a (polynomial-time deterministic) reconstruction 
heuristic on both sets of data functions and subsequently comparing the reconstructions. In practice such checks have shown to be able to detect production flaws even on very limited data and quite poor (and very fast) reconstructions algorithms.
%-------------------------------------------------------------------------------------------------

\section{Superresolution and discrete tomography}\label{sect:superresolution}

Electron tomography, pioneered originally in the life sciences (see~\cite{ETreview1, frankbook, oktembook}), is becoming an increasingly important tool in materials science for studying the three-dimensional morphologies and chemical compositions of nanostructures~\cite{alpersgardner13, DART3, DART1,jinschek-08}.  For various technical reasons, however, atomic resolution tomographic imaging as envision in~\cite{quantitem, sksbko-93} has not become a full reality yet (favorable instances are reported in~\cite{rafalET, batenburgnature}; see also the surveys~\cite{atomicETreview, HRTEMlimits3}).  
One of the challenges faced by current technology is that tomographic tilt series need to be properly aligned (see, e.g.,~\cite{HRTEMbook, houbenalignment}). Therefore, and also to prevent radiation damage, one might wonder whether is is possible to proceed in a multimodal scheme.

Suppose some reconstruction has been obtained from a (possibly technologically less-demanding) lower-resolution data set. Can one then use limited additional high-resolution data (for instance, acquired from only two directions) to enhance the resolution in a subsequent step?  As we will now see the tractability of this approach depends strongly on the reliability of the initial lower-resolution reconstruction. Details of the presented results can be found in~\cite{agsuperresolution}.

\subsection{Computational aspects}
\begin{figure}[htb]
\centering
\includegraphics[width=0.5\textwidth]{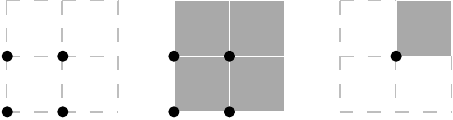}
\caption{Lattice points (left) and pixels (middle); right: pixel associated with its lattice point.}\label{fig:gridimage}
\end{figure}

We have already remarked that a function $\psi\in \CF^d$ can be viewed as a characteristic function that encodes a finite
lattice set. In a different, yet equivalent, model the function~$\psi$ can be viewed as representing a binary image. In this interpretation the points $x\in\mathbb{Z}^d$ represent the pixel/voxel coordinates while~$\psi(x)$ denotes their colors (typically, values~0 and~1 are considered to represent white and black pixels, respectively); see Fig.~\ref{fig:gridimage} for an illustration. 
Similarly, for $l\in \mathbb{N},$ a function $\rho\in \mathcal{F}(\Z^d,[l]_0)$ can be viewed as representing 
a gray-scale image with~$l+1$ different gray levels (values~$0$ and~$l$ typically representing the `gray level' white and black, respectively). 

For simplicity of the exposition we restrict our discussion to the case $d=2$. 
Now suppose we want to reconstruct a binary image $\psi\in \CF^2$ contained in an~$n_1\times n_2$ box 
from low-resolution gray scale information and high-resolution X-ray data.
The lower resolution is quantified by some $k\in \mathbb{N}\setminus \{1\}$, and we assume that~$n_1$ and $n_2$ are divisible by $k$. More precisely, 
we assume that an~$n_1/k\times n_2/k$ low-resolution (gray-scale) image $\rho \in \mathcal{F}(\Z^2,[k^2]_0)$ of $\psi$ 
is available, and the pixels~$x$ in $\psi$ result from a $k\times k$ subdivision of the pixels~$y$ of $\rho.$ 
Hence in any such subdivision~$B$ we have 
\[
\sum_{x\in B}\psi(x)=\rho(y).\label{eq:blockconstr}
\]
For given $\rho(y)$ and unknown $\psi(x),$ $x\in B,$ we call the above equation a \emph{$k\times k$ block constraint}. 
We say that, for some $\varepsilon \in \N_0$, a block constraint is \emph{satisfied within an error of $\varepsilon$}, if 
\[\rho(y)-\varepsilon\:\:\leq\:\: \sum_{x\in B}\psi(x) \:\:\leq\:\: \rho(y)+\varepsilon.\] 
We may think of $\rho$ as being the result of some lower-resolution reconstruction of~$\psi$. In order to increase the resolution we want to utilize additional high-resolution X-ray data~$X_{S}\psi$ that are available 
from the two coordinate directions $S_1$ and $S_2$, and we set $\CS=\{S_1,S_2\}$.
Relatively to $\rho$ the data $X_{S}\psi,$ $S\in\CS$, can be considered as $k$-times finer resolution X-ray data.

For given $k\geq 2$ and $\varepsilon\in\mathbb{N}_0$ the task of \emph{(noisy) superresolution} is as follows.\\

\hspace*{-4ex}\begin{problem}{\textsc{nSR}$(k,\varepsilon)$}
\item[Instance] \hspace*{-0ex} A gray-level image $\rho\in \mathcal{F}(\Z^2,[k^2]_0),$ \\
                \hspace*{-0ex} a subset $R$ of `reliable pixels' of $\rho,$ and\\
                \hspace*{-0ex} data functions $f_{S_1},f_{S_2}$ at a $k$-times finer resolution.
\item[Task] \hspace*{-0ex} Determine a function $\psi\in \CF^2$ such that \\[1ex]
\begin{minipage}[ht]{1\textwidth} 
\hspace*{2ex} $X_{S}\psi=f_{S}$ for $S\in \{S_1,S_2\},$ \\
\hspace*{2ex} all $k\times k$ block constraints for the pixels in~$R$ are satisfied, and\\
\hspace*{2ex} all other $k\times k$ block constraints are satisfied within an error of $\varepsilon,$ 
\end{minipage}\\[1ex]
\hspace*{0ex} or decide that no such function exists.
\end{problem}

\begin{figure}[htb]
\centering
\includegraphics[width=0.47\textwidth]{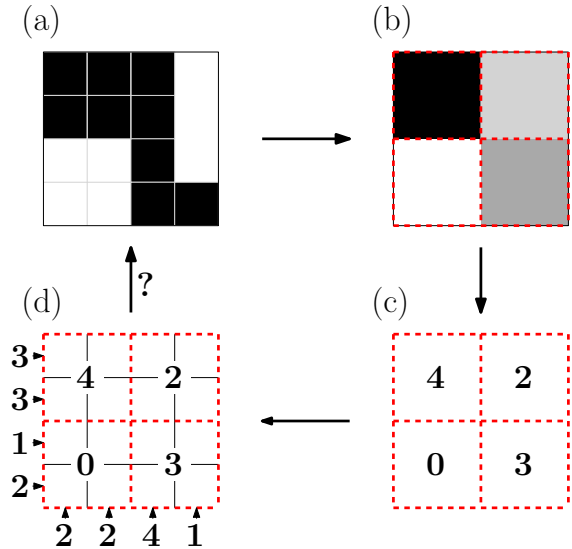}
\caption{(From~\cite{agsuperresolution}) The double-resolution imaging task \textsc{DR}. (a) Original (unknown) high-resolution image, (b) the corresponding low-resolution gray-scale image, (c) gray levels converted into block constraints, (d) taken in combination with double-resolution row and column sum data. The task is to reconstruct from~(d) the original binary image shown in~(a). } \label{super:fig:process}
\end{figure}

Since our focus is in the following on double-resolution imaging, i.e., on the case $k = 2,$ let us set $\textsc{nDR}(\varepsilon)=\textsc{nSR}(2,\varepsilon),$ for $\varepsilon>0.$ In the reliable situation, i.e., for $\varepsilon=0$ we simply speak of double-resolution and set $\textsc{DR}=\textsc{nSR}(2,0).$ (Then, of course, the set $R$ can be omitted from the input.)
An illustration is given in Fig.~\ref{super:fig:process}. 

As it turns out \textsc{DR} is tractable.

\begin{thm}[\cite{agsuperresolution}]  \label{super:thm:main1} \hfill
\textsc{DR} and also the corresponding uniqueness problem can be solved in polynomial time. 
\end{thm}

The algorithm presented in~\cite{agsuperresolution} is based on a decomposition into subproblems, which allows 
to treat the different gray levels separately.
If we view \textsc{DR} as the reconstruction problem for $m=|\mathcal{S}|=2$ with additionally block constraints we can compare Thm.~\ref{super:thm:main1} with Thm.~\ref{thm:complexity} and see that bock constraints impose fewer algorithmic difficulties than X-ray data from a third direction.

The next result, which deals with the case that some of the gray levels come with small uncertainties depicts 
(potentially unexpected) complexity jumps.

%\newpage

\begin{thm}[\cite{agsuperresolution}]\label{super:thm:main2} Let $k\geq 2$ and $\varepsilon>0.$  \hfill
\begin{enumerate}[label=(\roman*)]
\item \textsc{nSR}$(k,\varepsilon)$ is $\mathbb{N}\mathbb{P}$-hard.
\item  The problem of deciding whether a given solution of an instance
of \textsc{nSR}$(k,\varepsilon)$ has a non-unique solution is $\mathbb{N}\mathbb{P}$-complete.
\end{enumerate}
\end{thm}

To put it succinctly: noise in tomographic superresolution imaging does not only affect the quality of a reconstructed image but also the algorithmic tractability of the inverse problem itself.

\begin{figure}[htb]
\centering
\includegraphics[width=0.6\textwidth]{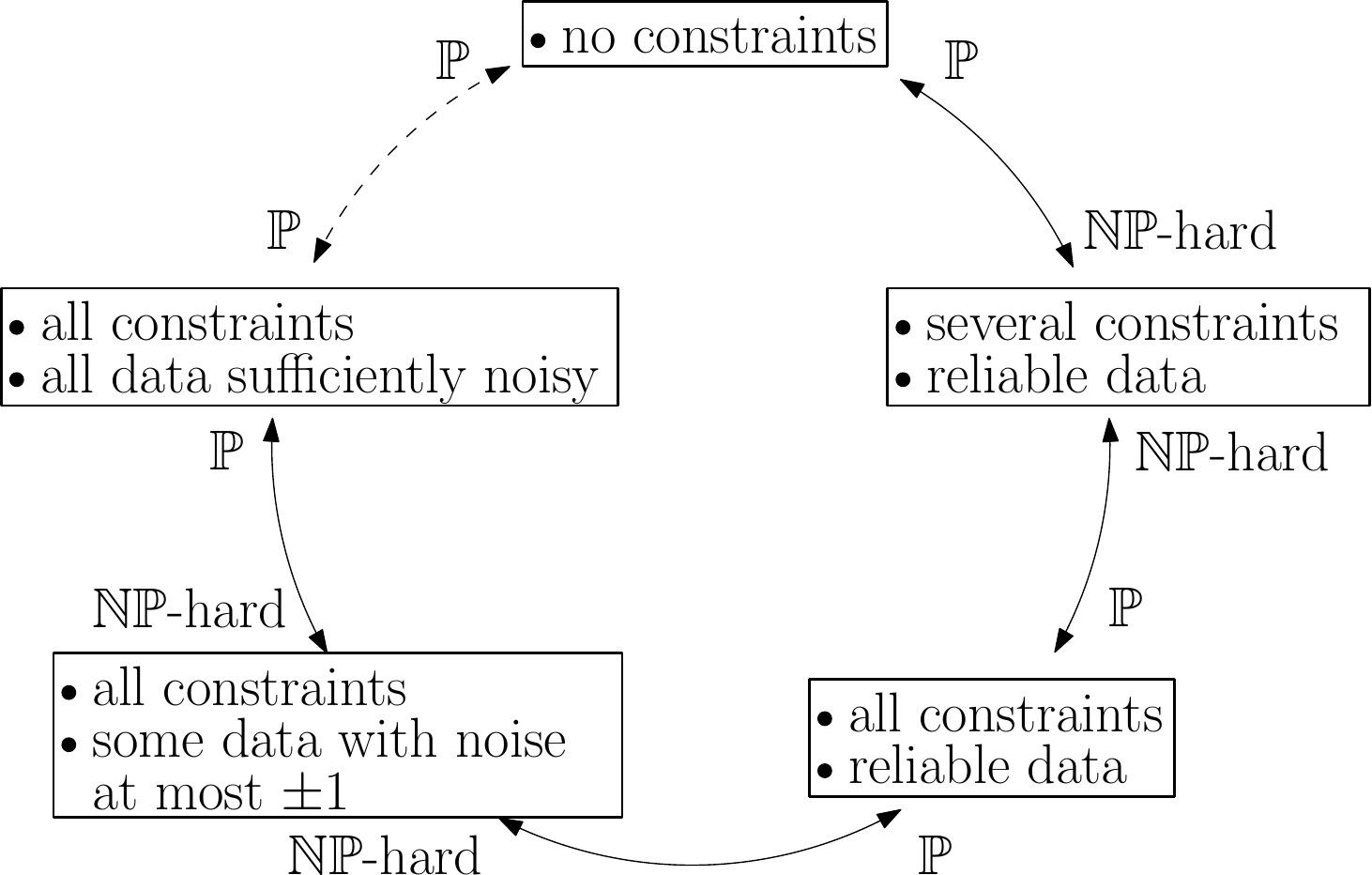}
\caption{(From \cite{agsuperresolution}) Overview of complexity jumps for the problem of reconstructing a binary image from row and column sums and additional $2\times2$ block constraints.} \label{super:fig:complexityjumps}
\end{figure}

$\textsc{DR}$ without any block constraints boils down to the reconstruction problem for $m=2$ and is hence solvable in polynomial-time.
$\textsc{DR}$ is, however, $\mathbb{N}\mathbb{P}$-hard if \emph{several} (but not all) block constraints (which are required to be satisfied with equality) are present (Thm.~\ref{super:thm:main2}).  Possibly less expectedly, if \emph{all} block constraints are included, 
then the problem becomes polynomial-time solvable again (Thm.~\ref{super:thm:main1}). 
If, on the other hand, from \emph{all} block constraints \emph{some} of the data come with \emph{noise} at most~$1,$ 
then the problem becomes again $\mathbb{N}\mathbb{P}$-hard (Thm.~\ref{super:thm:main2}). 
And yet again, if from \emph{all} block constraints \emph{all} of the data are \emph{sufficiently noisy}, then the problem is in~$\mathbb{P}$ (as this is again the problem of reconstructing binary images from X-ray data taken from two directions). Figure~\ref{super:fig:complexityjumps} gives an overview of these complexity jumps.

It does not seem likely that, but is still open, whether the tractability result of Thm.~\ref{super:thm:main1} persists for~$k\ge 3$.

\begin{question}[\cite{agsuperresolution}]
Is the problem \textsc{nSR}$(k,0)$ $\mathbb{N}\mathbb{P}$-hard for $k\geq 3$?
\end{question}

In the realm of dynamic discrete tomography (see Sect.~\ref{sect:dynamics}) block constraints play the role of 
special \emph{window constraints} which can be used to encoding velocity information for moving points.

For additional information on discrete tomography problems involving other kinds of  constraints, see~\cite[Sect.~4]{brualdibook}.

\subsection{Stability and instability}

Let us now turn to a discussion of the stability of the solutions to~\textsc{DR}. 

\begin{thm} \label{thm:instabDR}
Let $\CS=\{\textnormal{lin}(1,0)^T,\textnormal{lin}(0,1)^T\},$ and $\alpha \in \mathbb{N}$. Then there exist instances~$\mathcal{I}_1$ and $\mathcal{I}_2$ of \textsc{DR}
with the following properties:
\begin{enumerate}[label=(\roman*), itemindent=4ex]
%\begin{enumerate}[label=(\roman*), itemindent=1ex]
\item $F_1$ is the unique solution to~$\mathcal{I}_1;$
\item $F_2$ is the unique solution to~$\mathcal{I}_2;$
\item  $\Delta_{\CS}(F_1,F_2)=4;$
\item $|F_1|=|F_2| \ge \alpha$;
\item $|F_1 \cap F_2|=\frac{1}{2}|F_1|.$
\end{enumerate}
\end{thm}

\begin{figure}[htb] 
\centering
\subfigure[]{\includegraphics[width=0.3\textwidth]{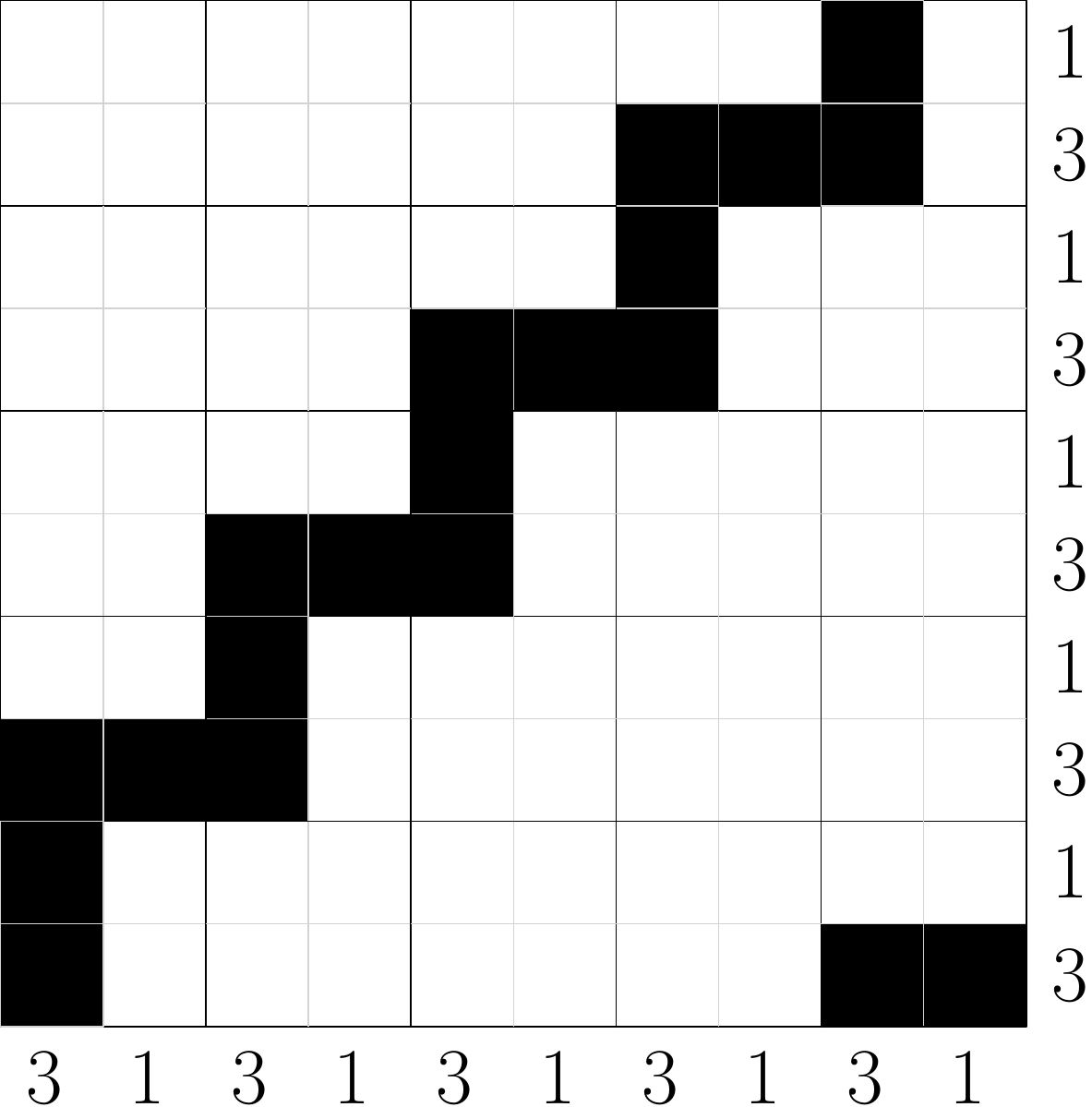}}\hspace*{8ex}
\subfigure[]{\includegraphics[width=0.3\textwidth]{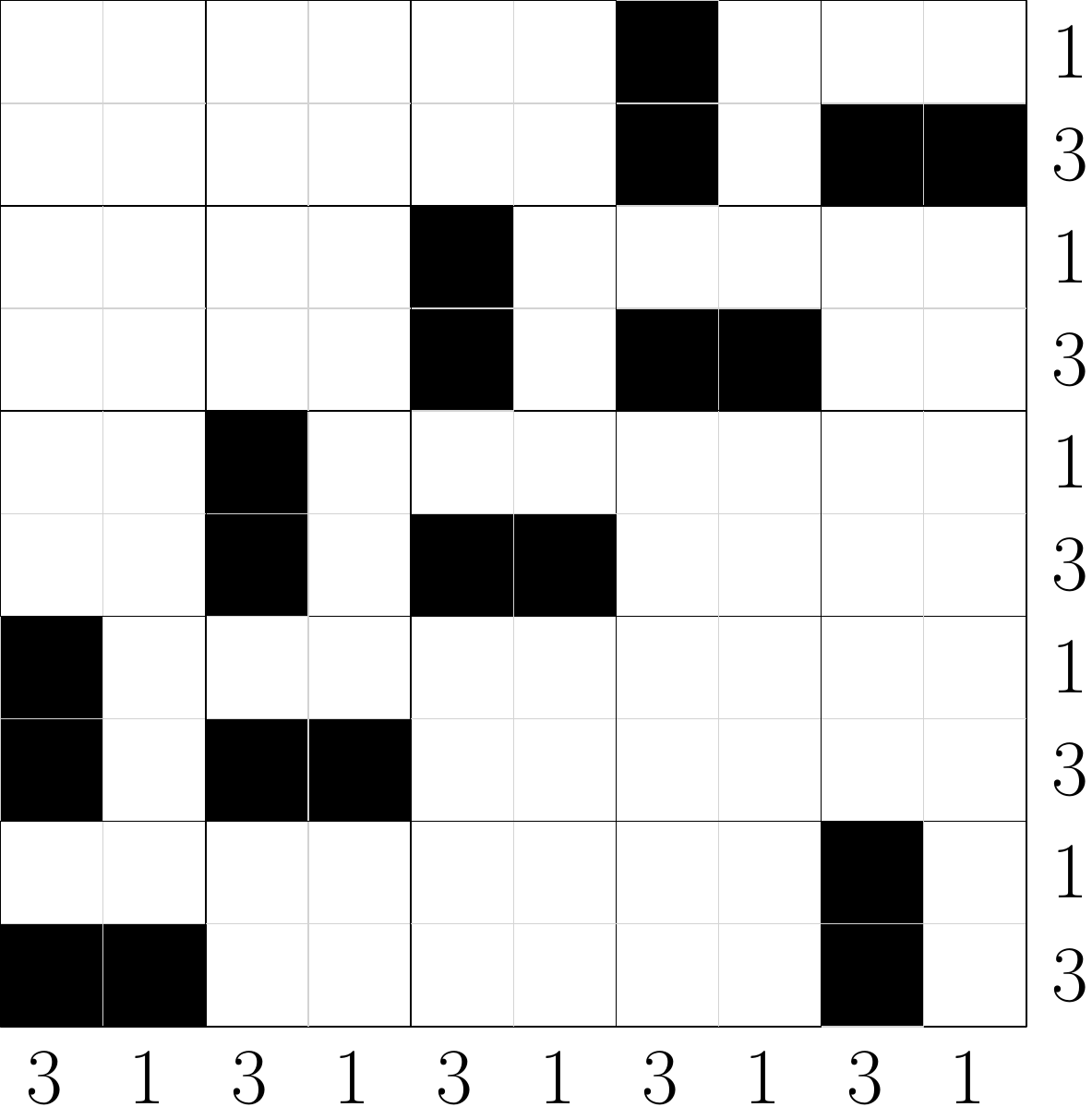}}
\subfigure[]{\includegraphics[width=0.3\textwidth]{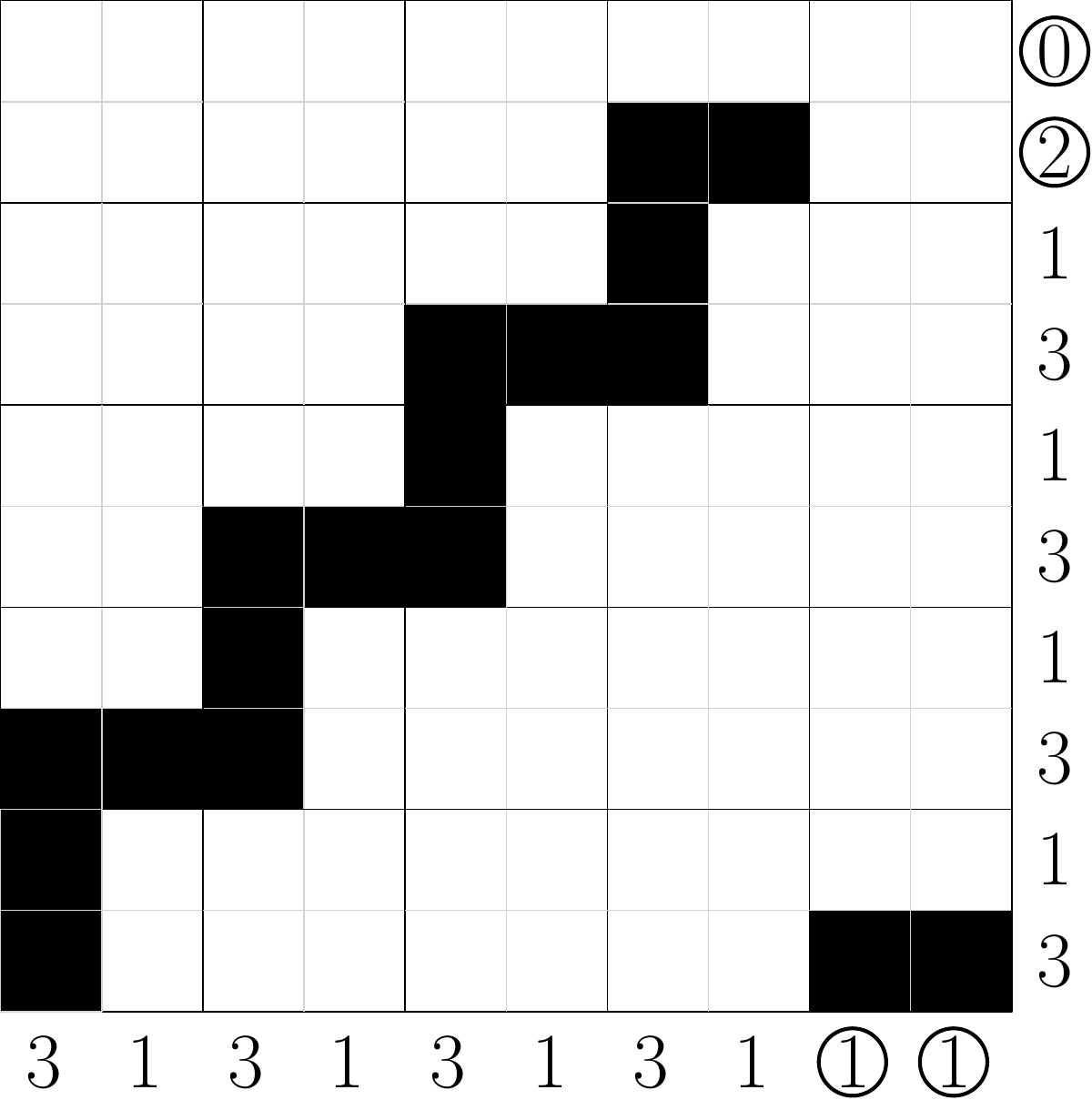}}\hspace*{8ex}
\subfigure[]{\includegraphics[width=0.3\textwidth]{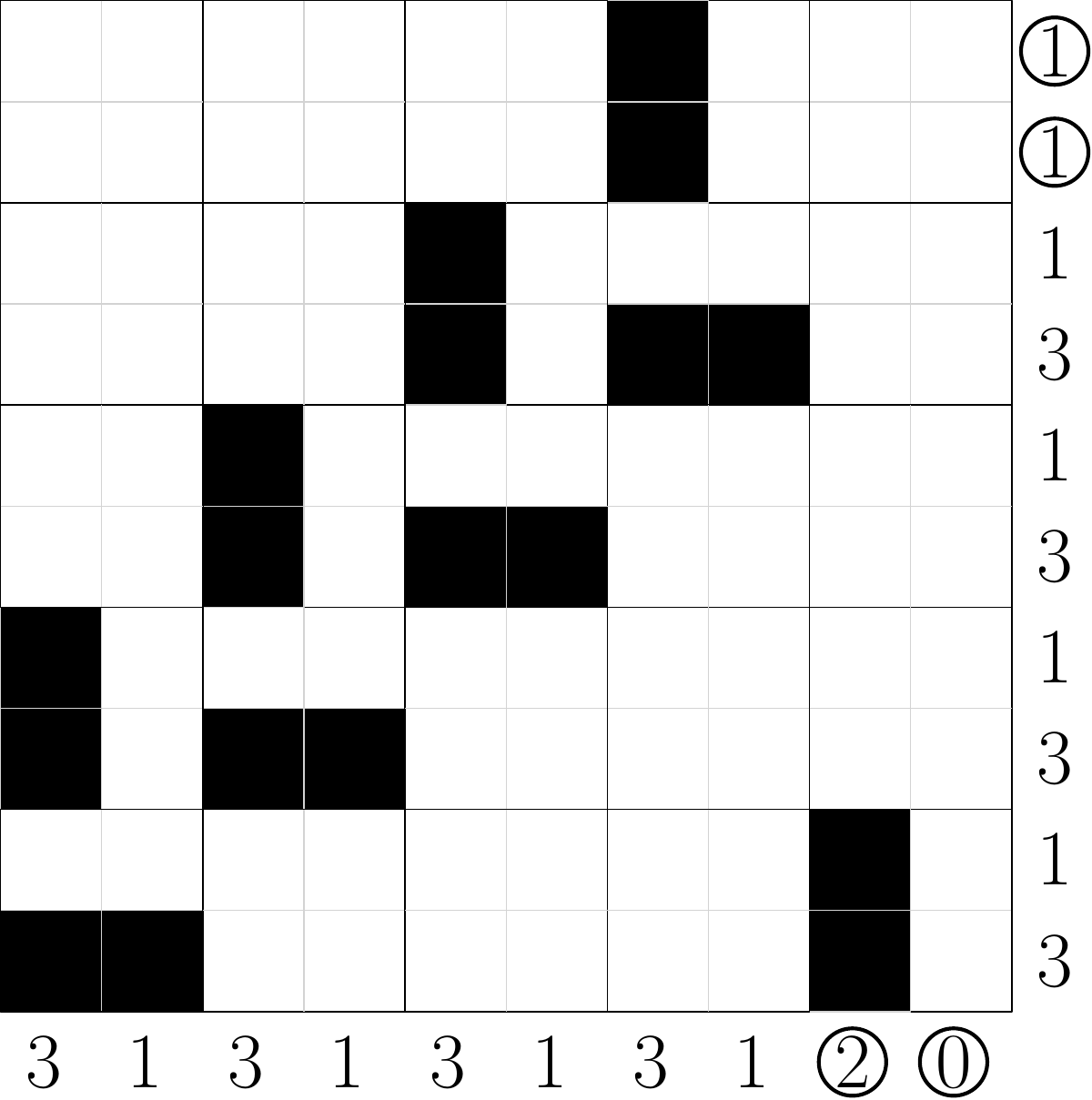}}
\caption{An example illustrating instability of~\textsc{DR}. (a) and (b): two solutions of the same problem instance; 
(c) and (d): uniquely determined solutions $F_1,$ $F_2$ to the two problem~instances obtained by deleting points of a block from (a) and (b), respectively. (The X-rays are indicated by the numbers to the bottom and right.) The X-rays differ in the circled numbers yielding an X-ray error $\Delta_{\CS}(F_1,F_2)=4.$}\label{fig:largeswitchingcomp}
\end{figure}

The proof is based on a construction in \cite{agsuperresolution} of an instance~$\mathcal{I}$ of~\textsc{DR} that admits precisely 
two solutions $F'_1 \neq F'_2$ with $|F'_1|=|F'_2|\geq \alpha+2.$ From these two solutions points in one block are deleted
to obtain $F_1$ and $F_2$; see Fig. \ref{fig:largeswitchingcomp} for an illustration. 

A small X-ray error of~$4$ can thus lead to quite different reconstructions (again, see Fig.~\ref{fig:largeswitchingcomp}).  It should be noted, however, that the set $F_2$ has a much larger \emph{total variation~(TV)} than~$F_1$ (for some background information see, e.g.,~\cite{chambolle}). Regularization by total variation minimization, as proposed in~\cite{agsuperresolution}, would therefore always favor the reconstruction~$F_1.$  

It is instructive to compare~\textsc{DR} (Thm.~\ref{thm:instabDR}) with its discrete tomography counterparts for $m=2$ (Thm.~\ref{thm:stabDTDalen}) and $m\geq 3$ (Thm.~\ref{thm:instabDT3}), which do not involve any block constraints. 

On the one hand, the reconstruction problem for $m=2$ is much more stable than its double-resolution counterpart. In fact, an easy calculation for $\beta=4$ shows that Thm.~\ref{thm:stabDTDalen} implies the bound \[|F_1\cap F_2|\geq |F_1|-5\sqrt{|F_1|}-9.\] 
Thus, if the original set $F_1$ is uniquely determined by its X-rays, then any reconstruction~$F_2$ from X-rays with error~$4$ needs to coincide with~$F_1$ by an asymptotically much larger fraction than the~$|F_1|/2$ provided in Thm.~\ref{thm:instabDR} for~\textsc{DR}. 

On the other hand, the instability result for~$m=3$ (see~Thm.~\ref{thm:instabDT3}) is stronger than that of Thm.~\ref{thm:instabDR} as for the former an X-ray error of~4 can lead to disjoint reconstructions. 

Hence in terms of \mbox{(in-)stabilities} the block constraints seem to play a somewhat weaker role than constraints modeling data from a third direction.

\section{Dynamics} \label{sect:dynamics}
Let us now turn to dynamic discrete tomography, which, in fact, represents rather recent developments in the 
field (see~\cite{agwindowconstraints, agdynamic, agms-15, glidingarc-15}). (For dynamic aspects of computerized tomography, see, e.g.,~\cite{siltanen1, hahn1} and the references cited therein.) 

We focus here on the task of \emph{tomographic particle (or point) tracking}, which amounts to determining the paths $\mathcal{P}_1,\ldots,\mathcal{P}_n$ of~$n$ points in space over a period of~$t\in \mathbb{N}$ moments in time from X-ray images taken from a fixed number~$m$ of directions. 

This problem comprises, in fact, two different but coupled basic underlying  tasks, the reconstruction of a finite set of points from few of their X-ray images (\emph{discrete tomography}) and the identification of the points over time (\emph{tracking}). 
The latter is closely related to topics in combinatorial optimization including matching and $k$-assigment problems; see~\cite{assignmentbook09} for a comprehensive survey on assignment problems.
 
Let us remark that particle tracking methods have been proven useful in many different fields such as fluid mechanics, 
geoscience, elementary particle physics, plasma physics, combustion, and biomedical imaging \cite{adrianlowdensity, biomedicalpiv, elementparticletracking, Reuss1986, ptvgeoscience, glidingarc-15} (see also the monograph~\cite{a2010} and the references cited therein). 
Most previous tomographic particle tracking methods (such as~\cite{DPS17, Elsinga2006, batenburg10, Williams2011}) can be considered as \emph{particle imaging velocimetry (PIV)} as they aim at capturing several statistical parameters of groups of particles instead of dealing with them individually. The individual tracking considered here is in the literature also sometimes referred to as \emph{particle tracking velocimetry (PTV)} or  \emph{low particle number density PIV}~\cite{adrianlowdensity}. For more general background information on particle tracking methods, see the monographs~\cite{a2010,piv2book}.

The exposition in this section will partly follow \cite{agdynamic}.

\subsection{Algorithmic problems}

We want to focus here on the interplay between discrete tomography and tracking. Therefore, we will distinguish the cases that for none, some or all of the~$\tau\in[t]$ moments in time, a solution $F^{(\tau)}\in\mathcal{F}^d$ of the discrete tomography task at time~$\tau$ is explicitly available (and is then considered \emph{the} correct solution regardless whether it is uniquely determined by its X-rays). The former case will be referred to as the \emph{(partially)} or \emph{(totally) tomographic} case while we speak of the latter as \emph{positionally determined}. It should be noted that the positionally determined case can be viewed as being the generic case in $\mathbb{R}^d,$ $d\geq3,$ because there any two (affine) lines in general position are disjoint, hence X-ray lines meet only in the points of~$F^{(\tau)}.$
  
For simplicity we assume in the following that there are no particles disappearing or reappearing within the tracked time interval. 
When $P=\{p_1,\dots,p_n\}$ denotes the (abstract) set of $n$ particles, we are in the tracking step thus interested in a one-to-one 
mapping $\pi^{(\tau)}:P\to F^{(\tau)},$ $\tau\in[t],$ that identifies the points of $F^{(\tau)}$ with the particles. 
The particle tracks are then given by $\mathcal{P}_i=(\pi^{(1)}(p_i),\dots,\pi^{(t)}(p_i)),$ $i\in[n].$ This identification 
is referred to as \emph{coupling}.  

In typical applications we would like to incorporate prior knowledge about `physically likely' paths. 
It seems most natural to input such information in terms of the cost $c(\CP_1,\ldots,\CP_n)$ of the feasible
particle tracks. Note, however, that the number of different particle tracks $(\CP_1,\ldots,\CP_n)$ is $(n!)^t$, hence {\em exponential} in~$n$ and~$t.$ This means that already for moderate problem sizes the costs of all potential tracks cannot be encoded explicitly.
There are various ways to deal with this problem. 
The most general approach is based on the assumption that `an expert knows a good solution if she or he sees it.' More technically
speaking, it is enough for an algorithm to have access to the cost $c(\CP_1,\ldots,\CP_n)$ only when the particle track 
$c(\CP_1,\ldots,\CP_n)$ is considered. Accordingly, \cite{agdynamic} suggest an oracular model, where
such knowledge is available through an algorithm $\CO$, called an {\em objective function oracle}, which computes 
for any solution~$(\CP_1,\ldots,\CP_n)$ its cost $c(\CP_1,\ldots,\CP_n)$ in time that is polynomial in all the other input data.  
Then the general problem of tomographic particle tracking for $\mathcal{S}\subset \mathcal{S}^d,$ can be formulated as follows.

\begin{problem}{{\sc TomTrac}${(\CO; \mathcal{S})}$}
\item[Instance] $t\in \N$ and  data functions $f^{(\tau)}_S$ with $\|f^{(\tau)}_S\|_{1}=n,$ for $S\in\mathcal{S},$ $\tau\in [t].$
\item[Task] Decide whether, for each $\tau\in [t]$, there exists a set 
$F^{(\tau)}\in \CF^d$ such that $X_{S}F^{(\tau)}=f_S^{(\tau)}$ for all $S\in \mathcal{S}.$
If so, find particle tracks $\CP_1,\ldots,\CP_n$ of
minimal cost for $\CO$ among all couplings of all 
tomographic solutions $F^{(1)}, \ldots,F^{(t)}.$
\end{problem}

In the positionally determined case the problem {\sc TomTrac}${(\CO; \mathcal{S})}$ reduces to the following tracking problem, which can be viewed as a \emph{$t$-dimensional assignment} problem. 

\begin{problem}{{\sc Trac}$(\CO;d)$}
\item[Instance] $t\in \N$ and sets $F^{(1)},\ldots,F^{(t)}\in \CF^d$ with
$|F^{(1)}|= \ldots = |F^{(t)}|=n$.
\item[Task] Find particle tracks $\CP_1,\ldots,\CP_n$ of
minimal cost for $\CO$ among all couplings of 
the sets $F^{(1)}, \ldots,F^{(t)}.$
\end{problem}

A priori knowledge may be available in various ways and may then lead to different objective function oracles;
see \cite{agdynamic}. Here we focus on information that is actually explicitly available. For instance, we 
speak of a {\em path value oracle} if the cost $c(\CP_1,\ldots,\CP_n)$ is just the sum $\sum_{i=1}^n w(\CP_i)$ 
of the weights of the individual paths $\CP_i,$ $i\in[n]$. Note that the number of different weights is bounded 
by $n^t$, and can hence be encoded explicitly for fixed (and small) $t$; see Thm. \ref{thm:matching}.
If, further, the weights are just the sums of all costs of 
assigning points between consecutive moments in time the objective function can be described by just $(t-1)n^2$,
i.e., polynomially many numbers. In this case, the objective function is of {\em Markov-type} as it reflects only 
memoryless dependencies. \emph{Combinatorial models}, which can be viewed as special choices of such parameters,
are based on the knowledge that the positions of the particles in the next time step lie in certain \emph{windows}.  
A particular such situation has be analyzed in Sect.~\ref{sect:superresolution}. For more results on
combinatorial models see \cite{agwindowconstraints, agdynamic}.

\subsection{Algorithms and complexity}\label{subsec:complexity}

We begin with a simple tractability result for the positionally determined case.

\begin{thm}[\cite{agdynamic}]\label{thm:ILP-posdet} 
For Markov-type objective function oracles~$\mathcal{O}$ the problem {\sc Trac}$(\mathcal{O};d)$ decomposes  
into uncoupled minimum weight perfect bipartite matching problems and can hence be solved in polynomial time.
\end{thm}

Although the reconstruction problem in discrete tomography for $|\mathcal{S}|=2$ directions can be solved in polynomial time (see Thm.~\ref{thm:complexity}) it turns out that there are  severe limitations of extending the previous result already for the following quite restricted partially tomographic case. In fact, the problem becomes hard even if there is only one time step, i.e.,~$t=2$, and~$F^{(1)}$ is  
explicitly known while the set $F^{(2)}$ of particle positions for $\tau=2$ is only accessible through
its two X-rays~$X_{S_1}F^{(2)}$ and $X_{S_2}F^{(2)}$.

\begin{thm}[\cite{agdynamic}]\label{thm:tomo-matching}
Even if all instances are restricted to the case $t=2$, where the solution 
$F^{(1)}$ is given explicitly, {\sc TomTrac}${(\mathcal{O};\mathcal{S})},$ $|\mathcal{S}|=2,$ for Markov-type objective function oracles~$\mathcal{O}$ is $\NP$-hard.
Also the corresponding uniqueness problem is $\NP$-complete and the counting problem 
is $\#\P$-complete.
\end{thm}

Unless $\mathbb{P}=\mathbb{N}\mathbb{P},$ there is thus, in general, no efficient algorithm that provides exact solutions to every instance of {\sc TomTrac}${(\mathcal{O};\mathcal{S})},$ $|\mathcal{S}|=2.$ A possible remedy is to resort to heuristics, which aim at providing approximate solutions. Before we discuss such a heuristic let us state two additional intractability results, which concern the positionally determined case for non Markov-type function oracles. 

\begin{thm}[\cite{agdynamic}]\label{thm:matching}
The problem {\sc Trac}$(\CO;d)$ is $\NP$-hard, even if all instances are restricted to a fixed $t\geq3,$ 
and $\CO$ is a path value oracle. The $\NP$-hardness persists if the objective function values 
provided by $\CO$ are all encoded explicitly.
\end{thm}

It turns out that even if the particles are expected to move along straight lines, this a priori knowledge cannot be exploited efficiently  
(unless $\mathbb{P}=\mathbb{N}\mathbb{P}$). 

\begin{thm}[\cite{agdynamic}] \label{cor:straightline}
For every fixed $d\geq 2$ and $t\geq 3$ it is an $\mathbb{N}\mathbb{P}$-complete problem to decide whether a solution of {\sc Trac}$(\CO;d)$ exists where all particles move along straight lines.
\end{thm}

The proof of Thm.~\ref{cor:straightline} given in~\cite{agdynamic} relies on the hardness of the particular variant {\sc A3ap} of {\sc $3$D-Matching} established
in~\cite{multitracking10}. For further results and a discussion of their practical implications see \cite{agdynamic}.

The previous complexity results show that even for $t=3$ and even if there is no tomography involved
the coupling becomes hard unless it is of the Markov-type, i.e., it only incorporates information that relate not more than two 
consecutive moments in time (Thm. \ref{thm:matching}). But even for $t=2$, which, of course, is of Markov-type, the problem is hard if 
tomography is involved at one point in time (Thm. \ref{thm:tomo-matching}). This means that there is not much room for efficient
algorithms or `self-suggesting' polynomial-time heuristics.

There are, however, quite involved heuristics for {\sc TomTrac}${(\mathcal{O};S_1,S_2)},$ which 
allow to incorporate varous different forms of a priori knowledge and different levels of `particle history'; see \cite{agdynamic}.
Here we focus only on one basic method, called {\sc Rolling Horizon Tomography},  which was introduced in~ \cite{agms-15} and applied to the study of the slip velocity of a gliding arc discharge in~\cite{glidingarc-15}.

The general idea in {\sc Rolling Horizon Tomography} is to model the time step from $\tau$ to $\tau+1$ as a linear program, based on the assumption that~$F^{(1)}$ is known (hence we are dealing with the partially tomographic case). 
The constraints encode the X-rays provided by the data functions
$f_1^{(\tau+1)}$, $f_2^{(\tau+1)}$. The variables correspond to the points in the 
grid $G^{(\tau+1)}$ and are collected in a vector~$x^{(\tau+1)}$.
The X-ray information is encoded by means of a totally unimodular matrix~$A^{(\tau+1)}$ and a right-hand side vector $b^{(\tau+1)}$. 
Further, each point~$g_i^{(\tau+1)} \in G^{(\tau+1)}$ carries a weight $\alpha_{i}^{(\tau+1)},$ which reflects the `distance' to a best 
point $g_i^{(\tau)} \in F^{(\tau)}$ (which is a likely `predecessor'). These weights are collect  in a vector~$a^{(\tau+1)}$. 
Various choices of weights are  discussed in~\cite{agms-15}, which, for instance, model knowledge on 
the velocity of the particles. The algorithm can then be described as follows.

Beginning with $F^{(1)},$ {\sc Rolling Horizon Tomography} solves successively for $\tau\in [t-1]$ the linear program
\[
\begin{array}{lrcl}
&\multicolumn{3}{c}{\min \,\,\bigl(a^{(\tau+1)}\bigr)^T x^{(\tau+1)}}\\[.1cm]
\textnormal{s.\,t. }&A^{(\tau+1)}x^{(\tau+1)} & = & b^{(\tau+1)},\\
&       x^{(\tau+1)} & \le &\1,\\
&       x^{(\tau+1)} & \ge & 0,\\
\end{array}
\] 
in order to determine $F^{(\tau+1)}$ (via its encoding as a $0$-$1$ incidence vector of a 
basic feasible solution of the linear program). 
Finally the paths $\CP_1,\ldots,\CP_n$ are obtained by a routine  that computes a perfect bipartite matching 
in the graph with vertices $F^{(\tau)}\cup F^{(\tau + 1)}$ and edges corresponding to the pairs of vertices that 
realize the distances~$\alpha_{i}^{(\tau)}.$

{\sc Rolling Horizon Tomography} runs in polynomial time, is exact in the sense that it is guaranteed
to return a solution which matches the data. It also allows to incorporate physical knowledge and it is reported 
to work quite well in practice (see \cite{agms-15, glidingarc-15}). However, (and with a view to Thm.~\ref{thm:tomo-matching} not surprisingly),
it is only a heuristic, which may fail to reconstruct the correct paths. The reason is that the weights used to measure the quality of
the assignment do not incorporate the requirement that no two particles can have originated from the
same location at the previous moment in time. Explicit example are given in \cite{agdynamic} which also gives generalizations that
combine the general rolling horizon approach with interpolation and backtracking techniques to provide algorithms that incorporate physical knowledge even better while still running in polynomial time.

\section{Tomographic grain mapping}\label{sec:tom-grain-mapping}

\emph{Tomographic grain mapping} deals with the problem of characterizing \emph{polycrystalline materials} from tomographic data. Polycrystalline materials consist of multiple crystals, called \emph{grains}. These grains, often $10-100$ micrometer in diameter, are of central interest in many areas of materials science as  most metals, ceramics and alloys are such polycrystalline materials.
In fact, the grains determine many of the material's physical, chemical, and mechanical properties (see, e.g., \cite{graindeformation, graindeformation2, grainnucleation, graingrowth} or the monographs~\cite{grainboundarybook2, grainboundarybook}). 

\subsection{Diffraction and indexing}

\begin{figure}[htb]
\centering
\subfigure[] { \includegraphics[width=0.45\textwidth]{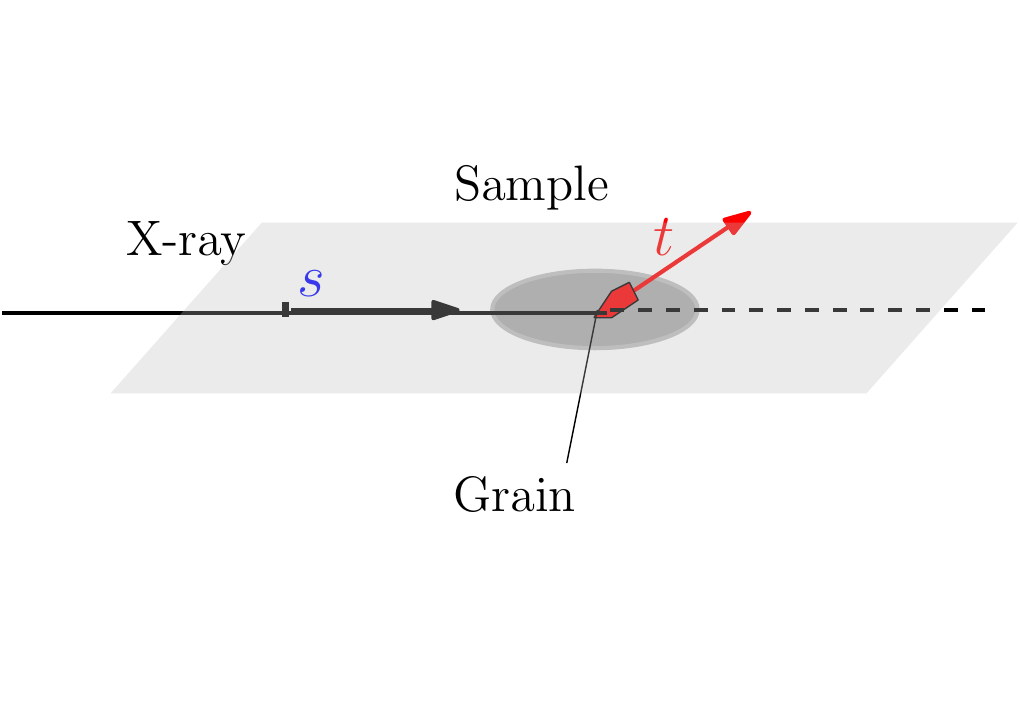}}\hspace*{1ex}
\subfigure[] { \includegraphics[width=0.45\textwidth]{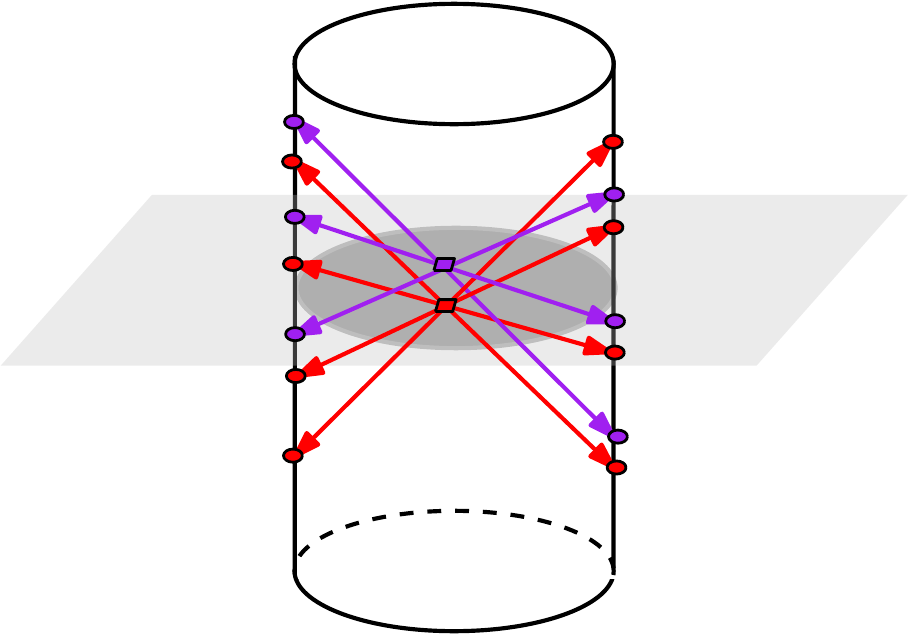}}
\caption{(a) Diffraction by a grain (incoming and diffraction directions are $s$ and $t,$ respectively); (b) the general indexing problem (determining the grains from diffraction spots).} \label{fig:indexingbragg}
\end{figure}

There are several non-trivial technological and algorithmical challenges involved in tomographic grain mapping on different scales. 
Typically, only high-energy X-rays will penetrate the material. In fact, the  required X-ray energies are often so large 
that current experiments need to be conducted at modern synchrotron facilities. For many applications the data are acquired by 
diffraction (as, e.g., in the 3-Dimensional X-ray Diffraction microscopy technique, 3DXRD~\cite{alpers-09, bahnhartbook, poulsenbook, 3dxrdintro} and in Diffraction Constrast Tomography, DCT~\cite{DCT}). Diffraction occurs, however, only if the grain is in a `favorable' position.
This is  governed by \emph{Bragg's law} which relates the unit vectors $t,s$ that signify the incoming and the diffraction directions
and the wavelength $\lambda$ of the X-ray with the crystalline structure of the grain encoded by its dual (or reciprocal) lattice 
$L^\circ$. More precicely, Bragg's law is as follows:
\[\frac{t-s}{\lambda}=\ell\in L^\circ\setminus\{0\};\] 
(see Fig.~\ref{fig:indexingbragg}(a) for an illustration).
Consequently, tomographic data are typically only available from a small number of directions (often, $8-10$). 

The limited number of data, and the fact that multiple grains are simultaneously imaged, poses major algorithmic challenges
at quite different scales, from the atomic to the macroscopic level.  
The problem, commonly referred to as \emph{indexing}~\cite{poulsenbook, graindex, friedel2}, is to group the tomographic 
data according to their grain of origin. This allows often the determination of grain parameters like the lattice (including its orientation),
or the center of mass; see Fig.~\ref{fig:indexingbragg}(b). 
Based on the tomographic data acquired for each single grain, the macroscopic geometric structure of the full collection of different grains is then
to be determined. Of course, such tasks can be highly interrelated, and there are also possible cases where it is favorable to reconstruct several of the grains simultaneously. More details can be found in~\cite{heise}.

\subsection{Macroscopic reconstruction}

As in Sect.~\ref{sect:superresolution}, a single grain~$g$ can be considered as a binary image $\psi_g\in\CF^3.$ 
The points of $\textnormal{supp}(\psi_g)$ correspond to the pixels that belong to~$g$.
The paper~\cite{fu} describes one of the first attempts of reconstructing multiple grains, the so-called \emph{grain map}. 
In this paper the ART algorithm is used, but it is found that often the reconstructions contain unrealistic void spaces 
between adjacent grains. To overcome this problem, a Monte-Carlo approach based on Gibbs priors was introduced in~\cite{apkh-06}. This approach was generalized in~\cite{ks5113} (see also~\cite{arpkh-07}) to deal with the task of 
reconstructing grain maps of moderately deformed grains. More stochastic approaches to grain map reconstruction 
can be found in~\cite{kahkrp-09, grainfollow4, grainfollow2, ks5113}. 
For alternative approaches, see~\cite{grainfollow6, grainfollow5, DARTgrain, grainfollow3, reischigsurvey, suter, grainfollow1}. 

In the following we describe a linear-programming based method, introduced in~\cite{philmag}, which returns 
approximations of grain maps. It is based on only a few input parameters for each grain:
(approximations of) its \emph{center-of-mass}, its \emph{volume} and, if available, its \emph{second-order moments}.
The centers-of-mass can be determined by the indexing procedure, the grain volume by integration of the respective 
X-ray data, and the second-order moments by backprojecting the projections acquired from the same grain.

The aim is to reconstruct what we call \emph{generalized balanced power diagrams~(GBPDs)}. These diagrams generalize \emph{power diagrams} (which are also known as  \emph{Laguerre} or \emph{Dirichlet tessellations}), which in turn generalize \emph{Voronoi diagrams}; see also \cite{aurenhammer87} and \cite[Sect.~6.2]{akl-13}. 

Any GBPD is specified by a set of distinct \emph{sites}~$S=\{s_1,\dots,s_l\}\subset \mathbb{R}^d,$ \emph{additive weights} $(\sigma_1,\dots,\sigma_l)^T\in\mathbb{R}^l,$ and positive definite matrices~$A_1,\dots,A_l\in\mathbb{R}^{d\times d}.$ The \emph{$j$th generalized balanced power cell} $P_j$ is then defined by 
\[
P_j=\{x \in \mathbb{R}^d\::\: ||x - s_j||_{A_j}^2-\sigma_j \leq ||x -s_k ||_{A_k}^2 -\sigma_k, \: \forall k\neq j\},
\] where $||\cdot||_{A_j},$ $j\in[l],$ denotes the \emph{ellipsoidal norm}
\[
|| x||_{A_j}=\sqrt{ x^T A_j  x}.
\]
The generalized balanced power diagram $P$ is the $l$-tuple $P=(P_1,\dots,P_l)$. The proposed method is able to find optimal $\sigma_1,\dots,\sigma_l$ that guarantee that the volumes of each cell are within prescribed ranges.

The concept of GBPDs can be viewed as structure-driven weight balanced clusterings; see~\cite{briedengritzmann12, Brieden2017}. 
For $j\in [l]$ let $s_j$ denote the center of the $j$th grain, and let $\kappa_j^-,$ $\kappa_j^+$ be lower and upper bounds for its volume, respectively.
Further, let $x_1,\dots,x_q$ be the points of the image that has to be partitioned into the grains,
and set $\gamma_{i,j}=||x_i-s_j||_{A_i}^2$ for all~$i,j$.

Then we can model the assignment problem by the following linear program:

\[
\begin{array}{lll}
\displaystyle \textnormal{(LP)}        &\displaystyle\min \sum_{i=1}^q\sum_{j=1}^l\gamma_{i,j}\xi_{i,j}     &\\
\displaystyle \textnormal{subject to}  &\displaystyle\sum_{j=1}^l\xi_{i,j}=1                                 &\displaystyle(i\in[q]),\\
\displaystyle                          &\displaystyle\kappa_j^-\leq\sum_{i=1}^q \xi_{i,j} \leq \kappa_j^+               &\displaystyle(j\in[l]),\\[4.0ex]
\displaystyle                          &\displaystyle\xi_{i,j}\geq 0                                         &\displaystyle(i\in[q];\: j\in[l]).
\end{array}
\]

In general, the variables $\xi_{i,j}$ specify the fraction of the point $x_i$ that is assigned to the center $s_j.$ 
Since, however, the coefficient matrix is totally unimodular all basic feasible solutions are binary, and we obtain 
an optimal assignment of pixels to grains in polynomial time.
 
\begin{figure}[htb]
\centering
\subfigure[]{\includegraphics[width=0.31\textwidth]{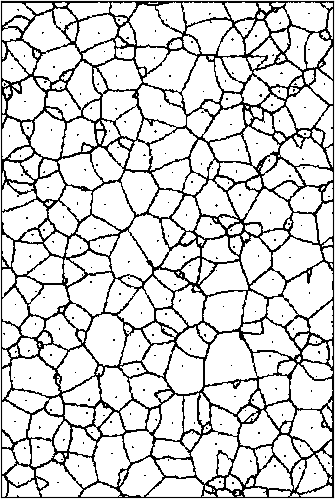}}\hspace*{2ex}
\subfigure[]{\includegraphics[width=0.31\textwidth]{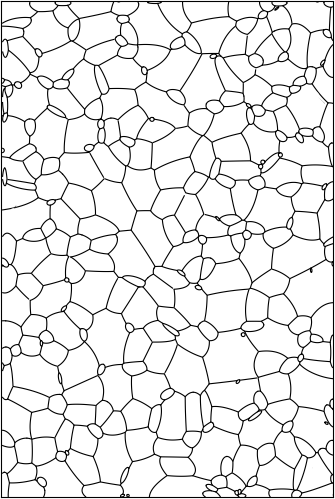}}\hspace*{2ex}
\subfigure[]{\includegraphics[width=0.31\textwidth]{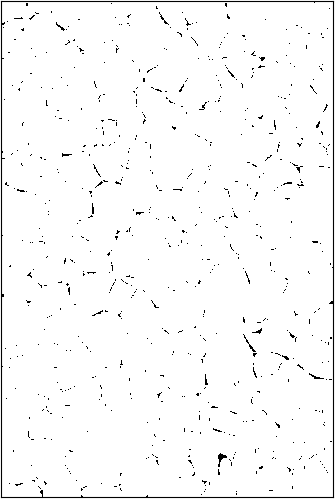}}
\caption{(a) Original image from~\cite[Fig.~9.7]{stoyanbook}. (Black dots represent grain centers.) (b)~Reconstructed generalized balanced power diagram. (c)~Difference map. (Black pixels indicate the pixels that are black in~(b) but white in~(a)). }\label{fig:stoyan}
\end{figure}

An example for the quality of reconstruction for planar grain maps (which are easier to visualize) is shown in Fig.~\ref{fig:stoyan}. 
Reports on the favorable performance of the presented approach on various (real-world) data sets can be found in the recent papers~\cite{stoyan, sedivy2, sedivy,spettl, teferrarowenhorst18}.

We remark that the clustering approach described above was previously applied (in an `isotropic' fashion) in the context of farmland consolidation~\cite{bbg14, farmland}. For an application in designing electoral districts where municipalities of a state have to be grouped into districts of nearly equal population while obeying certain politically motivated requirements, see~\cite{Brieden2017}.

\section{Switching components and a problem in number theory} \label{sect:numbertheory}

Switching components, i.e., pairs of tomographically equivalent sets as introduced in Sect.~\ref{subsec-uniqueness}, 
are strongly related to an old problem in Diophantine number theory, called the \emph{Prouhet-Tarry-Escott} or PTE-problem,
named after Eug{\`e}ne Prouhet~\cite{prouhet51},  Gaston Tarry~\cite{tarry}, and Edward B. Escott~\cite{escott}.

\begin{question}[Prouhet, 1851; Tarry, 1912; Escott, 1910]\hfill

\noindent Given $k,n\in\mathbb{N},$ find two different multisets $X=\{\xi_1,\dots,\xi_n\}\subset \mathbb{Z}$ and $Y=\{\eta_1,\dots,\eta_n\}\subset \mathbb{Z},$ such that
\[
\xi_1^j+\xi_2^j+\ldots+\xi_n^j = \eta_1^j+\eta_2^j+\ldots+\eta_n^j,\qquad \textnormal{for }j\in[k].
\]
\end{question}

Pairs $(X,Y)$ satisfying the above equation are called \emph{PTE solutions}. 
More precisely, we speak of \emph{ $(k,n)$-solutions}, and the numbers~$k$ and $n,$ respectively, are referred to as 
the \emph{degree} and \emph{size} of the PTE solution. Often the notation $X\stackrel{k}{=}Y$ is used to indicate 
that $(X,Y)$ is a degree~$k$ solution. For instance, as an elementary calculation shows,  
\[
\{0,14,28,56,70,84\}\stackrel{5}{=}\{4,6,40,44,78,80\}.
\] 

The PTE problem has connections to several other problems in number theory,  including the \emph{`easier' Waring problem} \cite{wright34, wright72}, \cite[Sect.~12]{borwein-02}, the \emph{Hilbert-Kamke problem} \cite{kamke, kleiman}, and a conjecture due to 
Erd{\H{o}}s and Szekeres \cite{purepr, erdoesszekeres, maltby97}, \cite[Sect.~13]{borwein-02}.  There are also connections to \emph{Ramanujan identities}~\cite{r1, r2}, other types of multigrade equations~\cite{newPTE,ptelikep}, problems in algebra~\cite{maltby00, myerson86}, geometry~\cite{newtonpolygons},  combinatorics~\cite{adler77, upolynomial, thuemorse, cerny2},  graph theory~\cite{integerroots}, and computer science~\cite{fewproductgates, cerny, codes}.
For background information see~\cite{alpers-h18, borwein-02, borweinrevisited, dickson, gloden}.

The PTE problem can be traced back to a correspondence between Goldbach and Euler. In his 1950 letter~\cite{goldbachletter} Goldbach states the identity
\begin{multline*}
(\alpha+\beta+\delta)^2+(\alpha+\gamma+\delta)^2+(\beta+\gamma+\delta)^2+\delta^2\\
=(\alpha+\delta)^2+(\beta+\delta)^2+(\gamma+\delta)^2+(\alpha+\beta+\gamma+\delta)^2,\label{eq:classic}\end{multline*} 
 where $\alpha,\beta,\gamma,\delta\in\mathbb{Z}.$ In other words,  
\[
\{\alpha+\beta+\delta, \alpha+\gamma+\delta, \beta+\gamma+\delta,\delta\}\stackrel{2}{=}\{\alpha+\delta,\beta+\delta,\gamma+\delta,\alpha+\beta+\gamma+\delta\}.
\]  

It was already known to Prouhet, Tarry, and Escott that there exist $(k,2^k)$-solutions for every~$k$ (see~\cite{wright59} and \cite[Sect.~24]{dickson}). Such solutions can be generated as follows. Express each $p\in [2^{k+1}-1]_0$ as a binary number. If this binary expression of~$p$ contains an even number of $1$'s, then assign~$p$ to the set~$X,$ otherwise to~$Y.$ Then,  $(X,Y)$ with $X=\{\xi_1,\dots,\xi_{2^k}\}$ and $Y=\{\eta_1,\dots,\eta_{2^k}\}$ is a~$(k,2^k)$-solution. Proofs of this result can be found  in~\cite{nguyen16, wright59}. For generalizations, see~\cite{lehmer, sinha}.  

On the other hand, there are no $(k,n)$-solutions whenever $n<k+1.$  This result, commonly attributed to Bastien~\cite{bastien}, can be derived from the Newton's identities \cite[Sect.~21.9]{hardywright08}. A $(k,n)$-solution is called \emph{ideal} if $n=k+1.$ 

The following is a long-standing open question (see~\cite{wright35} and~\cite[Sect.~11]{borwein-02}).

\begin{question}
Do there exist ideal PTE solutions for every~$k?$
\end{question}

Presently, ideal solutions are only known for $k\in[11]\setminus\{10\}.$  
Concerning upper bounds on $n,$ the currently best bound (of \cite{melzak}) guarantees that for any $k$ there exists a  
$(k,n)$-solution with $n\leq \frac{1}{2}(k^2-3)$ if~$k$ is odd and $n \leq \frac{1}{2}(k^2-4)$ if~$k$ is even. The proofs are non-constructive. In fact, all currently known constructive proofs yield bounds that are exponential in~$k.$

\subsection{PTE solutions from switching components}
The following explicit connection between the PTE problem and switching components first appeared in \cite[Sect.~6]{alpers-d03}. Following~\cite{alpers-tijdeman-07}, we will focus on the case~$d=2$ (for general~$d$ see~\cite{ghiglione}).

For given $M\subset \mathbb{Z}^d$ and $c\in \mathbb{Z}^d,$ let  $\Pi_c(M)$ denote the multiset
\[
\Pi_c(M)=\{c^Tx\::\: x\in X\}.
\] Clearly, $\Pi_c(M)\subset \mathbb{Z}.$ Perhaps more surprisingly, the following result holds if we insert the  points of a switching component.

\begin{thm}[{\cite{alpers-tijdeman-07}}] \label{thm:PT1} 
If $(X,Y)$ is an $(m+1)$-switching component in $\mathbb{Z}^2$ and $c\in\mathbb{Z}^2$ such that $\Pi_c(X)\neq\Pi_c(Y),$ then $(\Pi_c(X),\Pi_c(Y))$ is a degree~$m$ solution of the PTE problem.
\end{thm}

\begin{figure}[htb]
\begin{center} 
\subfigure[]{\includegraphics[width=0.15\textwidth]{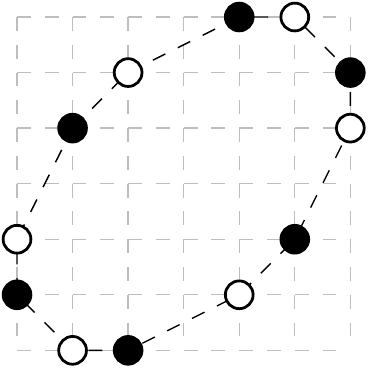}} \hspace*{2ex}
\subfigure[]{\includegraphics[width=0.195\textwidth]{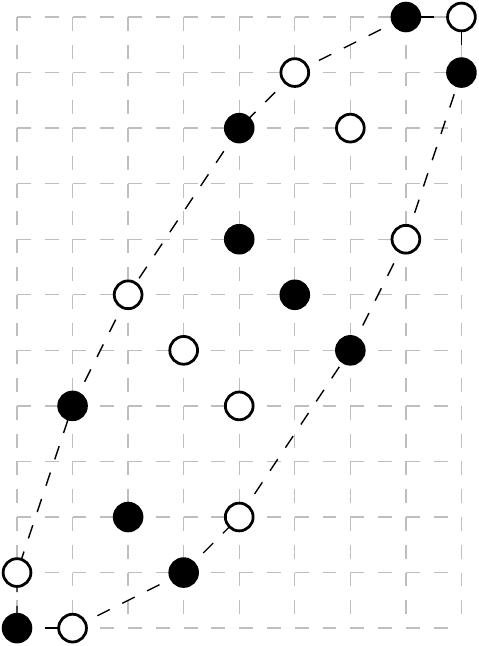}} \hspace*{2ex}
\subfigure[]{\includegraphics[width=0.22\textwidth]{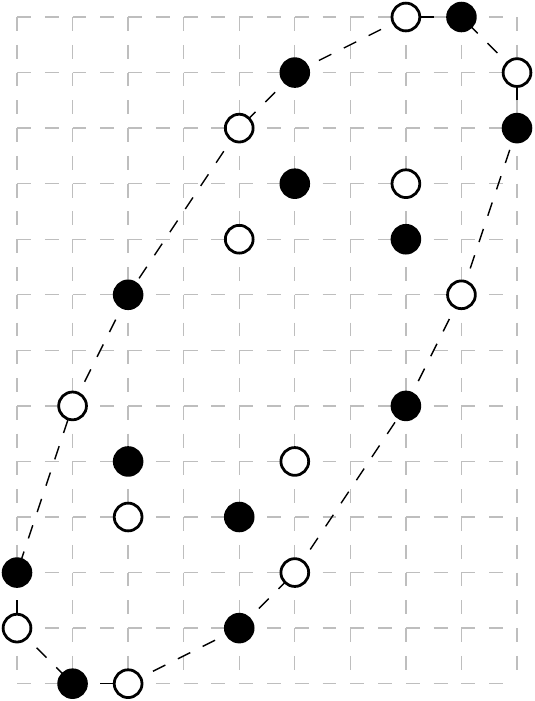}} \hspace*{2ex}
\subfigure[]{\includegraphics[width=0.255\textwidth]{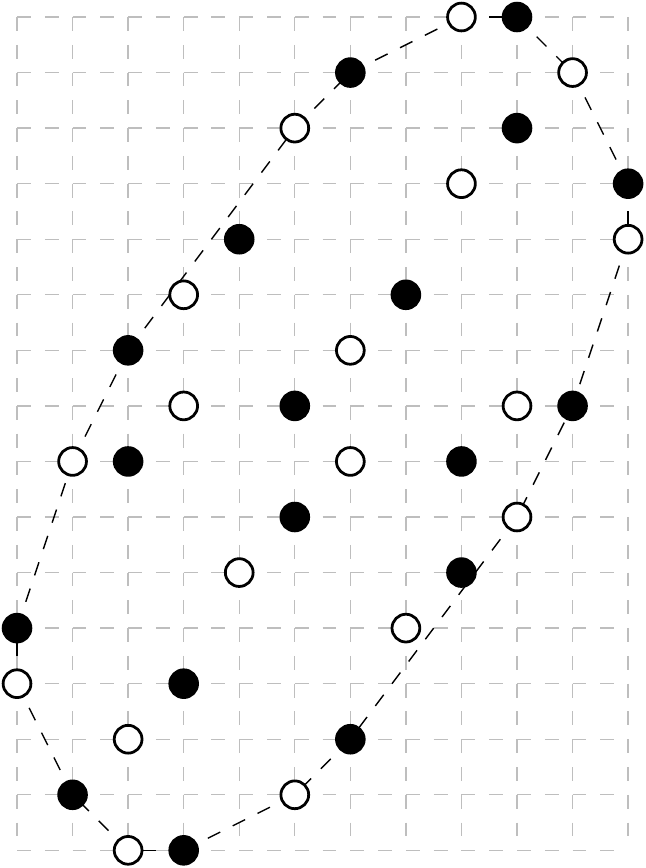}} \hspace*{2ex}
%\subfigure[]{\includegraphics[width=0.38\textwidth]{s5.pdf}} 
\caption{Examples of small switching components for (a) $m=6,$ (b) $m=7,$ (c) $m=8,$ (d) $m=9,$ directions (indicated as polygon edges). 
The switching components are the sets of $6,$ $10,$ $12,$ and $18$ black and white points, respectively. }\label{fig:smallswitch}
\end{center}
\end{figure}

This construction of PTE-solutions from switching components can be exemplified, say, for the switching components depicted in 
Fig. \ref{fig:smallswitch}.
 
For instance, if in Fig.~\ref{fig:smallswitch}(a) the origin is located in the lower left lattice point (which, of course, is an 
arbitrary choice) the sets~$X$ and~$Y$ of black and white points are
\[
\begin{array}{lll}
X&=&\left\{\left(\begin{array}{c}0\\2\end{array}\right), \left(\begin{array}{c}1\\0\end{array}\right), \left(\begin{array}{c}2\\5\end{array}\right), \left(\begin{array}{c}4\\1\end{array}\right), \left(\begin{array}{c}5\\6\end{array}\right),\left(\begin{array}{c}6\\4\end{array}\right)\right\},\\[2ex]
Y&=&\left\{\left(\begin{array}{c}0\\1\end{array}\right), \left(\begin{array}{c}1\\4\end{array}\right), \left(\begin{array}{c}2\\0\end{array}\right), \left(\begin{array}{c}4\\6\end{array}\right), \left(\begin{array}{c}5\\2\end{array}\right),\left(\begin{array}{c}6\\5\end{array}\right)\right\}.
\end{array} 
\]
For $c^T=(1,2)$ we obtain the PTE solution 
\[\Pi_c(X)=\{1,4,6,12,14,17\} \stackrel{5}{=} \{2,2,9,9,16,16\}=\Pi_c(Y)\]
of degree~5.
As a basic ingredient the standard proof of Thm.~\ref{thm:PT1} uses the encoding of points by polynomials mentioned 
in connection with Thm.~\ref{thm:hajdutijdeman}.

Thm.~\ref{thm:PT1} allows to derive explicit constructions of families of PTE solutions; \cite{alpers-tijdeman-07}).
As an example let us consider the result of Prouhet, Tarry, and Escott that $(k,2^k)$-solutions exist for every~$k.$ 
The proof given in~\cite{wright35} extends over two half-pages. 
The geometric shortcut via Thm.~\ref{thm:PT1} just uses the construction of switching components $(X,Y)$ with $|X|=|Y|\leq2^{k}$ from 
Fig.~\ref{fig:zonotopeconstruction}. 

\subsection{Generalizations}
The geometric point of view also helps in studying other variants of the PTE problem. 
Naturally, PTE can be considered over arbitrary rings $R$. For $R=\Z^d$ we obtain  \emph{PTE$_d$} which can be viewed as a $d$-dimensional or multinomial version of the original PTE problem.

\newpage

\begin{question}[PTE$_d$]\hfill

\noindent Given $d,k,n\in\mathbb{N},$ find two different multisets 
$X=\{\xi_1,\dots,\xi_n\},$ $Y=\{\eta_1,\dots,\eta_n \}\subset \mathbb{Z}^d$ 
with $\xi_l=(\xi_{l1}, \dots, \xi_{ld})^T,$ $\eta_l=(\eta_{l1}, \dots, \eta_{ld})^T$ for $l\in[n]$  such that
\[  \sum_{l=1}^n \xi_{l1}^{j_1}\xi_{l2}^{j_2} \cdot \ldots \cdot\xi_{ld}^{j_d}=\sum_{l=1}^n \eta_{l1}^{j_1}\eta_{l2}^{j_2} \cdot \ldots \cdot \eta_{ld}^{j_d} \]
for all non-negative integers  $j_1, \dots, j_d$ with $j_1+j_2+\ldots +j_d \leq k.$
\end{question} 

There are trivial ways of generating PTE$_d$-solution from PTE$_1$-solutions. For instance, if $\{\alpha_1,\dots,\alpha_n\}\stackrel{k}{=}\{\beta_1,\dots,\beta_n\}$ is a~PTE$_1$-solution, then \[\{(\alpha_1,\dots,\alpha_1)^T,\dots,(\alpha_n,\dots,\alpha_n)^T\}\stackrel{k}{=}\{(\beta_1,\dots,\beta_1)^T,\dots,(\beta_n,\dots,\beta_n)^T\}\] is a solution to PTE$_d.$ 
A general method of generating non-trivial solutions for PTE$_2$ is provided by \cite{alpers-tijdeman-07}.

\begin{thm}[{\cite{alpers-tijdeman-07}}] \label{prpproC}
Every $(m+1)$-switching component $(X,Y)$  in $\mathbb{Z}^2,$ is a degree~$m$ solution of the PTE$_2$ problem. 
\end{thm}

For instance, for the sets $X$ and $Y$ corresponding to Fig.~\ref{fig:smallswitch}(a) it is elementary to verify that
\[\begin{aligned}
0^i2^j+1^i0^j+2^i5^j+&4^i1^j+5^i6^j+6^i4^j\\
&= 0^i1^j+1^i4^j+2^i0^j+4^i6^j+5^i2^j+6^i5^j\end{aligned}
\] 
for all non-negative $i,j$ with $i+j\leq 5.$

Applying Thm.~\ref{prpproC} to the known smallest size switching components (an example for~$k+1=6$ is 
depicted in Fig.~\ref{fig:smallswitch}(a)), one sees that for every degree $k\in\{1,2,3,5\}$ there exist ideal PTE$_2$ solutions; \cite{alpers-tijdeman-07}.  

The following problem is, however, open already for $k=4.$

\begin{question} Do there exist ideal PTE$_2$ for every degree~$k$?
\end{question}

\section{Concluding remarks}\label{sec:concluding-remarks}

The present paper tried to support the following conviction of the authors: 
Discrete tomography is a broad and interesting field, both, in terms of its methods and its applications.
Discrete tomography has strong links to various areas within mathematics which have the potential
to provide new insight in older problems. 
Discrete tomography has a variety of applications to various other scientific fields and to
relevant real-world problems. And, finally, discrete tomography  is a rich source of scientific challenges. 

\section*{Acknowledgements}
This work was supported in part by  the {\em Deutsche Forschungsgemeinschaft} Grant GR 993/10-2 and
the {\em European COST Network}  MP1207.
The authors are grateful to Fabian Klemm for his help with producing Fig.~\ref{fig:stoyan} and both, Fabian Klemm 
and Viviana Ghiglione, for helpful discussions. 

 \newcommand{\noop}[1]{}


\begin{thebibliography}{100}
\item[]\hspace{-\labelwidth}\hspace{-\labelsep}\textbf{General reading: Introductions, surveys, books}:
\bibitem{a2010}
R.~J. Adrian and J.~Westerweel.
\newblock {\em Particle Image Velocimetry}.
\newblock Cambridge University Press, New York, NY, 2010.

\bibitem{alpers-d03}
A.~Alpers.
\newblock {\em Instability and Stability in Discrete Tomography}.
\newblock {PhD} thesis, Technische Universit{\"a}t M{\"u}nchen, Zentrum
  Mathematik, 2003.
\newblock (published by Shaker Verlag, ISBN 3-8322-2355-X).

\bibitem{alpers-09}
A.~Alpers.
\newblock A short introduction to tomographic grain map reconstruction.
\newblock {\em PU.M.A.}, 20(1-2):157--163, 2009.
\newblock URL: \url{http://puma.dimai.unifi.it/20_1_2/10_Alpers_7.pdf}.

\bibitem{alpers-h18}
A.~Alpers.
\newblock {\em On the Tomography of Discrete Structures: Mathematics,
  Complexity, Algorithms, and its Applications in Materials Science and Plasma
  Physics}.
\newblock {H}abilitation thesis, Technische Universit{\"a}t M{\"u}nchen,
  Zentrum Mathematik, 2018.
\newblock URL: \url{http://mediatum.ub.tum.de/1441933}.

\bibitem{arpkh-07}
A.~Alpers, L.~Rodek, H.~F. Poulsen, E.~Knudsen, and G.~T. Herman.
\newblock Discrete tomography for generating grain maps of polycrystals.
\newblock In G.~T. Herman and A.~Kuba, editors, {\em Advances in Discrete
  Tomography and its Applications}, pages 271--301. Birkh\"auser, Boston, 2007.
\newblock \href {http://dx.doi.org/10.1007/978-0-8176-4543-4_13}
  {\path{doi:10.1007/978-0-8176-4543-4_13}}.

\bibitem{discreteinversepbook2}
R.~C. Aster, B.~Borchers, and C.~H. Thurber.
\newblock {\em Parameter Estimation and Inverse Problems}.
\newblock Academic Press, Boston, MA, 2nd edition edition, 2013.
\newblock \href {http://dx.doi.org/10.1016/B978-0-12-385048-5.00030-6}
  {\path{doi:10.1016/B978-0-12-385048-5.00030-6}}.

\bibitem{aurenhammer87}
F.~Aurenhammer.
\newblock Power diagrams: {P}roperties, algorithms and applications.
\newblock {\em SIAM J. Comput.}, 16(1):78--96, 1987.
\newblock \href {http://dx.doi.org/10.1137/0216006}
  {\path{doi:10.1137/0216006}}.

\bibitem{akl-13}
F.~Aurenhammer, R.~Klein, and D.-T. Lee.
\newblock {\em Voronoi Diagrams and {D}elaunay Triangulations}.
\newblock World Scientific, Singapore, 2013.

\bibitem{atomicETreview}
S.~Bals, B.~Goris, A.~{De Backer}, S.~{Van Aert}, and G.~{Van Tendeloo}.
\newblock Atomic resolution electron tomography.
\newblock {\em MRS Bulletin}, 41(7):525--530, 2016.
\newblock \href {http://dx.doi.org/10.1557/mrs.2016.138}
  {\path{doi:10.1557/mrs.2016.138}}.

\bibitem{bahnhartbook}
J.~Banhart, editor.
\newblock {\em Advanced Tomographic Methods in Materials Research and
  Engineering}.
\newblock Oxford University Press, Oxford, 2008.

\bibitem{bbg14}
S.~Borgwardt, A.~Brieden, and P.~Gritzmann.
\newblock Geometric clustering for the consolidation of farmland and woodland.
\newblock {\em Math. Intell.}, 36(2):37--44, 2014.
\newblock \href {http://dx.doi.org/10.1007/s00283-014-9448-2}
  {\path{doi:10.1007/s00283-014-9448-2}}.

\bibitem{borwein-02}
P.~Borwein.
\newblock {\em Computational Excursions in Analysis and Number Theory}.
\newblock Springer, New York, NY, 2002.
\newblock \href {http://dx.doi.org/10.1007/978-0-387-21652-2}
  {\path{doi:10.1007/978-0-387-21652-2}}.

\bibitem{borweinrevisited}
P.~Borwein and C.~Ingalls.
\newblock The {P}rouhet-{T}arry-{E}scott problem revisited.
\newblock {\em Enseign. Math.}, 40(1-2):3--27, 1994.
\newblock \href {http://dx.doi.org/10.5169/seals-61102}
  {\path{doi:10.5169/seals-61102}}.

\bibitem{brualdibook}
R.~A. Brualdi.
\newblock {\em Combinatorial Matrix Classes}.
\newblock Cambridge University Press, Cambridge, 2006.

\bibitem{assignmentbook09}
R.~Burkard, M.~{Dell'Amico}, and S.~Martello.
\newblock {\em Assignment Problems}.
\newblock SIAM, Philadelphia, PA, 2009.
\newblock \href {http://dx.doi.org/10.1137/1.9781611972238}
  {\path{doi:10.1137/1.9781611972238}}.

\bibitem{stoyanbook}
S.~N. Chiu, D.~Stoyan, W.~S. Kendall, and J.~Mecke.
\newblock {\em Stochastic Geometry and its Applications}.
\newblock Wiley, Chichester, 3rd edition edition, 2013.

\bibitem{polyominoes-99}
A.~{Del Lungo} and M.~Nivat.
\newblock Reconstruction of connected sets from two projections.
\newblock In G.~T. Herman and A.~Kuba, editors, {\em Discrete Tomography},
  pages 163--188. Birkh\"auser, Boston, 1999.
\newblock \href {http://dx.doi.org/10.1007/978-1-4612-1568-4_7}
  {\path{doi:10.1007/978-1-4612-1568-4_7}}.

\bibitem{dickson}
L.~E. Dickson.
\newblock {\em History of the Theory of Numbers}, volume~2.
\newblock Dover Publications, Mineola, NY, 1920.

\bibitem{ETreview1}
L.~Gan and G.~J. Jensen.
\newblock Electron tomography of cells.
\newblock {\em Q. Rev. Biophys.}, 45(1):27--56, 2012.
\newblock \href {http://dx.doi.org/10.1017/S0033583511000102}
  {\path{doi:10.1017/S0033583511000102}}.

\bibitem{gardnerbook}
R.~J. Gardner.
\newblock {\em Geometric Tomography}.
\newblock Cambridge University Press, New York, NY, 2nd edition edition, 2006.
\newblock \href {http://dx.doi.org/10.1017/CBO9781107341029}
  {\path{doi:10.1017/CBO9781107341029}}.

\bibitem{gardnergritzmann99}
R.~J. Gardner and P.~Gritzmann.
\newblock Uniqueness and complexity in discrete tomography.
\newblock In G.~T. Herman and A.~Kuba, editors, {\em Discrete Tomography},
  pages 85--113. Birkh\"auser, Boston, 1999.
\newblock \href {http://dx.doi.org/10.1007/978-1-4612-1568-4_4}
  {\path{doi:10.1007/978-1-4612-1568-4_4}}.

\bibitem{gareyjohnsonbook}
M.~R. Garey and D.~S. Johnson.
\newblock {\em Computers and Intractability}.
\newblock W. H. Freeman and Co., San Francisco, CA, 1979.

\bibitem{gloden}
A.~Gloden.
\newblock {\em Mehrgradige Gleichungen}.
\newblock P. Noordhoff, Groningen, 2nd edition edition, 1943.

\bibitem{grimmgritzmannhuck}
U.~Grimm, P.~Gritzmann, and C.~Huck.
\newblock Discrete tomography of model sets: {R}econstruction and uniqueness.
\newblock In M.~Baake and U.~Grimm, editors, {\em Aperiodic Order}, volume~2,
  pages 39--72. Cambridge University Press, Cambridge, 2017.
\newblock \href {http://dx.doi.org/10.1017/9781139033862.004}
  {\path{doi:10.1017/9781139033862.004}}.

\bibitem{gritzmann-97}
P.~Gritzmann.
\newblock On the reconstruction of finite lattice sets from their {X}-rays.
\newblock In E.~Ahronovitz and C.~Fiorio, editors, {\em Discrete Geometry for
  Computer Imagery}, pages 19--32. Springer, Berlin, 1997.
\newblock \href {http://dx.doi.org/10.1007/BFb0024827}
  {\path{doi:10.1007/BFb0024827}}.

\bibitem{gr-10}
P.~Gritzmann.
\newblock Discrete tomography: From battleship to nanotechnology.
\newblock In Aignern M. and E.~Behrends, editors, {\em Mathematics Everywhere},
  pages 81--98. American Mathematical Society, 2010.
\newblock \href {http://dx.doi.org/10.1136/bmj.c3793}
  {\path{doi:10.1136/bmj.c3793}}.

\bibitem{gritzmann-devries-2001}
P.~Gritzmann and S.~{de Vries}.
\newblock Reconstructing crystalline structures from few images under high
  resolution transmission electron microscopy.
\newblock In W.~J{\"a}ger and H.~J. Krebs, editors, {\em Mathematics: Key
  Technology for the Future}, pages 441--459. Springer, Berlin, 2003.
\newblock \href {http://dx.doi.org/10.1007/978-3-642-55753-8_36}
  {\path{doi:10.1007/978-3-642-55753-8_36}}.

\bibitem{hadamard-23}
J.~Hadamard.
\newblock {\em Lectures on the {C}auchy Problem in Linear Partial Differential
  Equations}.
\newblock Dover Publications Inc., Mineola, NY, 1952.
\newblock Reprint of the 1923 original.

\bibitem{hajdutijdeman-07}
L.~Hajdu and R.~Tijdeman.
\newblock Algebraic discrete tomography.
\newblock In G.~T. Herman and A.~Kuba, editors, {\em Advances in Discrete
  Tomography and its Applications}, pages 55--81. Birkh\"auser, Boston, 2007.
\newblock \href {http://dx.doi.org/10.1007/978-0-8176-4543-4_4}
  {\path{doi:10.1007/978-0-8176-4543-4_4}}.

\bibitem{hansenbook}
P.~C. Hansen.
\newblock {\em Discrete Inverse Problems: Insight and Algorithms}.
\newblock SIAM, Philadelphia, PA, 2010.
\newblock \href {http://dx.doi.org/10.1137/1.9780898718836}
  {\path{doi:10.1137/1.9780898718836}}.

\bibitem{hardywright08}
G.~H. Hardy and E.~M. Wright.
\newblock {\em An Introduction to the Theory of Numbers}.
\newblock Oxford University Press, Oxford, 6th edition edition, 2008.

\bibitem{hermanbook}
G.~T. Herman.
\newblock {\em Fundamentals of Computerized Tomography: Image Reconstruction
  from Projections}.
\newblock Springer, London, 2nd edition edition, 2009.
\newblock \href {http://dx.doi.org/10.1007/978-1-84628-723-7}
  {\path{doi:10.1007/978-1-84628-723-7}}.

\bibitem{hermanbookchapter}
G.~T. Herman.
\newblock Iterative reconstruction techniques and their superiorization for the
  inversion of the {R}adon transform.
\newblock In O.~Scherzer and R.~Ramlau, editors, {\em The Radon Transform: The
  First 100 Years and Beyond}, pages ??--?? De Gruyter, Berlin, 2019.

\bibitem{frankbook}
G.~T. Herman and J.~Frank, editors.
\newblock {\em Computational Methods for Three-Dimensional Microscopy
  Reconstruction}.
\newblock Springer, New York, NY, 2014.
\newblock \href {http://dx.doi.org/10.1007/978-1-4614-9521-5}
  {\path{doi:10.1007/978-1-4614-9521-5}}.

\bibitem{kubaherman1}
G.~T. Herman and A.~Kuba, editors.
\newblock {\em Discrete Tomography: Foundations, Algorithms, and Applications}.
\newblock Birkh\"auser, Boston, MA, 1999.
\newblock \href {http://dx.doi.org/10.1007/978-1-4612-1568-4}
  {\path{doi:10.1007/978-1-4612-1568-4}}.

\bibitem{kubaherman2}
G.~T. Herman and A.~Kuba, editors.
\newblock {\em Advances in Discrete Tomography and its Applications}.
\newblock Birkh\"auser, Boston, MA, 2007.
\newblock \href {http://dx.doi.org/10.1007/978-0-8176-4543-4}
  {\path{doi:10.1007/978-0-8176-4543-4}}.

\bibitem{katzbook}
M.~B. Katz.
\newblock {\em Questions of Uniqueness and Resolution in Reconstruction from
  Projections}.
\newblock Springer, Berlin, 1978.
\newblock \href {http://dx.doi.org/10.1007/978-3-642-45507-0}
  {\path{doi:10.1007/978-3-642-45507-0}}.

\bibitem{kirschbook}
A.~Kirsch.
\newblock {\em An Introduction to the Mathematical Theory of Inverse Problems}.
\newblock Springer, New York, NY, 2011.
\newblock \href {http://dx.doi.org/10.1007/978-1-4419-8474-6}
  {\path{doi:10.1007/978-1-4419-8474-6}}.

\bibitem{grainboundarybook2}
U.~F. Kocks, C.~N. Tom{\'e}, and H.-R. Wenk.
\newblock {\em Texture and Anisotropy: Preferred Orientations in Polycrystals
  and their Effect on Materials Properties}.
\newblock Cambridge University Press, Cambridge, 2000.

\bibitem{louisbookchapter}
A.~K. Louis.
\newblock Uncertainty, ghosts and resolution in {R}adon problems.
\newblock In O.~Scherzer and R.~Ramlau, editors, {\em The Radon Transform: The
  First 100 Years and Beyond}, pages ??--?? De Gruyter, Berlin, 2019.

\bibitem{markoebook}
A.~Markoe.
\newblock {\em Analytic Tomography}.
\newblock Cambridge University Press, New York, NY, 2014.
\newblock \href {http://dx.doi.org/10.1017/CBO9780511530012}
  {\path{doi:10.1017/CBO9780511530012}}.

\bibitem{siltanenbook}
J.~L. Mueller and S.~Siltanen.
\newblock {\em Linear and Nonlinear Inverse Problems with Practical
  Applications}.
\newblock SIAM, Philadelphia, PA, 2012.
\newblock \href {http://dx.doi.org/10.1137/1.9781611972344}
  {\path{doi:10.1137/1.9781611972344}}.

\bibitem{nattererbook}
F.~Natterer.
\newblock {\em The Mathematics of Computerized Tomography}.
\newblock SIAM, Philadelphia, PA, 2001.
\newblock \href {http://dx.doi.org/10.1137/1.9780898719284}
  {\path{doi:10.1137/1.9780898719284}}.

\bibitem{nattererbook2}
F.~Natterer and F.~W{\"u}bbeling.
\newblock {\em Mathematical Methods in Image Reconstruction}.
\newblock SIAM, Philadelphia, PA, 2001.
\newblock \href {http://dx.doi.org/10.1137/1.9780898718324}
  {\path{doi:10.1137/1.9780898718324}}.

\bibitem{oktembook}
O.~{\"O}ktem.
\newblock Mathematics of electron tomography.
\newblock In O.~Scherzer, editor, {\em Handbook of Mathematical Methods in
  Imaging}, pages 937--1031. Springer, New York, NY, 2nd edition edition, 2015.
\newblock \href {http://dx.doi.org/10.1007/978-3-642-27795-5_43-2}
  {\path{doi:10.1007/978-3-642-27795-5_43-2}}.

\bibitem{compcomplbook}
C.~H. Papadimitriou.
\newblock {\em Computational Complexity}.
\newblock Addison-Wesley, Reading, MA, 1995.

\bibitem{poulsenbook}
H.~F. Poulsen.
\newblock {\em Three-Dimensional {X}-ray Diffraction Microscopy}.
\newblock Springer, Berlin, 2004.

\bibitem{3dxrdintro}
H.~F. Poulsen.
\newblock An introduction to three-dimensional {X}-ray diffraction microscopy.
\newblock {\em J. Appl. Cryst.}, 45(6):1084--1097, 2012.
\newblock \href {http://dx.doi.org/10.1107/S0021889812039143}
  {\path{doi:10.1107/S0021889812039143}}.

\bibitem{grainfollow6}
H.~F. Poulsen, S.~Schmidt, D.~Juul~Jensen, H.~O. S{\o}rensen, E.~M. Lauridsen,
  U.~L. Olsen, W.~Ludwig, A.~King, J.~P. Wright, and G.~B.~M. Vaughan.
\newblock {3D} {X}-ray diffraction microscopy.
\newblock In R.~Barabash and G.~Ice, editors, {\em Strain and Dislocation
  Gradients from Diffraction: Spatially-Resolved Local Structure and Defects},
  pages 205--253. World Scientific, Singapore, 2014.

\bibitem{grainboundarybook}
L.~Priester.
\newblock {\em Grain Boundaries: From Theory to Engineering}.
\newblock Springer, Dordrecht, 2013.
\newblock \href {http://dx.doi.org/10.1007/978-94-007-4969-6}
  {\path{doi:10.1007/978-94-007-4969-6}}.

\bibitem{ryser}
H.~J. Ryser.
\newblock Combinatorial properties of matrices of zeros and ones.
\newblock {\em Canad. J. Math.}, 9(1):371--377, 1957.
\newblock \href {http://dx.doi.org/10.1007/978-0-8176-4842-8_18}
  {\path{doi:10.1007/978-0-8176-4842-8_18}}.

\bibitem{ryserbook}
H.~J. Ryser.
\newblock {\em Combinatorial Mathematics}.
\newblock MAA, Washington, DC, 1963.

\bibitem{schrijverbook}
A.~Schrijver.
\newblock {\em Theory of Linear and Integer Programming}.
\newblock Wiley, Chichester, 1986.

\bibitem{piv2book}
A.~Schr\"oder and C.~E. Willert, editors.
\newblock {\em Particle Image Velocimetry}.
\newblock Springer, Berlin, 2008.
\newblock \href {http://dx.doi.org/10.1007/978-3-540-73528-1}
  {\path{doi:10.1007/978-3-540-73528-1}}.

\bibitem{HRTEMlimits3}
D.~J. Smith.
\newblock Progress and perspectives for atomic-resolution electron microscopy.
\newblock {\em Ultramicroscopy}, 108(3):159--166, 2018.
\newblock \href {http://dx.doi.org/10.1016/j.ultramic.2007.08.015}
  {\path{doi:10.1016/j.ultramic.2007.08.015}}.

\bibitem{HRTEMbook}
J.~C.~H. Spence.
\newblock {\em High-Resolution Electron Microscopy}, volume~4.
\newblock Oxford Univ. Press, Oxford, 2013.
\newblock \href {http://dx.doi.org/10.1093/acprof:oso/9780199668632.001.0001}
  {\path{doi:10.1093/acprof:oso/9780199668632.001.0001}}.

\bibitem{tarantolabook}
A.~Tarantola.
\newblock {\em Inverse Problem Theory and Methods for Model Parameter
  Estimation}.
\newblock SIAM, Philadelphia, PA, 2005.

\item[]\hspace{-\labelwidth}\hspace{-\labelsep}\textbf{Citations in tomography}:

\bibitem{aharoniherman}
R.~Aharoni, G.~T. Herman, and A.~Kuba.
\newblock Binary vectors partially determined by linear equation systems.
\newblock {\em Discrete Math.}, 171(1-3):1--16, 1997.
\newblock \href {http://dx.doi.org/10.1016/S0012-365X(96)00068-4}
  {\path{doi:10.1016/S0012-365X(96)00068-4}}.

\bibitem{ab-05}
A.~Alpers and S.~Brunetti.
\newblock On the stability of reconstructing lattice sets from {X}-rays along
  two directions.
\newblock In {\em Discrete Geometry for Computer Imagery}, LNCS 3429, pages
  92--103. Springer, Berlin, 2005.
\newblock \href {http://dx.doi.org/10.1007/978-3-540-31965-8_9}
  {\path{doi:10.1007/978-3-540-31965-8_9}}.

\bibitem{alpersgardner13}
A.~Alpers, R.~J. Gardner, S.~K{\"o}nig, R.~S. Pennington, C.~B. Boothroyd,
  L.~Houben, R.~E. Dunin-Borkowski, and K.~J. Batenburg.
\newblock Geometric reconstruction methods for electron tomography.
\newblock {\em Ultramicroscopy}, 128(C):42--54, 2013.
\newblock \href {http://dx.doi.org/10.1016/j.ultramic.2013.01.002}
  {\path{doi:10.1016/j.ultramic.2013.01.002}}.

\bibitem{ghiglione}
A.~Alpers, V.~Ghiglione, and P.~Gritzmann.
\newblock On the geometry of switching components.
\newblock in preparation, 2018.

\bibitem{ag06}
A.~Alpers and P.~Gritzmann.
\newblock On stability, error correction, and noise compensation in discrete
  tomography.
\newblock {\em SIAM J. Discrete Math.}, 20(1):227--239, 2006.
\newblock \href {http://dx.doi.org/10.1137/040617443}
  {\path{doi:10.1137/040617443}}.

\bibitem{agwindowconstraints}
A.~Alpers and P.~Gritzmann.
\newblock Reconstructing binary matrices under window constraints from their
  row and column sums.
\newblock {\em Fund. Inf.}, 155(4):321--340, 2017.
\newblock \href {http://dx.doi.org/10.3233/FI-2017-1588}
  {\path{doi:10.3233/FI-2017-1588}}.

\bibitem{agdynamic}
A.~Alpers and P.~Gritzmann.
\newblock Dynamic discrete tomography.
\newblock {\em Inverse Probl.}, 34(3):034003 (26pp), 2018.
\newblock \href {http://dx.doi.org/10.1088/1361-6420/aaa202}
  {\path{doi:10.1088/1361-6420/aaa202}}.

\bibitem{agsuperresolution}
A.~Alpers and P.~Gritzmann.
\newblock On double-resolution imaging and discrete tomography.
\newblock {\em SIAM J. Discrete Math.}, 32(2):1369--1399, 2018.
\newblock \href {http://dx.doi.org/10.1137/17M1115629}
  {\path{doi:10.1137/17M1115629}}.

\bibitem{agt-01}
A.~Alpers, P.~Gritzmann, and L.~Thorens.
\newblock Stability and instability in discrete tomography.
\newblock In {\em Digital and Image Geometry}, volume 2243 of {\em LNCS 2243},
  pages 175--186. Springer, Berlin, 2001.
\newblock \href {http://dx.doi.org/10.1007/3-540-45576-0_11}
  {\path{doi:10.1007/3-540-45576-0_11}}.

\bibitem{alpers-larman15}
A.~Alpers and D.~G. Larman.
\newblock The smallest sets of points not determined by their {X}-rays.
\newblock {\em Bull. London Math. Soc.}, 47(1):171--176, 2015.
\newblock \href {http://dx.doi.org/10.1112/blms/bdu111}
  {\path{doi:10.1112/blms/bdu111}}.

\bibitem{colorpicapix}
A.~Bains and T.~Biedl.
\newblock Reconstructing $hv$-convex multi-coloured polyominoes.
\newblock {\em Theor. Comput. Sci.}, 411(34-36):3123--3128, 2010.
\newblock \href {http://dx.doi.org/10.1016/j.tcs.2010.04.041}
  {\path{doi:10.1016/j.tcs.2010.04.041}}.

\bibitem{DART3}
S.~Bals, K.~J. Batenburg, J.~Verbeeck, J.~Sijbers, and G.~{Van Tendeloo}.
\newblock Quantitative three-dimensional reconstruction of catalyst particles
  for bamboo-like carbon-nanotubes.
\newblock {\em Nano Lett.}, 7(12):3669--3674, 2007.
\newblock \href {http://dx.doi.org/10.1021/nl071899m}
  {\path{doi:10.1021/nl071899m}}.

\bibitem{barcucciconvex}
E.~Barcucci, A.~{Del Lungo}, M.~Nivat, and R.~Pinzani.
\newblock X-rays characterizing some classes of discrete sets.
\newblock {\em Linear Algebra Appl.}, 339(1-3):3--21, 2001.
\newblock \href {http://dx.doi.org/10.1016/S0024-3795(01)00431-1}
  {\path{doi:10.1016/S0024-3795(01)00431-1}}.

\bibitem{bdfr-17}
E.~Barcucci, P.~Dulio, A.~Frosini, and S.~Rinaldi.
\newblock Ambiguity results in the characterization of {$hv$}-convex
  polyominoes from projections.
\newblock In {\em Discrete Geometry for Computer Imagery}, LNCS 10502, pages
  147--158. Springer, Berlin, 2017.
\newblock \href {http://dx.doi.org/10.1007/978-3-319-66272-5_13}
  {\path{doi:10.1007/978-3-319-66272-5_13}}.

\bibitem{DART1}
K.~J. Batenburg, S.~Bals, S.~Sijbers, C.~Kuebel, P.~A. Midgley, J.~C.
  Hernandez, U.~Kaiser, E.~R. Encina, E.~A. Coronado, and G.~{Van Tendeloo}.
\newblock {3D} imaging of nanomaterials by discrete tomography.
\newblock {\em Ultramicroscopy}, 109(6):730--740, 2009.
\newblock \href {http://dx.doi.org/10.1016/j.ultramic.2009.01.009}
  {\path{doi:10.1016/j.ultramic.2009.01.009}}.

\bibitem{nonograms}
K.~J. Batenburg and W.~A. Kosters.
\newblock A discrete tomography approach to {J}apanese puzzles.
\newblock In {\em Proceedings of the 16th Belgium-Netherlands Conference on
  Artificial Intelligence (BNAIC)}, pages 243--250, 2004.

\bibitem{DART2}
K.~J. Batenburg and J.~Sijbers.
\newblock {DART}: {A} practical reconstruction algorithm for discrete
  tomography.
\newblock {\em IEEE Trans. Image Process.}, 20(9):2542--2553, 2011.
\newblock \href {http://dx.doi.org/10.1109/TIP.2011.2131661}
  {\path{doi:10.1109/TIP.2011.2131661}}.

\bibitem{bianchilonginetti}
G.~Bianchi and M.~Longinetti.
\newblock Reconstructing plane sets from projections.
\newblock {\em Discrete Comput. Geom.}, 5(3):223--242, 1990.
\newblock \href {http://dx.doi.org/10.1007/BF02187787}
  {\path{doi:10.1007/BF02187787}}.

\bibitem{pathpuzzles}
J.~Bosboom, E.~D. Demaine, M.~L. Demaine, A.~Hesterberg, R.~Kimball, and
  J.~Kopinsky.
\newblock Path puzzles: {D}iscrete tomography with a path constraint is hard.
\newblock 2018.
\newblock URL: \url{http://arxiv.org/abs/1803.01176}.

\bibitem{qconvex}
S.~Brunetti and A.~Daurat.
\newblock An algorithm reconstructing convex lattice sets.
\newblock {\em Theor. Comput. Sci.}, 304(1-3):35--57, 2003.
\newblock \href {http://dx.doi.org/10.1016/S0304-3975(03)00050-1}
  {\path{doi:10.1016/S0304-3975(03)00050-1}}.

\bibitem{brunettidaurat08}
S.~Brunetti and A.~Daurat.
\newblock Reconstruction of convex lattice sets from tomographic projections in
  quartic time.
\newblock {\em Theor. Comput. Sci.}, 406(1-2):55--62, 2008.
\newblock \href {http://dx.doi.org/10.1016/j.tcs.2008.06.003}
  {\path{doi:10.1016/j.tcs.2008.06.003}}.

\bibitem{permutationmatrices08}
S.~Brunetti, A.~{Del Lungo}, P.~Gritzmann, and S.~{de Vries}.
\newblock On the reconstruction of binary and permutation matrices under
  (binary) tomographic constraints.
\newblock {\em Theor. Comput. Sci.}, 406(1-2):63--71, 2008.
\newblock \href {http://dx.doi.org/10.1016/j.tcs.2008.06.014}
  {\path{doi:10.1016/j.tcs.2008.06.014}}.

\bibitem{brunettighosts}
S.~Brunetti, P.~Dulio, L.~Hajdu, and C.~Peri.
\newblock Ghosts in discrete tomography.
\newblock {\em J. Math. Imaging Vision}, 53(2):210--224, 2015.
\newblock \href {http://dx.doi.org/10.1007/s10851-015-0571-2}
  {\path{doi:10.1007/s10851-015-0571-2}}.

\bibitem{brunettiboundedsets}
S.~Brunetti, P.~Dulio, and C.~Peri.
\newblock Discrete tomography determination of bounded sets in
  {$\mathbb{Z}^n$}.
\newblock {\em Discrete Appl. Math.}, 183:20--30, 2015.
\newblock \href {http://dx.doi.org/10.1016/j.dam.2014.01.016}
  {\path{doi:10.1016/j.dam.2014.01.016}}.

\bibitem{siltanen1}
M.~Burger, H.~Dirks, L.~Frerking, T.~Hauptmann, A.~Helin, and S.~Siltanen.
\newblock A variational reconstruction method for undersampled dynamic {X}-ray
  tomography based on physical motion models.
\newblock {\em Inverse Probl.}, 33(12):124008, 2017.
\newblock \href {http://dx.doi.org/10.1088/1361-6420/aa99cf}
  {\path{doi:10.1088/1361-6420/aa99cf}}.

\bibitem{chambolle}
A.~Chambolle.
\newblock An algorithm for total variation minimization and applications.
\newblock {\em J. Math. Imaging Vision}, 20(1):89--97, 2004.
\newblock \href {http://dx.doi.org/10.1023/B:JMIV.0000011325.36760.1e}
  {\path{doi:10.1023/B:JMIV.0000011325.36760.1e}}.

\bibitem{chang}
S-K. Chang.
\newblock The reconstruction of binary patterns from their projections.
\newblock {\em Comm. ACM}, 14(1):21--25, 1971.
\newblock \href {http://dx.doi.org/10.1145/362452.362471}
  {\path{doi:10.1145/362452.362471}}.

\bibitem{sudoku}
M.~J. Chlond.
\newblock Classroom exercises in {IP} modeling: {S}u {D}oku and the {L}og
  {P}ile.
\newblock {\em INFORMS Trans. Ed.}, 5(2):77--79, 2005.
\newblock \href {http://dx.doi.org/10.1287/ited.5.2.77}
  {\path{doi:10.1287/ited.5.2.77}}.

\bibitem{daurat05}
A.~Daurat.
\newblock Determination of {Q}-convex sets by {X}-rays.
\newblock {\em Theor. Comput. Sci.}, 332(1-3):19--45, 2005.
\newblock \href {http://dx.doi.org/10.1016/j.tcs.2004.10.001}
  {\path{doi:10.1016/j.tcs.2004.10.001}}.

\bibitem{graphsinDT}
D.~{de Werra}, M.~C. Costa, C.~Picouleau, and B.~Ries.
\newblock On the use of graphs in discrete tomography.
\newblock {\em 4OR}, 6(2):101--123, 2008.
\newblock \href {http://dx.doi.org/10.1007/s10288-008-0077-5}
  {\path{doi:10.1007/s10288-008-0077-5}}.

\bibitem{ferrara}
J.~Diemunsch, M.~Ferrara, S.~Jahanbekam, and J.~M. Shook.
\newblock Extremal theorems for degree sequence packing and the two-color
  discrete tomography problem.
\newblock {\em SIAM J. Discrete Math.}, 29(4):2088--2099, 2015.
\newblock \href {http://dx.doi.org/10.1137/140987912}
  {\path{doi:10.1137/140987912}}.

\bibitem{duliogardner}
P.~Dulio, R.~J. Gardner, and C.~Peri.
\newblock Discrete point {X}-rays.
\newblock {\em SIAM J. Discrete Math.}, 20(1):171--188, 2006.
\newblock \href {http://dx.doi.org/10.1137/040621375}
  {\path{doi:10.1137/040621375}}.

\bibitem{planepartitions}
P.~Dulio and C.~Peri.
\newblock Discrete tomography and plane partitions.
\newblock {\em Adv. Appl. Math.}, 50(3):390--408, 2013.
\newblock \href {http://dx.doi.org/10.1016/j.aam.2012.10.005}
  {\path{doi:10.1016/j.aam.2012.10.005}}.

\bibitem{discretetomographyapp}
C.~D\"urr.
\newblock Discrete tomography applets.
\newblock Accessed: 2018-09.
\newblock URL: \url{http://www-desir.lip6.fr/~durrc/Xray/Complexity/#DGM}.

\bibitem{duerr-Guinez-matamala-12}
C.~D{\"u}rr, F.~Gui{\~n}ez, and M.~Matamala.
\newblock Reconstructing 3-colored grids from horizontal and vertical
  projections is {NP}-hard: {A} solution to the 2-atom problem in discrete
  tomography.
\newblock {\em SIAM J. Discrete Math.}, 26(1):330--352, 2012.
\newblock \href {http://dx.doi.org/10.1137/100799733}
  {\path{doi:10.1137/100799733}}.

\bibitem{flrs-91}
P.~C Fishburn, J.~C Lagarias, J.~A. Reeds, and L.~A. Shepp.
\newblock Sets uniquely determined by projections on axes {II}: {D}iscrete
  case.
\newblock {\em Discrete Math.}, 91(2):149--159, 1991.
\newblock \href {http://dx.doi.org/10.1016/0012-365X(91)90106-C}
  {\path{doi:10.1016/0012-365X(91)90106-C}}.

\bibitem{gale57}
D.~Gale.
\newblock A theorem on flows in networks.
\newblock {\em Pacific J. Math.}, 7(2):1073--1082, 1957.

\bibitem{geometrictomographyapp}
R.~J. Gardner.
\newblock Geometric tomography website.
\newblock Accessed: 2018-09.
\newblock URL: \url{http://www.geometrictomography.com/}.

\bibitem{gardnergritzmann97}
R.~J. Gardner and P.~Gritzmann.
\newblock Discrete tomography: {D}etermination of finite sets by {$X$}-rays.
\newblock {\em Trans. Amer. Math. Soc.}, 349(6):2271--2295, 1997.
\newblock \href {http://dx.doi.org/10.1090/S0002-9947-97-01741-8}
  {\path{doi:10.1090/S0002-9947-97-01741-8}}.

\bibitem{h-GGP96}
R.~J. Gardner, P.~Gritzmann, and D.~Prangenberg.
\newblock On the reconstruction of binary images from their discrete {R}adon
  transform.
\newblock In R.~A. Melter, A.~Y. Wu, and L.~Latecki, editors, {\em Vision
  Geometry V}, SPIE Proc. 2826, pages 121--132. Society of Photo-Optical
  Instrumentation Engineers, Denver, CO, 1996.
\newblock \href {http://dx.doi.org/10.1117/12.251785}
  {\path{doi:10.1117/12.251785}}.

\bibitem{ggp-99}
R.~J. Gardner, P.~Gritzmann, and D.~Prangenberg.
\newblock On the computational complexity of reconstructing lattice sets from
  their {X}-rays.
\newblock {\em Discrete Math.}, 202(1-3):45--71, 1999.
\newblock \href {http://dx.doi.org/10.1016/S0012-365X(98)00347-1}
  {\path{doi:10.1016/S0012-365X(98)00347-1}}.

\bibitem{ggp-2000}
R.~J. Gardner, P.~Gritzmann, and D.~Prangenberg.
\newblock On the computational complexity of determining polyatomic structures
  by {X}-rays.
\newblock {\em Theoret. Comput. Sci.}, 233(1-2):91--106, 2000.
\newblock \href {http://dx.doi.org/10.1016/S0304-3975(97)00298-3}
  {\path{doi:10.1016/S0304-3975(97)00298-3}}.

\bibitem{GarM80}
R.~J. Gardner and P.~McMullen.
\newblock On {H}ammer's {X}-ray problem.
\newblock {\em J. London Math. Soc. (2)}, 21(1):171--175, 1980.
\newblock \href {http://dx.doi.org/10.1112/jlms/s2-21.1.171}
  {\path{doi:10.1112/jlms/s2-21.1.171}}.

\bibitem{gritzmann-devries-2002}
P.~Gritzmann and S.~{de Vries}.
\newblock On the algorithmic inversion of the discrete {R}adon transform.
\newblock {\em Theor. Comput. Sci.}, 281(1-2):455--469, 2002.
\newblock \href {http://dx.doi.org/10.1016/S0304-3975(02)00023-3}
  {\path{doi:10.1016/S0304-3975(02)00023-3}}.

\bibitem{glw11}
P.~Gritzmann, B.~Langfeld, and M.~Wiegelmann.
\newblock Uniqueness in discrete tomography: {T}hree remarks and a corollary.
\newblock {\em SIAM J. Discrete Math.}, 25(4):1589--1599, 2011.
\newblock \href {http://dx.doi.org/10.1137/100803262}
  {\path{doi:10.1137/100803262}}.

\bibitem{gpvw-98}
P.~Gritzmann, D.~Prangenberg, S.~{de Vries}, and M.~Wiegelmann.
\newblock Success and failure of certain reconstruction and uniqueness
  algorithms in discrete tomography.
\newblock {\em Int. J. Imaging Syst. Technol.}, 9(2-3):101--109, 1998.
\newblock \href
  {http://dx.doi.org/10.1002/(SICI)1098-1098(1998)9:2/3<101::AID-IMA6>3.0.CO;2-F}
  {\path{doi:10.1002/(SICI)1098-1098(1998)9:2/3<101::AID-IMA6>3.0.CO;2-F}}.

\bibitem{hahn1}
B.~Hahn.
\newblock Reconstruction of dynamic objects with affine deformations in
  computerized tomography.
\newblock {\em J. Inverse Ill-Posed Probl.}, 22(3):323--339, 2014.
\newblock \href {http://dx.doi.org/10.1515/jip-2012-0094}
  {\path{doi:10.1515/jip-2012-0094}}.

\bibitem{hajdutijdeman}
L.~Hajdu and R.~Tijdeman.
\newblock Algebraic aspects of discrete tomography.
\newblock {\em J. Reine Angew. Math.}, 534:119--128, 2001.
\newblock \href {http://dx.doi.org/10.1515/crll.2001.037}
  {\path{doi:10.1515/crll.2001.037}}.

\bibitem{heppes}
A.~Heppes.
\newblock On the determination of probability distributions of more dimensions
  by their projections.
\newblock {\em Acta Math. Acad. Sci. Hung.}, 7(3-4):403--410, 1956.
\newblock \href {http://dx.doi.org/10.1007/BF02020535}
  {\path{doi:10.1007/BF02020535}}.

\bibitem{bart2}
G.~T. Herman.
\newblock Reconstruction of binary patterns from a few projections.
\newblock In A.~G\"unther, B.~Levrat, and H.~Lipps, editors, {\em International
  Computing Symposium 1973}, pages 371--378. North-Holland, Amsterdam, 1974.

\bibitem{houbenalignment}
L.~Houben and M.~Bar~Sadan.
\newblock Refinement procedure for the image alignment in high-resolution
  electron tomography.
\newblock {\em Ultramicroscopy}, 111(9-10):1512--1520, 2011.
\newblock \href {http://dx.doi.org/10.1016/j.ultramic.2011.06.001}
  {\path{doi:10.1016/j.ultramic.2011.06.001}}.

\bibitem{huckspiess13}
C.~Huck.
\newblock Solution of a uniqueness problem in the discrete tomography of
  algebraic {D}elone sets.
\newblock {\em J. Reine Angew. Math.}, 677:199--224, 2013.
\newblock \href {http://dx.doi.org/10.1515/crelle.2012.026}
  {\path{doi:10.1515/crelle.2012.026}}.

\bibitem{irving-jerrum-94}
R.~W. Irving and M.~R. Jerrum.
\newblock Three-dimensional statistical data security problems.
\newblock {\em SIAM J. Comput.}, 23(1):170--184, 1994.
\newblock \href {http://dx.doi.org/10.1137/S0097539790191010}
  {\path{doi:10.1137/S0097539790191010}}.

\bibitem{rafalET}
C.~L. Jia, S.~B. Mi, J.~Barthel, D.~W. Wang, R.~E. Dunin-Borkowski, K.~W.
  Urban, and A.~Thust.
\newblock Determination of the {3D} shape of a nanoscale crystal with atomic
  resolution from a single image.
\newblock {\em Nat. Mater.}, 13:1044--1049, 2014.
\newblock \href {http://dx.doi.org/10.1038/nmat4087}
  {\path{doi:10.1038/nmat4087}}.

\bibitem{jinschek-08}
J.~R. Jinschek, K.~J. Batenburg, H.~A. Calderon, R.~Kilaas, V.~Radmilovic, and
  C.~Kisielowski.
\newblock 3-{D} reconstruction of the atomic positions in a simulated gold
  nanocrystal based on discrete tomography: {P}rospects of atomic resolution
  electron tomography.
\newblock {\em Ultramicroscopy}, 108(6):589--604, 2008.
\newblock \href {http://dx.doi.org/10.1016/j.ultramic.2007.10.002}
  {\path{doi:10.1016/j.ultramic.2007.10.002}}.

\bibitem{quantitem}
C.~Kisielowski, P.~Schwander, F.~H. Baumann, M.~Seibt, Y.~Kim, and A.~Ourmazd.
\newblock An approach to quantitative high-resolution transmission electron
  microscopy of crystalline materials.
\newblock {\em Ultramicroscopy}, 58(2):131--155, 1995.
\newblock \href {http://dx.doi.org/10.1016/0304-3991(94)00202-X}
  {\path{doi:10.1016/0304-3991(94)00202-X}}.

\bibitem{snark09}
J.~Klukowska, R.~Davidi, and G.~T. Herman.
\newblock {SNARK09 - A} software package for reconstruction of {2D} images from
  {1D} projections.
\newblock {\em Comput. Meth. Prog. Bio.}, 110(3):424--440, 2013.
\newblock \href {http://dx.doi.org/10.1016/j.cmpb.2013.01.003}
  {\path{doi:10.1016/j.cmpb.2013.01.003}}.

\bibitem{kongherman98}
T.~Y. Kong and G.~T. Herman.
\newblock On which grids can tomographic equivalence of binary pictures be
  characterized in terms of elementary switching operations?
\newblock {\em Int. J. Imaging Syst. Technol.}, 9(2-3):118--125, 1998.
\newblock \href
  {http://dx.doi.org/10.1002/(SICI)1098-1098(1998)9:2/3<118::AID-IMA8>3.0.CO;2-E}
  {\path{doi:10.1002/(SICI)1098-1098(1998)9:2/3<118::AID-IMA8>3.0.CO;2-E}}.

\bibitem{kubavolcic88}
A.~Kuba and A.~Vol{\v{c}}i{\v{c}}.
\newblock Characterisation of measurable plane sets which are reconstructable
  from their two projections.
\newblock {\em Inverse Probl.}, 4(2):513--527, 1988.
\newblock \href {http://dx.doi.org/10.1088/0266-5611/4/2/014}
  {\path{doi:10.1088/0266-5611/4/2/014}}.

\bibitem{stabilitylonginetti}
M.~Longinetti.
\newblock Some questions of stability in the reconstruction of plane convex
  bodies from projections.
\newblock {\em Inverse Probl.}, 1(1):87--97, 1985.
\newblock \href {http://dx.doi.org/10.1088/0266-5611/1/1/008}
  {\path{doi:10.1088/0266-5611/1/1/008}}.

\bibitem{lorentz49}
G.~G. Lorentz.
\newblock A problem of plane measure.
\newblock {\em Amer. J. Math.}, 71(2):417--426, 1949.
\newblock \href {http://dx.doi.org/10.2307/2372255}
  {\path{doi:10.2307/2372255}}.

\bibitem{pmaass}
P.~Maass.
\newblock The {X}-ray transform: {S}ingular value decomposition and resolution.
\newblock {\em Inverse Probl.}, 3(4):729--741, 1987.
\newblock \href {http://dx.doi.org/10.1088/0266-5611/3/4/016}
  {\path{doi:10.1088/0266-5611/3/4/016}}.

\bibitem{matousek08}
J.~Matou{\v{s}}ek, A.~P{\v{r}}{\'i}v{\v{e}}tiv{\'y}, and
  P.~{\v{S}}kovro{\v{n}}.
\newblock How many points can be reconstructed from k projections?
\newblock {\em SIAM J. Discrete Math.}, 22(4):1605--1623, 2008.
\newblock \href {http://dx.doi.org/10.1016/j.endm.2007.07.069}
  {\path{doi:10.1016/j.endm.2007.07.069}}.

\bibitem{renyi52}
A.~R{\'e}nyi.
\newblock On projections of probability distributions.
\newblock {\em Acta Math. Acad. Sci. Hungar.}, 3(3):131--142, 1952.
\newblock \href {http://dx.doi.org/10.1007/BF02022515}
  {\path{doi:10.1007/BF02022515}}.

\bibitem{sksbko-93}
P.~Schwander, C.~Kisielowski, F.~H. Baumann, Y.~Kim, and A.~Ourmazd.
\newblock Mapping projected potential, interfacial roughness, and composition
  in general crystalline solids by quantitative transmission electron
  microscopy.
\newblock {\em Phys. Rev. Lett.}, 71(25):4150--4153, 1993.
\newblock \href {http://dx.doi.org/10.1103/PhysRevLett.71.4150}
  {\path{doi:10.1103/PhysRevLett.71.4150}}.

\bibitem{shilferstein}
A.~Shliferstein and Y.~T. Chien.
\newblock Switching components and the ambiguity problem in the reconstruction
  of pictures from their projections.
\newblock {\em Pattern Recogn.}, 10(5-6):327--340, 1978.
\newblock \href {http://dx.doi.org/10.1016/0031-3203(78)90004-3}
  {\path{doi:10.1016/0031-3203(78)90004-3}}.

\bibitem{datacompression}
A.~R. Shliferstein and Y.~T. Chien.
\newblock Some properties of image-processing operations on projection sets
  obtained from digital pictures.
\newblock {\em IEEE Trans. Comput.}, C-26(10):958--970, 1977.
\newblock \href {http://dx.doi.org/10.1109/TC.1977.1674731}
  {\path{doi:10.1109/TC.1977.1674731}}.

\bibitem{slump-gerbrands-82}
C.~H. Slump and J.~J. Gerbrands.
\newblock A network flow approach to reconstruction of the left ventricle from
  two projections.
\newblock {\em Comput. Vision. Graph.}, 18(1):18--36, 1982.
\newblock \href {http://dx.doi.org/0.1016/0146-664X(82)90097-1}
  {\path{doi:0.1016/0146-664X(82)90097-1}}.

\bibitem{solomonwagner}
K.~T. Smith, D.~C. Solmon, and S.~L. Wagner.
\newblock Practical and mathematical aspects of the problem of reconstructing
  objects from radiographs.
\newblock {\em Bull. Amer. Math. Soc.}, 83(6):1227--1270, 1977.
\newblock URL: \url{http://projecteuclid.org/euclid.bams/1183539851}.

\bibitem{svalbe3}
I.~Svalbe and M.~Ceko.
\newblock Maximal {$N$}-ghosts and minimal information recovery from {$N$}
  projected views of an array.
\newblock In {\em Discrete Geometry for Computer Imagery}, LNCS 10502, pages
  135--146. Springer, Cham, 2017.
\newblock \href {http://dx.doi.org/10.1007/978-3-319-66272-5_12}
  {\path{doi:10.1007/978-3-319-66272-5_12}}.

\bibitem{svalbe1}
I.~Svalbe and S.~Chandra.
\newblock Growth of discrete projection ghosts created by iteration.
\newblock In {\em Discrete Geometry for Computer Imagery}, LNCS 6607, pages
  406--416. Springer, Heidelberg, 2011.
\newblock \href {http://dx.doi.org/10.1007/978-3-642-19867-0_34}
  {\path{doi:10.1007/978-3-642-19867-0_34}}.

\bibitem{svalbe0}
I.~Svalbe, N.~Nazareth, N.~Normand, and S.~Chandra.
\newblock On constructing minimal ghosts.
\newblock In {\em 2010 International Conference on Digital Image Computing:
  Techniques and Applications}, pages 276--281. IEEE, Los Alamitos, 2010.
\newblock \href {http://dx.doi.org/10.1109/DICTA.2010.56}
  {\path{doi:10.1109/DICTA.2010.56}}.

\bibitem{svalbe2}
I.~Svalbe and N.~Normand.
\newblock Properties of minimal ghosts.
\newblock In {\em Discrete Geometry for Computer Imagery}, LNCS 6607, pages
  417--428. Springer, Heidelberg, 2011.
\newblock \href {http://dx.doi.org/10.1007/978-3-642-19867-0_35}
  {\path{doi:10.1007/978-3-642-19867-0_35}}.

\bibitem{astratoolbox}
W.~{Van Aarle}, W.~J. Palenstijn, J.~{De Beenhouwer}, T.~Altantzis, S.~Bals,
  K.~J. Batenburg, and J.~Sijbers.
\newblock The {ASTRA} toolbox: {A} platform for advanced algorithm development
  in electron tomography.
\newblock {\em Ultramicroscopy}, 157:35--47, 2015.
\newblock URL: \url{www.astra-toolbox.com}, \href
  {http://dx.doi.org/10.1016/j.ultramic.2015.05.002}
  {\path{doi:10.1016/j.ultramic.2015.05.002}}.

\bibitem{batenburgnature}
S.~{Van Aert}, K.~J. Batenburg, M.~D. Rossell, R.~Erni, and G.~{Van Tendeloo}.
\newblock Three-dimensional atomic imaging of crystalline nanoparticles.
\newblock {\em Nature}, 470(7334):374--376, 2011.
\newblock \href {http://dx.doi.org/10.1038/nature09741}
  {\path{doi:10.1038/nature09741}}.

\bibitem{dalen}
B.~{Van~Dalen}.
\newblock Stability results for uniquely determined sets from two directions in
  discrete tomography.
\newblock {\em Discrete Math.}, 309(12):3905--3916, 2009.
\newblock \href {http://dx.doi.org/10.1016/j.disc.2008.11.018}
  {\path{doi:10.1016/j.disc.2008.11.018}}.

\bibitem{volcicwellposed}
A.~Vol{\v{c}}i{\v{c}}.
\newblock Well-posedness of the {G}ardner-{M}c{M}ullen reconstruction problem.
\newblock In {\em Proceedings of Conference on Measure Theory, Oberwolfach,
  1983, LNM 1089}, pages 199--210, 1984.
\newblock \href {http://dx.doi.org/10.1007/BFb0072615}
  {\path{doi:10.1007/BFb0072615}}.

\bibitem{TVRDART}
X.~Zhuge, W.~J. Palenstijn, and K.~J. Batenburg.
\newblock A more robust algorithm for discrete tomography from limited
  projection data with automated gray value estimation.
\newblock {\em IEEE Trans. Image Process.}, 25(1):455--468, 2016.
\newblock \href {http://dx.doi.org/10.1109/TIP.2015.2504869}
  {\path{doi:10.1109/TIP.2015.2504869}}.

\bibitem{zopf}
S.~Zopf.
\newblock Construction of switching components.
\newblock In {\em Discrete Geometry for Computer Imagery}, LNCS 4245, pages
  157--168. Springer, Berlin, 2006.
\newblock \href {http://dx.doi.org/10.1007/11907350_14}
  {\path{doi:10.1007/11907350_14}}.

\item[]\hspace{-\labelwidth}\hspace{-\labelsep}\textbf{Further reading: Particle tracking}:

\bibitem{adrianlowdensity}
R.~J. Adrian.
\newblock Particle-imaging techniques for experimental fluid mechanics.
\newblock {\em Annu. Rev. Fluid Mech.}, 23:261--304, 1991.
\newblock \href {http://dx.doi.org/10.1146/annurev.fl.23.010191.001401}
  {\path{doi:10.1146/annurev.fl.23.010191.001401}}.

\bibitem{agms-15}
A.~Alpers, P.~Gritzmann, D.~Moseev, and M.~Salewski.
\newblock {3D} particle tracking velocimetry using dynamic discrete tomography.
\newblock {\em Comput. Phys. Commun.}, 187(1):130--136, 2015.
\newblock \href {http://dx.doi.org/10.1016/j.cpc.2014.10.022}
  {\path{doi:10.1016/j.cpc.2014.10.022}}.

\bibitem{DPS17}
R.~Dalitz, S.~Petra, and C.~Schn{\"o}rr.
\newblock {C}ompressed {M}otion {S}ensing.
\newblock In {\em Proc.~SSVM}, volume 10302 of {\em LNCS}, pages 602--613.
  Springer, 2017.
\newblock \href {http://dx.doi.org/10.1007/978-3-319-58771-4_48}
  {\path{doi:10.1007/978-3-319-58771-4_48}}.

\bibitem{Elsinga2006}
G.~E. Elsinga, F.~Scarano, B.~Wieneke, and B.~W. Oudheusden.
\newblock Tomographic particle image velocimetry.
\newblock {\em Exp. Fluids}, 41(6):933--947, 2006.
\newblock \href {http://dx.doi.org/10.1007/s00348-006-0212-z}
  {\path{doi:10.1007/s00348-006-0212-z}}.

\bibitem{biomedicalpiv}
R.~A. Jamison, A.~Fouras, and R.~J. Bryson-Richardson.
\newblock Cardiac-phase filtering in intracardiac particle image velocimetry.
\newblock {\em J. Biomed. Opt.}, 17(3):036007, 2012.
\newblock \href {http://dx.doi.org/10.1117/1.JBO.17.3.036007}
  {\path{doi:10.1117/1.JBO.17.3.036007}}.

\bibitem{batenburg10}
M.~Novara, K.~J. Batenburg, and F.~Scarano.
\newblock Motion tracking-enhanced {MART} for tomographic {PIV}.
\newblock {\em Meas. Sci. Technol.}, 21(3):035401, 2010.
\newblock \href {http://dx.doi.org/10.1088/0957-0233/21/3/035401}
  {\path{doi:10.1088/0957-0233/21/3/035401}}.

\bibitem{elementparticletracking}
J.~F. Pusztaszeri, P.~E. Rensing, and T.~M. Liebling.
\newblock Tracking elementary particles near their primary vertex: {A}
  combinatorial approach.
\newblock {\em J. Global Optim.}, 9(1):41--64, 1996.
\newblock \href {http://dx.doi.org/10.1007/BF00121750}
  {\path{doi:10.1007/BF00121750}}.

\bibitem{Reuss1986}
D.~Reuss, R.~Adrian, and C.~Landreth.
\newblock Two-dimensional velocity measurements in a laminar flame using
  particle image velocimetry.
\newblock {\em Combust. Sci. Technol.}, 67(4-6):73--83, 1986.
\newblock \href {http://dx.doi.org/10.1080/00102208908924062}
  {\path{doi:10.1080/00102208908924062}}.

\bibitem{multitracking10}
F.~C.~R. Spieksma and G.~J. Woeginger.
\newblock Geometric three-dimensional assignment problems.
\newblock {\em European J. Oper. Res.}, 91(3):611--618, 1996.
\newblock \href {http://dx.doi.org/10.1016/0377-2217(95)00003-8}
  {\path{doi:10.1016/0377-2217(95)00003-8}}.

\bibitem{ptvgeoscience}
M.~Umeyama and S.~Matsuki.
\newblock Measurements of velocity and trajectory of water particle for
  internal waves in two density layers.
\newblock {\em Geophys. Res. Lett.}, 38(3):L03612, 2011.
\newblock \href {http://dx.doi.org/10.1029/2010GL046419}
  {\path{doi:10.1029/2010GL046419}}.

\bibitem{Williams2011}
J.~Williams.
\newblock Application of tomographic particle image velocimetry to studies of
  transport in complex (dusty) plasma.
\newblock {\em Phys. Plasmas}, 18(5):050702, 2011.
\newblock \href {http://dx.doi.org/10.1063/1.3587090}
  {\path{doi:10.1063/1.3587090}}.

\bibitem{glidingarc-15}
J.~Zhu, J.~Gao, A.~Ehn, M.~Ald{\'e}n, Z.~Li, D.~Moseev, Y.~Kusano, M.~Salewski,
  A.~Alpers, P.~Gritzmann, and M.~Schwenk.
\newblock Measurements of {3D} slip velocities and plasma column lengths of a
  gliding arc discharge.
\newblock {\em Appl. Phys. Lett.}, 106(4):044101, 2015.
\newblock \href {http://dx.doi.org/10.1063/1.4906928}
  {\path{doi:10.1063/1.4906928}}.

\item[]\hspace{-\labelwidth}\hspace{-\labelsep}\textbf{Further reading: Tomographic grain mapping}:

\bibitem{heise}
A.~Alpers, P.~Gritzmann, C.~G. Heise, and A.~Taraz.
\newblock On the mathematics of grain reconstruction~{I}: {M}odeling and
  computational complexity.
\newblock in preparation, 2018.

\bibitem{apkh-06}
A.~Alpers, H.~F. Poulsen, E.~Knudsen, and G.~T. Herman.
\newblock A discrete tomography algorithm for improving the quality of {3DXRD}
  grain maps.
\newblock {\em J. Appl. Crystallogr.}, 39(4):582--588, 2006.
\newblock \href {http://dx.doi.org/10.1107/S002188980601939X}
  {\path{doi:10.1107/S002188980601939X}}.

\bibitem{grainfollow5}
N.~R. Barton and J.~V. Bernier.
\newblock A method for intragranular orientation and lattice strain
  distribution determination.
\newblock {\em J. Appl. Crystallogr.}, 45(6):1145--1155, 2012.
\newblock \href {http://dx.doi.org/10.1107/S0021889812040782}
  {\path{doi:10.1107/S0021889812040782}}.

\bibitem{DARTgrain}
K.~J. Batenburg, J.~Sijbers, H.~F. Poulsen, and E.~Knudsen.
\newblock {DART}: {A} robust algorithm for fast reconstruction of
  three-dimensional grain maps.
\newblock {\em J. Appl. Crystallogr.}, 43(6):1464--1473, 2010.
\newblock \href {http://dx.doi.org/10.1107/S0021889810034114}
  {\path{doi:10.1107/S0021889810034114}}.

\bibitem{grainfollow3}
Y.~Hayashi, Y.~Hirose, and Y.~Seno.
\newblock Polycrystal orientation mapping using scanning three-dimensional
  {X}-ray diffraction microscopy.
\newblock {\em J. Appl. Crystallogr.}, 48(4):1094--1101, 2015.
\newblock \href {http://dx.doi.org/10.1107/S1600576715009899}
  {\path{doi:10.1107/S1600576715009899}}.

\bibitem{graindeformation}
B.~Jakobsen, H.~F. Poulsen, U.~Lienert, J.~Almer, S.~D. Shastri, H.~O.
  Sorensen, C.~Gundlach, and W.~Pantleon.
\newblock Formation and subdivision of deformation structures during plastic
  deformation.
\newblock {\em Science}, 312(5775):889--892, 2006.
\newblock \href {http://dx.doi.org/10.1126/science.1124141}
  {\path{doi:10.1126/science.1124141}}.

\bibitem{kahkrp-09}
A.~K. Kulshreshth, A.~Alpers, G.~T. Herman, E.~Knudsen, L.~Rodek, and H.~F.
  Poulsen.
\newblock A greedy method for reconstructing polycrystals from
  three-dimensional {X}-ray diffraction data.
\newblock {\em Inverse Probl. Imaging}, 3(1):69--85, 2009.
\newblock \href {http://dx.doi.org/10.3934/ipi.2009.3.69}
  {\path{doi:10.3934/ipi.2009.3.69}}.

\bibitem{graindex}
E.~M. Lauridsen, S.~Schmidt, R.~M. Suter, and H.~F. Poulsen.
\newblock Tracking: {A} method for structural characterization of grains in
  powders or polycrystals.
\newblock {\em J. Appl. Crystallogr.}, 34:744--750, 2001.
\newblock \href {http://dx.doi.org/10.1107/S0021889801014170}
  {\path{doi:10.1107/S0021889801014170}}.

\bibitem{grainfollow4}
H.~Li, N.~Chawla, and Y.~Jiao.
\newblock Reconstruction of heterogeneous materials via stochastic optimization
  of limited-angle {X}-ray tomographic projections.
\newblock {\em Scr. Mater.}, 86(1):48--51, 2014.
\newblock \href {http://dx.doi.org/10.1016/j.scriptamat.2014.05.002}
  {\path{doi:10.1016/j.scriptamat.2014.05.002}}.

\bibitem{grainfollow2}
H.~Li, S.~Kaira, N.~Chawla, and Y.~Jiao.
\newblock Accurate stochastic reconstruction of heterogeneous microstructures
  by limited {X}-ray tomographic projections.
\newblock {\em J. Microsc.}, 264(3):339--350, 2016.
\newblock \href {http://dx.doi.org/10.1111/jmi.12449}
  {\path{doi:10.1111/jmi.12449}}.

\bibitem{DCT}
W.~Ludwig, S.~Schmidt, E.~M. Lauridsen, and H.~F. Poulsen.
\newblock X-ray diffraction contrast tomography: {A} novel technique for
  three-dimensional grain mapping of polycrystals. {I}. {D}irect beam case.
\newblock {\em J. Appl. Crystallogr.}, 41(2):302--309, 2008.
\newblock \href {http://dx.doi.org/10.1107/S002188980800168}
  {\path{doi:10.1107/S002188980800168}}.

\bibitem{graindeformation2}
L.~Margulies, G.~Winther, and H.~F. Poulsen.
\newblock In situ measurement of grain rotation during deformation of
  polycrystals.
\newblock {\em Science}, 291(5512):2392--2394, 2001.
\newblock \href {http://dx.doi.org/10.1126/science.1057956}
  {\path{doi:10.1126/science.1057956}}.

\bibitem{friedel2}
M.~Moscicki, P.~Kenesei, J.~Wright, H.~Pinto, T.~Lippmann, A.~Borbely, and
  A.~R. Pyzalla.
\newblock Friedel-pair based indexing method for characterization of single
  grains with hard {X}-rays.
\newblock {\em Mater. Sci. Eng. A}, 524(1-2):64--68, 2009.
\newblock \href {http://dx.doi.org/10.1016/j.msea.2009.05.002}
  {\path{doi:10.1016/j.msea.2009.05.002}}.

\bibitem{grainnucleation}
S.~E. Offerman, N.~H. {van Dijk}, J.~Sietsma, S.~Grigull, E.~M. Lauridsen,
  L.~Margulies, H.~F. Poulsen, M.~T. Rekveldt, and S.~{van der Zwaag}.
\newblock Grain nucleation and growth during phase transformations.
\newblock {\em Science}, 298(5595):1003--1005, 2002.
\newblock \href {http://dx.doi.org/10.1126/science.1076681}
  {\path{doi:10.1126/science.1076681}}.

\bibitem{fu}
H.~F. Poulsen and X.~Fu.
\newblock Generation of grain maps by an algebraic reconstruction technique.
\newblock {\em J. Appl. Crystallogr.}, 36(4):1062--1068, 2003.
\newblock \href {http://dx.doi.org/10.1107/S0021889803011063}
  {\path{doi:10.1107/S0021889803011063}}.

\bibitem{reischigsurvey}
P.~Reischig, A.~King, L.~Nervo, N.~Vigan{\`{o}}, Y.~Guilhem, W.~J. Palenstijn,
  K.~J. Batenburg, M.~Preuss, and W.~Ludwig.
\newblock Advances in {X}-ray diffraction contrast tomography: {F}lexibility in
  the setup geometry and application to multiphase materials.
\newblock {\em J. Appl. Crystallogr.}, 46(2):297--311, 2013.
\newblock \href {http://dx.doi.org/10.1107/S002188981300260}
  {\path{doi:10.1107/S002188981300260}}.

\bibitem{ks5113}
L.~Rodek, H.~F. Poulsen, E.~Knudsen, and G.~T. Herman.
\newblock A stochastic algorithm for reconstruction of grain maps of moderately
  deformed specimens based on {X}-ray diffraction.
\newblock {\em J. Appl. Crystallogr.}, 40(2):313--321, 2007.
\newblock \href {http://dx.doi.org/10.1107/S0021889807001288}
  {\path{doi:10.1107/S0021889807001288}}.

\bibitem{graingrowth}
S.~Schmidt, S.~F. Nielsen, C.~Gundlach, L.~Margulies, X.~Huang, and
  D.~Juul~Jensen.
\newblock Watching the growth of bulk grains during recrystallization of
  deformed metals.
\newblock {\em Science}, 305(5681):229--232, 2004.
\newblock \href {http://dx.doi.org/10.1126/science.1098627}
  {\path{doi:10.1126/science.1098627}}.

\bibitem{suter}
R.~M. Suter, D.~Hennessy, C.~Xiao, and U.~Lienert.
\newblock Forward modeling method for microstructure reconstruction using
  {X}-ray diffraction microscopy: {S}ingle-crystal verification.
\newblock {\em Rev. Sci. Instr.}, 77(12):123905, 2006.
\newblock \href {http://dx.doi.org/10.1063/1.2400017}
  {\path{doi:10.1063/1.2400017}}.

\bibitem{grainfollow1}
N.~Vigan{\`{o}}, W.~Ludwig, and K.~J. Batenburg.
\newblock Reconstruction of local orientation in grains using a discrete
  representation of orientation space.
\newblock {\em J. Appl. Crystallogr.}, 47(6):1826--1840, 2014.
\newblock \href {http://dx.doi.org/10.1107/S1600576714020147}
  {\path{doi:10.1107/S1600576714020147}}.

\item[]\hspace{-\labelwidth}\hspace{-\labelsep}\textbf{Further reading: Macroscopic grain mapping}:

\bibitem{philmag}
A.~Alpers, A.~Brieden, P.~Gritzmann, A.~Lyckegaard, and H.~F. Poulsen.
\newblock Generalized balanced power diagrams for {3D} representations of
  polycrystals.
\newblock {\em Phil. Mag.}, 95(9):1016--1028, 2015.
\newblock \href {http://dx.doi.org/10.1080/14786435.2015.1015469}
  {\path{doi:10.1080/14786435.2015.1015469}}.

\bibitem{farmland}
A.~Brieden and P.~Gritzmann.
\newblock A quadratic optimization model for the consolidation of farmland by
  means of lend-lease agreements.
\newblock In D.~Ahr, R.~Fahrion, M.~Oswald, and G.~Reinelt, editors, {\em
  Operations Research Proceedings 2003: Selected Papers of the International
  Conference on Operations Research (OR 2003)}, pages 324--331. Springer,
  Heidelberg, 2004.
\newblock \href {http://dx.doi.org/10.1007/978-3-642-17022-5_42}
  {\path{doi:10.1007/978-3-642-17022-5_42}}.

\bibitem{briedengritzmann12}
A.~Brieden and P.~Gritzmann.
\newblock On optimal weighted balanced clusterings: {G}ravity bodies and power
  diagrams.
\newblock {\em SIAM J. Discrete Math.}, 26(2):415--434, 2012.
\newblock \href {http://dx.doi.org/10.1137/110832707}
  {\path{doi:10.1137/110832707}}.

\bibitem{Brieden2017}
A.~Brieden, P.~Gritzmann, and F.~Klemm.
\newblock Constrained clustering via diagrams: {A} unified theory and its
  applications to electoral district design.
\newblock {\em Eur. J. Oper. Res.}, 263(1):18--34, 2017.
\newblock \href {http://dx.doi.org/10.1016/j.ejor.2017.04.018}
  {\path{doi:10.1016/j.ejor.2017.04.018}}.

\bibitem{stoyan}
S.~N. Chiu, D.~Stoyan, W.~Kendall, and J.~Mecke.
\newblock Accompanying web page for the book: {S}tochastic {G}eometry and its
  {A}pplications, 3rd edition.
\newblock [Accessed: 2018-09].
\newblock URL: \url{http://www.math.hkbu.edu.hk/~snchiu/cskm/cskm2013.html}.

\bibitem{sedivy2}
O.~{\v{S}}ediv{\'y}, T.~Brereton, D.~Westhoff, L.~Pol{\'i}vka, V.~Bene{\v{s}},
  V.~Schmidt, and A.~J{\"a}ger.
\newblock 3{D} reconstruction of grains in polycrystalline materials using a
  tessellation model with curved grain boundaries.
\newblock {\em Phil. Mag.}, 96(18):1926--1949, 2016.
\newblock \href {http://dx.doi.org/10.1080/14786435.2016.1183829}
  {\path{doi:10.1080/14786435.2016.1183829}}.

\bibitem{sedivy}
O.~{\v{S}}ediv{\'y}, J.~Dake, C.~E. Krill~III, V.~Schmidt, and A.~J{\"a}ger.
\newblock Description of the {3D} morphology of grain boundaries in aluminum
  alloys using tessellation models generated by ellipsoids.
\newblock {\em Image Anal. Stereol.}, 36(1):5--13, 2017.
\newblock \href {http://dx.doi.org/10.5566/ias.1656}
  {\path{doi:10.5566/ias.1656}}.

\bibitem{spettl}
A.~Spettl, T.~Brereton, Q.~Duan, T.~Werz, C.~E. Krill~III, D.~P. Kroese, and
  V.~Schmidt.
\newblock Fitting {L}aguerre tessellation approximations to tomographic image
  data.
\newblock {\em Phil. Mag.}, 96(2):166--189, 2016.
\newblock \href {http://dx.doi.org/10.1080/14786435.2015.1125540}
  {\path{doi:10.1080/14786435.2015.1125540}}.

\bibitem{teferrarowenhorst18}
K.~Teferra and D.~J. Rowenhorst.
\newblock Direct parameter estimation for generalized balanced power diagrams.
\newblock {\em Phil. Mag. Lett.}, 98(2):79--87, 2018.
\newblock \href {http://dx.doi.org/10.1080/09500839.2018.1472399}
  {\path{doi:10.1080/09500839.2018.1472399}}.

\item[]\hspace{-\labelwidth}\hspace{-\labelsep}\textbf{Further reading: The Prouhet-Tarry-Escott problem}:

\bibitem{adler77}
A.~Adler and S-Y.~R. Li.
\newblock Magic cubes and {P}rouhet sequences.
\newblock {\em Amer. Math. Monthly}, 84(8):618--627, 1977.
\newblock \href {http://dx.doi.org/10.2307/2321011}
  {\path{doi:10.2307/2321011}}.

\bibitem{upolynomial}
J.~Aliste-Prieto, A.~{de Mier}, and J.~Zamora.
\newblock On trees with the same restricted {U}-polynomial and the
  {P}rouhet-{T}arry-{E}scott problem.
\newblock {\em Discrete Math.}, 340(6):1435--1441, 2017.
\newblock \href {http://dx.doi.org/10.1016/j.disc.2016.09.019}
  {\path{doi:10.1016/j.disc.2016.09.019}}.

\bibitem{alpers-tijdeman-07}
A.~Alpers and R.~Tijdeman.
\newblock The two-dimensional {P}rouhet-{T}arry-{E}scott problem.
\newblock {\em J. Number Theory}, 123(2):403--412, 2007.
\newblock \href {http://dx.doi.org/10.1016/j.jnt.2006.07.001}
  {\path{doi:10.1016/j.jnt.2006.07.001}}.

\bibitem{bastien}
L.~Bastien.
\newblock Impossibilit{\'e} de {$u+ v \stackrel{3}{=}x+ y+ z$}.
\newblock {\em Sphinx-Oedipe}, 8(1):171--172, 1913.

\bibitem{thuemorse}
E.~D. Bolker, C.~Offner, R.~Richman, and C.~Zara.
\newblock The {P}rouhet-{T}arry-{E}scott problem and generalized {T}hue-{M}orse
  sequences.
\newblock {\em J. Comb.}, 7(1):117--133, 2016.
\newblock \href {http://dx.doi.org/10.4310/JOC.2016.v7.n1.a5}
  {\path{doi:10.4310/JOC.2016.v7.n1.a5}}.

\bibitem{fewproductgates}
B.~Borchert, P.~Mc{K}enzie, and K.~Reinhardt.
\newblock Few product gates but many zeroes.
\newblock {\em Chicago J. Theoret. Comput. Sci.}, 2013(2):1--22, 2013.
\newblock \href {http://dx.doi.org/10.1007/978-3-642-03816-7_15}
  {\path{doi:10.1007/978-3-642-03816-7_15}}.

\bibitem{cerny}
A.~{\v{C}}ern{\'y}.
\newblock On {P}rouhet's solution to the equal powers problem.
\newblock {\em Theor. Comput. Sci.}, 491(17):33--46, 2013.
\newblock \href {http://dx.doi.org/10.1016/j.tcs.2013.04.001}
  {\path{doi:10.1016/j.tcs.2013.04.001}}.

\bibitem{cerny2}
A.~{\v{C}}ern{\'y}.
\newblock Solutions to the multi-dimensional {P}rouhet-{T}arry-{E}scott
  problem resulting from composition of balanced morphisms.
\newblock {\em Inf. Comput.}, 253(3):424--435, 2017.
\newblock \href {http://dx.doi.org/10.1016/j.ic.2016.06.008}
  {\path{doi:10.1016/j.ic.2016.06.008}}.

\bibitem{newPTE}
A.~Choudhry.
\newblock A new approach to the {T}arry-{E}scott problem.
\newblock {\em Int. J. Number Theory}, 13(2):393--417, 2017.
\newblock \href {http://dx.doi.org/10.1142/S1793042117500233}
  {\path{doi:10.1142/S1793042117500233}}.

\bibitem{purepr}
M.~Cipu.
\newblock Upper bounds for norms of products of binomials.
\newblock {\em LMS J. Comput. Math.}, 7(1):37--49, 2004.
\newblock \href {http://dx.doi.org/10.1112/S1461157000001030}
  {\path{doi:10.1112/S1461157000001030}}.

\bibitem{erdoesszekeres}
P.~Erd{\H o}s and G.~Szekeres.
\newblock On the product {$\prod_{k=1}^n(1-z^{a_k})$}.
\newblock {\em Acad. Serbe Sci. Publ. Inst. Math.}, 13(1):29--34, 1959.
\newblock URL: \url{http://eudml.org/doc/271936}.

\bibitem{escott}
E.~B. Escott.
\newblock The calculation of logarithms.
\newblock {\em Quart. J. Math.}, 41(2):147--167, 1910.

\bibitem{newtonpolygons}
M.~Filaseta and M.~Markovich.
\newblock Newton polygons and the {P}rouhet-{T}arry-{E}scott problem.
\newblock {\em J. Number Theory}, 174(1):384--400, 2017.
\newblock \href {http://dx.doi.org/10.1016/j.jnt.2016.10.009}
  {\path{doi:10.1016/j.jnt.2016.10.009}}.

\bibitem{codes}
V.~Gandikota, B.~Ghazi, and E.~Grigorescu.
\newblock {NP}-hardness of {R}eed-{S}olomon decoding, and the
  {P}rouhet-{T}arry-{E}scott problem.
\newblock In {\em IEEE 57th Annual Symposium on Foundations of Computer Science
  (FOCS)}, volume~1, pages 760--769. IEEE, 2016.
\newblock \href {http://dx.doi.org/10.1109/FOCS.2016.86}
  {\path{doi:10.1109/FOCS.2016.86}}.

\bibitem{goldbachletter}
C.~Goldbach.
\newblock Letter to {E}uler, {J}uly 18, 1750.
\newblock In {\em Corresp. Math. Phys. (ed. Fuss)}, volume~1, pages 525--526.
  St. Petersburg, 1843.
\newblock URL: \url{http://eulerarchive.maa.org/correspondence/}.

\bibitem{integerroots}
S.~Hern{\'a}ndez and F.~Luca.
\newblock Integer roots chromatic polynomials of nonchordal graphs and the
  {P}rouhet-{T}arry-{E}scott problem.
\newblock {\em Graphs Combin.}, 21(3):319--323, 2005.
\newblock \href {http://dx.doi.org/10.1007/s00373-005-0617-0}
  {\path{doi:10.1007/s00373-005-0617-0}}.

\bibitem{kamke}
E.~Kamke.
\newblock Verallgemeinerungen des {W}aring-{H}ilbertschen {S}atzes.
\newblock {\em Math. Ann.}, 83(1-2):85--112, 1921.

\bibitem{kleiman}
H.~Kleiman.
\newblock A note on the {T}arry-{E}scott problem.
\newblock {\em J. Reine Angew. Math.}, 278-279:48--51, 1975.
\newblock \href {http://dx.doi.org/10.1515/crll.1975.278-279.48}
  {\path{doi:10.1515/crll.1975.278-279.48}}.

\bibitem{lehmer}
D.~H. Lehmer.
\newblock The {T}arry-{E}scott problem.
\newblock {\em Scripta Math.}, 13(1):37--41, 1947.

\bibitem{maltby97}
R.~Maltby.
\newblock Pure product polynomials and the {P}rouhet-{T}arry-{E}scott problem.
\newblock {\em Math. Comp.}, 66(219):1323--1340, 1997.
\newblock \href {http://dx.doi.org/10.1090/S0025-5718-97-00865-X}
  {\path{doi:10.1090/S0025-5718-97-00865-X}}.

\bibitem{maltby00}
R.~Maltby.
\newblock A combinatorial identity of subset-sum powers in rings.
\newblock {\em Rocky Mountain J. Mat.}, 30(1):325--329, 2000.
\newblock \href {http://dx.doi.org/10.1216/rmjm/1022008994}
  {\path{doi:10.1216/rmjm/1022008994}}.

\bibitem{r1}
J.~Mc{L}aughlin.
\newblock An indentity motivated by an amazing identity of {R}amanujan.
\newblock {\em Fibonacci. Quart.}, 48(1):34--38, 2010.

\bibitem{melzak}
Z.~A. Melzak.
\newblock A note on the {T}arry-{E}scott problem.
\newblock {\em Canad. Math. Bull.}, 4(3):233--237, 1961.
\newblock \href {http://dx.doi.org/10.4153/CMB-1961-025-1}
  {\path{doi:10.4153/CMB-1961-025-1}}.

\bibitem{myerson86}
G.~Myerson.
\newblock How small can a sum of roots of unity be?
\newblock {\em Amer. Math. Monthly}, 93(6):457--459, 1986.
\newblock \href {http://dx.doi.org/10.2307/2323469}
  {\path{doi:10.2307/2323469}}.

\bibitem{nguyen16}
H.~D. Nguyen.
\newblock A new proof of the {P}rouhet-{T}arry-{E}scott problem.
\newblock {\em Integers}, 16(A1):1--9, 2016.

\bibitem{r2}
P.~A. Panzone.
\newblock On a formula of {S}.~{R}amanujan.
\newblock {\em Amer. Math. Monthly}, 122(1):65--69, 2015.
\newblock \href {http://dx.doi.org/10.4169/amer.math.monthly.122.01.65}
  {\path{doi:10.4169/amer.math.monthly.122.01.65}}.

\bibitem{prouhet51}
M.~E. Prouhet.
\newblock M{\'e}moire sur quelques relations entre les puissances des nombres.
\newblock {\em C. R. Math. Acad. Sci. Paris}, 33:225, 1851.

\bibitem{ptelikep}
T.~N. Sinha.
\newblock A relation between the coefficients and roots of two equations and
  its application to diophantine problems.
\newblock {\em J. Res. Nat. Bur. Standards Sect. B}, 74B(1):31--36, 1970.
\newblock \href {http://dx.doi.org/10.6028/jres.074B.002}
  {\path{doi:10.6028/jres.074B.002}}.

\bibitem{sinha}
T.~N. Sinha.
\newblock A note on a theorem of {L}ehmer.
\newblock {\em J. London Math. Soc.}, s2-4(3):541--544, 1972.
\newblock \href {http://dx.doi.org/10.1112/jlms/s2-4.3.541}
  {\path{doi:10.1112/jlms/s2-4.3.541}}.

\bibitem{tarry}
G.~Tarry.
\newblock Question 4100.
\newblock {\em Interm{\'e}d. Math.}, 19(1):200, 1912.

\bibitem{wright34}
E.~M. Wright.
\newblock An easier {W}aring's problem.
\newblock {\em J. London Math. Soc.}, 9(4):267--272, 1934.
\newblock \href {http://dx.doi.org/10.1112/jlms/s1-9.4.267}
  {\path{doi:10.1112/jlms/s1-9.4.267}}.

\bibitem{wright35}
E.~M. Wright.
\newblock On {T}arry's problem {(I)}.
\newblock {\em Quart. J. Math.}, 6(1):261--267, 1935.
\newblock \href {http://dx.doi.org/10.1093/qmath/os-6.1.261}
  {\path{doi:10.1093/qmath/os-6.1.261}}.

\bibitem{wright59}
E.~M. Wright.
\newblock Prouhet's 1851 solution of the {T}arry-{E}scott problem of 1910.
\newblock {\em Amer. Math. Monthly}, 66(3):199--201, 1959.
\newblock \href {http://dx.doi.org/10.2307/2309513}
  {\path{doi:10.2307/2309513}}.

\bibitem{wright72}
E.~M. Wright.
\newblock The {T}arry-{E}scott and the {``}easier{''} {W}aring problem.
\newblock {\em J. Reine Angew. Math.}, 309:170--173, 1979.
\newblock \href {http://dx.doi.org/10.1515/crll.1979.311-312.170}
  {\path{doi:10.1515/crll.1979.311-312.170}}.

\end{thebibliography}
\end{document}